\documentclass{aastex631}

\usepackage{multirow}
\usepackage{makecell}
\def\cof {$^{12}\mathrm{CO}~(J=1\rightarrow0)$}
\def\cos {$^{13}\mathrm{CO}~(J=1\rightarrow0)$}

\def\largestQ1{G034.4$+$01.0}
\def\cot {$\mathrm{C}^{18}\mathrm{O}~(J=1\rightarrow0)$}

\def\cofs {$^{12}\mathrm{CO}$}
\def\coss {$^{13}\mathrm{CO}$}
\def\cots{$\mathrm{C}^{18}\mathrm{O}$}
\def\deg  {\ifmmode {^\circ}\else {$^\circ$}\fi}

\def\kms     {km~s$^{-1}$}
\newcommand{\HI}{\mbox{H\,\textsc{i}}}%
\newcommand{\HII}{\mbox{H\,\textsc{ii}}}%

\usepackage{booktabs}
\usepackage{chngcntr}

\date{\today}
\submitjournal{ApJS}

\shorttitle{MWISP Data Release 1 }
\shortauthors{Yang et al.} 

\begin{document}

\title{The Milky Way Imaging Scroll Painting Survey: Data Release 1}

\correspondingauthor{Ji Yang}
\email{jiyang@pmo.ac.cn}

\author[0000-0001-7768-7320]{Ji Yang}
\affil{Purple Mountain Observatory and Key Laboratory of Radio Astronomy,\\
 Chinese Academy of Sciences, 10 Yuanhua Road, Qixia District, Nanjing 210033, People's Republic of China}

\author[0000-0003-4586-7751]{Qing-Zeng Yan}
\affil{Purple Mountain Observatory and Key Laboratory of Radio Astronomy,\\
 Chinese Academy of Sciences, 10 Yuanhua Road, Qixia District, Nanjing 210033, People's Republic of China}

  \author[0000-0002-0197-470X]{Yang Su}
\affil{Purple Mountain Observatory and Key Laboratory of Radio Astronomy,\\
 Chinese Academy of Sciences, 10 Yuanhua Road, Qixia District, Nanjing 210033, People's Republic of China}

   \author[0000-0003-2549-7247]{Shaobo Zhang}
\affil{Purple Mountain Observatory and Key Laboratory of Radio Astronomy,\\
 Chinese Academy of Sciences, 10 Yuanhua Road, Qixia District, Nanjing 210033, People's Republic of China}

   \author[0000-0003-2418-3350]{Xin Zhou}
\affil{Purple Mountain Observatory and Key Laboratory of Radio Astronomy,\\
 Chinese Academy of Sciences, 10 Yuanhua Road, Qixia District, Nanjing 210033, People's Republic of China}

   \author[0000-0002-3904-1622]{Yan Sun}
\affil{Purple Mountain Observatory and Key Laboratory of Radio Astronomy,\\
 Chinese Academy of Sciences, 10 Yuanhua Road, Qixia District, Nanjing 210033, People's Republic of China}


 \author[0000-0003-3139-2724]{Yiping Ao}
\affil{Purple Mountain Observatory and Key Laboratory of Radio Astronomy,\\
 Chinese Academy of Sciences, 10 Yuanhua Road, Qixia District, Nanjing 210033, People's Republic of China}

  \author[0000-0003-3151-8964]{Xuepeng Chen}
\affil{Purple Mountain Observatory and Key Laboratory of Radio Astronomy,\\
 Chinese Academy of Sciences, 10 Yuanhua Road, Qixia District, Nanjing 210033, People's Republic of China}

  \author[0000-0003-0849-0692]{Zhiwei Chen}
\affil{Purple Mountain Observatory and Key Laboratory of Radio Astronomy,\\
 Chinese Academy of Sciences, 10 Yuanhua Road, Qixia District, Nanjing 210033, People's Republic of China}
 
  \author[0000-0002-7489-0179]{Fujun Du}
\affil{Purple Mountain Observatory and Key Laboratory of Radio Astronomy,\\
 Chinese Academy of Sciences, 10 Yuanhua Road, Qixia District, Nanjing 210033, People's Republic of China}
 
 \author[0000-0001-8060-1321]{Min Fang}
\affil{Purple Mountain Observatory and Key Laboratory of Radio Astronomy,\\
 Chinese Academy of Sciences, 10 Yuanhua Road, Qixia District, Nanjing 210033, People's Republic of China}

 \author[0000-0002-3866-414X]{Yan Gong}
\affil{Purple Mountain Observatory and Key Laboratory of Radio Astronomy,\\
 Chinese Academy of Sciences, 10 Yuanhua Road, Qixia District, Nanjing 210033, People's Republic of China}

 \author[0000-0002-5920-031X]{Zhibo Jiang}
\affil{Purple Mountain Observatory and Key Laboratory of Radio Astronomy,\\
 Chinese Academy of Sciences, 10 Yuanhua Road, Qixia District, Nanjing 210033, People's Republic of China}

   \author{Shengyu Jin}
\affil{Purple Mountain Observatory and Key Laboratory of Radio Astronomy,\\
 Chinese Academy of Sciences, 10 Yuanhua Road, Qixia District, Nanjing 210033, People's Republic of China}
     \author{Binggang Ju}
\affil{Purple Mountain Observatory and Key Laboratory of Radio Astronomy,\\
 Chinese Academy of Sciences, 10 Yuanhua Road, Qixia District, Nanjing 210033, People's Republic of China}
     \author[0000-0003-2218-3437]{Chong Li}
\affil{Purple Mountain Observatory and Key Laboratory of Radio Astronomy,\\
 Chinese Academy of Sciences, 10 Yuanhua Road, Qixia District, Nanjing 210033, People's Republic of China}

  \author[0000-0001-7526-0120]{Yingjie Li}
\affil{Purple Mountain Observatory and Key Laboratory of Radio Astronomy,\\
 Chinese Academy of Sciences, 10 Yuanhua Road, Qixia District, Nanjing 210033, People's Republic of China}

 \author[0000-0001-7801-3272]{Yi Liu}
\affil{Purple Mountain Observatory and Key Laboratory of Radio Astronomy,\\
 Chinese Academy of Sciences, 10 Yuanhua Road, Qixia District, Nanjing 210033, People's Republic of China}

   \author{Dengrong Lu}
\affil{Purple Mountain Observatory and Key Laboratory of Radio Astronomy,\\
 Chinese Academy of Sciences, 10 Yuanhua Road, Qixia District, Nanjing 210033, People's Republic of China}

  \author{Chunsheng Luo}
\affil{Purple Mountain Observatory and Key Laboratory of Radio Astronomy,\\
 Chinese Academy of Sciences, 10 Yuanhua Road, Qixia District, Nanjing 210033, People's Republic of China}

  \author[0000-0002-8051-5228]{Yuehui Ma}
\affil{Purple Mountain Observatory and Key Laboratory of Radio Astronomy,\\
 Chinese Academy of Sciences, 10 Yuanhua Road, Qixia District, Nanjing 210033, People's Republic of China}

    \author[0000-0002-5551-6501]{Ruiqing Mao}
\affil{Purple Mountain Observatory and Key Laboratory of Radio Astronomy,\\
 Chinese Academy of Sciences, 10 Yuanhua Road, Qixia District, Nanjing 210033, People's Republic of China}

      \author{Jixian Sun}
\affil{Purple Mountain Observatory and Key Laboratory of Radio Astronomy,\\
 Chinese Academy of Sciences, 10 Yuanhua Road, Qixia District, Nanjing 210033, People's Republic of China}

     \author[0000-0001-8923-7757]{Chen Wang}
\affil{Purple Mountain Observatory and Key Laboratory of Radio Astronomy,\\
 Chinese Academy of Sciences, 10 Yuanhua Road, Qixia District, Nanjing 210033, People's Republic of China}

\author[0000-0003-0746-7968]{Hongchi Wang}
\affil{Purple Mountain Observatory and Key Laboratory of Radio Astronomy,\\
 Chinese Academy of Sciences, 10 Yuanhua Road, Qixia District, Nanjing 210033, People's Republic of China}

   \author[0000-0003-4767-6668]{Min Wang}
\affil{Purple Mountain Observatory and Key Laboratory of Radio Astronomy,\\
 Chinese Academy of Sciences, 10 Yuanhua Road, Qixia District, Nanjing 210033, People's Republic of China}

 \author{Min Wang (Qinghai)}
\affil{Purple Mountain Observatory and Key Laboratory of Radio Astronomy,\\
 Chinese Academy of Sciences, 10 Yuanhua Road, Qixia District, Nanjing 210033, People's Republic of China}

 \author{Xindong Wang}
\affil{Purple Mountain Observatory and Key Laboratory of Radio Astronomy,\\
 Chinese Academy of Sciences, 10 Yuanhua Road, Qixia District, Nanjing 210033, People's Republic of China}

\author{Wenting Xu}
\affil{Purple Mountain Observatory and Key Laboratory of Radio Astronomy,\\
 Chinese Academy of Sciences, 10 Yuanhua Road, Qixia District, Nanjing 210033, People's Republic of China}

\author[0000-0001-5602-3306]{Ye Xu}
\affil{Purple Mountain Observatory and Key Laboratory of Radio Astronomy,\\
 Chinese Academy of Sciences, 10 Yuanhua Road, Qixia District, Nanjing 210033, People's Republic of China}

\author{Kun Yan}
\affil{Purple Mountain Observatory and Key Laboratory of Radio Astronomy,\\
 Chinese Academy of Sciences, 10 Yuanhua Road, Qixia District, Nanjing 210033, People's Republic of China}

\author{Ping Yan}
\affil{Purple Mountain Observatory and Key Laboratory of Radio Astronomy,\\
 Chinese Academy of Sciences, 10 Yuanhua Road, Qixia District, Nanjing 210033, People's Republic of China}

  \author[0000-0003-0804-9055]{Lixia Yuan}
\affil{Purple Mountain Observatory and Key Laboratory of Radio Astronomy,\\
 Chinese Academy of Sciences, 10 Yuanhua Road, Qixia District, Nanjing 210033, People's Republic of China}

    \author[0000-0002-6388-649X]{Miaomiao Zhang}
\affil{Purple Mountain Observatory and Key Laboratory of Radio Astronomy,\\
 Chinese Academy of Sciences, 10 Yuanhua Road, Qixia District, Nanjing 210033, People's Republic of China}

    \author{Yongxing Zhang}
\affil{Purple Mountain Observatory and Key Laboratory of Radio Astronomy,\\
 Chinese Academy of Sciences, 10 Yuanhua Road, Qixia District, Nanjing 210033, People's Republic of China}



\begin{abstract}  
We present the first data release (DR1) of the Milky Way Imaging Scroll Painting (MWISP) survey, a mapping in the $J=1\rightarrow0$ transition lines of \cofs, \coss, and \cots\ toward the northern Galactic plane during 2011-2022. The MWISP survey was conducted using the PMO 13.7~m telescope at a  spatial resolution of approximately 50\arcsec\ and a velocity resolution of 0.16 \kms\ at 115 GHz. DR1 fully covered 2310 deg$^2$ within the Galactic   longitude ($l$) and latitude ($b$) range of $9.75\deg\leqslant l \leqslant229.75\deg$ and $|b|\leqslant5.25\deg$. The surveyed area was divided into cell units of $30\arcmin \times30 \arcmin$ for practical purposes and On-The-Fly (OTF) mapping was performed toward each target cell unit. The data were regridded into a regular 3D datacube in $l$–$b$–$V_{\rm LSR}$ with a  pixel size of 30\arcsec\ in $l$–$b$ axes and 0.16 \kms\ in the $V_{\rm LSR}$ axis. The median rms noise is 0.47 K, 0.25 K, and 0.25 K for \cofs, \coss, and \cots, respectively. The equivalent 3$\sigma$ sensitivity in \cofs\  luminosity is approximately 0.23 K \kms, making MWISP the most sensitive survey of its kind.  In this paper, we describe the survey data, including the calibration, data cleaning, data mosaic processes, and the data products. The final mosaicked data cubes contain about $3.33\times$10$^{7}$ spectra (pixels) for each CO isotopologue line. Color composite images, made from the intensities of the isotopologue lines, and some concise descriptions are provided. We constructed a molecular cloud catalog based on the  mosaicked \cofs\ data cube using the clustering algorithm DBSCAN, detecting 103,517 molecular clouds, 10,790 of which exhibit \coss\ emission and 304 of which show \cots\ emission. Based on the histogram of voxel brightness temperature, we estimated a total \cofs\ flux of 7.69$\pm$0.38$\times$$10^7$ K \kms\ arcmin$^2$, 82\% of which is captured by the DBSCAN algorithm.  The properties of molecular clouds show a large dynamic range, facilitating more accurate statistics. The data, together with the cloud sample, provide unique information on molecular gas in the northern Milky Way.

\end{abstract}



\section{Introduction} \label{sec:intro}

Molecular gas is one of the major components of the interstellar medium in the Milky Way. It is widely involved in the fundamental processes within the interstellar medium and star formation. It also serves as an effective indicator of the large-scale structure and motion of our Galaxy. Molecular gas is also deeply coupled with the Galactic ecological  processes driven by various energetic interactions such as ionized \HII\  regions, wind bubbles, and expanding supernova remnants. Although the main component of molecular clouds is $\rm H_2$, due to its low mass and the lack of a permanent electric dipole moment, observations of the bulk molecular clouds often rely on alternative, less abundant molecules, primarily CO \citep{1977IAUS...75...37T, 2015ARA&A..53..583H}.

Since the first detection of space CO emission \citep{1970ApJ...161L..43W}, CO surveys have been continuously conducted
\citep[see][for reviews]{2015ARA&A..53..583H, 2021MNRAS.500.3064S}. In the early days, surveys along the Galactic plane were made across a wide Galactic longitude range with more attention paid to the inner Galaxy  \citep[e.g.,][]{1977ApJ...216..381B,1984ApJ...276..182S}. The surveys provided fundamental knowledge of  molecular cloud distribution and physical properties  \citep[e.g.,][]{1979IAUS...84...35S, 1987ApJ...319..730S,1998ApJS..115..241H,2001ApJ...547..792D}. With the improvement of instruments, a number of new surveys were performed along both the northern and southern Galactic planes, including the CO High-Resolution Survey  \citep[COHRS,][]{2013ApJS..209....8D,2023ApJS..264...16P}, the Mopra Southern Galactic Plane CO Survey \citep{2015PASA...32...20B,2018PASA...35...29B,2023PASA...40...47C}, the Three-mm Ultimate Mopra Milky Way Survey \citep[ThrUMMS,][]{2015ApJ...812....6B}, the CO Heterodyne Inner Milky Way Plane Survey \citep[CHIMPS,][]{2016MNRAS.456.2885R}, the FOREST unbiased Galactic plane imaging survey with the Nobeyama 45 m telescope \citep[FUGIN,][]{2017PASJ...69...78U}, the Forgotten Quadrant Survey \citep[FQS,][]{2020A&A...633A.147B,2021A&A...654A.144B}, and the Structure, Excitation, and Dynamics of the Inner Galactic InterStellar Medium \citep[SEDIGISM,][]{2021MNRAS.500.3064S,2022A&A...658A..54C}.  These surveys provide new findings and more comprehensive details on molecular emission in the Galaxy, making a profound impact on a wide range of areas in astronomy.

Because of the overwhelming amount of molecular gas information carried by CO emission, we at Purple Mountain Observatory (PMO) launched the Milky Way Imaging Scroll Painting (MWISP, PI Ji Yang) project. Compared with the existing CO surveys, the MWISP survey is characterized by several key features,  including multi-line tracers, high sensitivity, large survey area,  full sampling, and moderately high angular resolution. Among the wealth of molecular cloud surveys, \cof, the line with the most extended emission, remains fundamental to molecular cloud studies. Furthermore, the combination of \cofs, \coss, and \cots\ provides an effective tool to probe a wide dynamic range of physical conditions,  such as a density range from $\sim$10$^2$ cm$^{-3}$ to $\sim$10$^4$ cm$^{-3}$. Specifically, the effective critical density of \cofs\ is low ($\sim$10$^2$ cm$^{-3}$), allowing the detection of faint and more extended CO emission, and its large optical depth makes it sensitive to gas temperature. In comparison, the lower optical depth of \coss\  makes it a good tracer of column densities. Furthermore, the even lower optical depth of \cots\ provides a sensor of higher volume densities. The combination of these three isotopologue lines provides a unique combination to census a wide range of physical conditions of molecular clouds.

The MWISP project facilitates a more comprehensive census of molecular clouds in the northern Galactic plane through observations of CO and its isotopologues. The scientific goals  can be summarized in five key aspects: (1) Investigating the distribution and physical properties of molecular clouds; (2) Exploring the internal structure and dynamics of molecular clouds; (3) Studying star formation activities and stellar feedback; (4) Examining structures and components of the Milky Way; (5) Providing molecular data for the multi-wavelength studies of the recycling process of the interstellar matter.

 \begin{deluxetable}{ccccc}
\tablecaption{Individual regions of the MWISP survey. \label{Tab:mwispregion}}
 
\tablehead{
 \colhead{$l$ range} & \colhead{$b$ range}  & \colhead{Angular area} & \colhead{Regions} &  \colhead{Reference} \\
\colhead{(deg) } & \colhead{(deg)} &  \colhead{(deg$^2$)}     &  &  
} 
\startdata
 \makecell{[9.75, 20]}   &   [-5.25, 5.25]   & $\sim$100 & L291, M17, M16, NGC 6604  &   \\  
  \makecell{[20, 45]}   &  [-5.25, 5.25]  & $\sim$250  &   \makecell{Aquila, W44, Phoenix, River}  &  \citet{2017ApJS..230...17S,2020ApJ...893...91S}  \\
 \makecell{[45, 60]}   &   [-5.25, 5.25]   & $\sim$100 & W51,Vul Rift &   \\  
 \makecell{[60, 75]}   &   [-5.25, 5.25]   & $\sim$150 & Vulpecula Rift &  \citet{2023ApJS..267...30L} \\  
 \makecell{[75, 105]}   &   [-5.25, 5.25]   & $\sim$300 &  \makecell{Cyg West, Cyg Rift, Sh2-106, Cyg X,\\ NGC 7027,NGC 7000, Cyg OB7,\\ IC 1396,Sh2-132}  & \citet{2020ApJS..248...15Z}  \\
 \makecell{[105, 120]}   &   [-5.25, 5.25]   & $\sim$150 & \makecell{Sh2-147, Cepheus OB3,\\ NGC 7538, Cas A} &  \citet{2021ApJS..254....3M}  \\   
 \makecell{[120, 130]}   &   [-5.25, 5.25]   & $\sim$100 &  \makecell{Cepheus OB4, Sh2-187} &   \citet{2016ApJS..224....7D}, \citet{2025arXiv250814547Z} \\    
 \makecell{[130, 140]}   &   [-5.25, 5.25]   & $\sim$100 &  W3, W4, W5& \citet{2020ApJS..246....7S}   \\    
 \makecell{[140, 150]}   &   [-5.25, 5.25]   & $\sim$100 &   AFGL 490 &   \citet{2017ApJS..229...24D}  \\    
  \makecell{[150, 170]}   &   [-5.25, 5.25]   & $\sim$200 &  Sh2-212  &  \citet{2017ApJ...838...49X,2022ApJ...938...44G}  \\    
  \makecell{[170, 180]}   &   [-5.25, 5.25]   & $\sim$100 &  Sh2-235, SIM 147  &   \citet{2019ApJ...880...88X}  \\ 
  \makecell{[180, 190]}   &   [-5.25, 5.25]   & $\sim$100 &  GGD4  &    \citet{2017ApJS..230....5W}   \\ 
  \makecell{[190, 200]}   &   [-5.25, 5.25]   & $\sim$100 &  Gem OB1,IC 443B, Sh2-258    &  \citet{2017ApJS..230....5W}   \\ 
  \makecell{[200, 210]}   &   [-5.25, 5.25]   & $\sim$100 & Mon OB1, Rosette &  \citet{2022ApJS..262...16M,2023AJ....165..106W}   \\ 
 \makecell{[210, 220]}   &   [-5.25, 5.25]   & $\sim$100 & Maddalena&   \citet{2019ApJ...885...19Y}   \\ 
  \makecell{[220, 229.75]}   &   [-5.25, 5.25]   & $\sim$100 & CMa OB1  &  \citet{2021ApJS..252...20L}   \\ 
\enddata 
\end{deluxetable}

 The MWISP CO survey provides enhanced molecular cloud samples, enabling detailed statistical studies, including morphology \citep{2021ApJS..257...51Y}, column density distribution \citep{2022ApJS..262...16M}, internal structures \citep{2024AJ....167..207Y,2024ApJ...968L..14Y}, stellar feedback \citep{2023ApJS..268...61Z}, and Galactic-scale molecular gas distribution \citep[e.g.,][]{2015ApJ...798L..27S,2017ApJS..229...24D,2024AJ....167..220Z}. A compilation of individual studies (larger than 10 deg$^2$) from MWISP is summarized in Table \ref{Tab:mwispregion}.

This work presents details on the first data release (DR1) of the MWISP phase I. The DR1 data cover a total angular area of about 2310 deg$^2$ in the northern sky. In addition to the MWISP DR1 data, we also provide a \cofs\ cloud catalog that contains  103,517 samples. The rest of this paper is organized as follows. In Section \ref{sec:data}, we describe the data acquisition process and assess the data quality. In Section \ref{sec:clean}, we illustrate the data reduction process and the data-cleaning procedure. Single cell units and mosaicked data cubes are portrayed in Sections \ref{sec:single} and \ref{sec:mosaic}. A number of CO images from the surveyed areas with descriptions are presented in Section \ref{sec:images}, and the constructed cloud catalog is described in Section \ref{sec:co12cat}. 
In Section \ref{sec:release}, we briefly explain the file structure for the data release. We discuss surface filling factors of the three CO isotopologue lines as well as the completeness of flux in Section \ref{sec:discuss}. The content of this data release is summarized in Section \ref{sec:summary}.

\section{Data Acquisition and Quality Assessment}
\label{sec:data}
The MWISP survey, with features of high sensitivity, multi-line spectroscopy, and wide-sky coverage, was conducted using the PMO 13.7~m millimeter telescope equipped with the Superconducting Spectroscopic Array Receiver (SSAR) system. At the frequency of 110$-$115~GHz, the telescope's half-power beam width (HPBW) is about 49$\arcsec$$-$51$\arcsec$. The SSAR system is composed of a 3$\times$3 sideband-separating mixer array in the front-end, followed by 3$\times$3$\times$2 (=18) IF chains, and an array of Fast Fourier Transform Spectrometer (FFTS) with 16,384 channels for each IF chain of 1~GHz instantaneous bandwidth, achieving a spectral resolution of 61 kHz, equivalent to  $\sim$0.16 \kms\ at 115 GHz. Details on the technical aspects of the SSAR system can be found in  \citet{2012ITTST...2..593S}.

As a multi-line survey, the $J=1\rightarrow0$ transition lines of \cofs, \coss, and \cots\ at  rest frequencies of 115.271~GHz, 110.201~GHz, and 109.782~GHz, respectively, are simultaneously detected by SSAR.  With the carefully chosen LO frequency at 112.6 GHz and IF center frequency at 2.64~GHz, the \cofs\ line is put into the upper sideband, and the \coss\ and \cots\ lines into the lower sideband.  The first LO maintains real-time Doppler tracking at the \cofs\ line frequency.  A fine-tuned frequency tracking is further done at the 2nd LO to match the observed frequencies of \coss\ and \cots\ lines. The simultaneous observation of the three lines by a single telescope provides a unique advantage in canceling the possible instrument effects and minimizing the fluctuation among the isotopologue line intensities \citep{1999AcApS..19...55Y}. 

\begin{figure} 
 \plotone{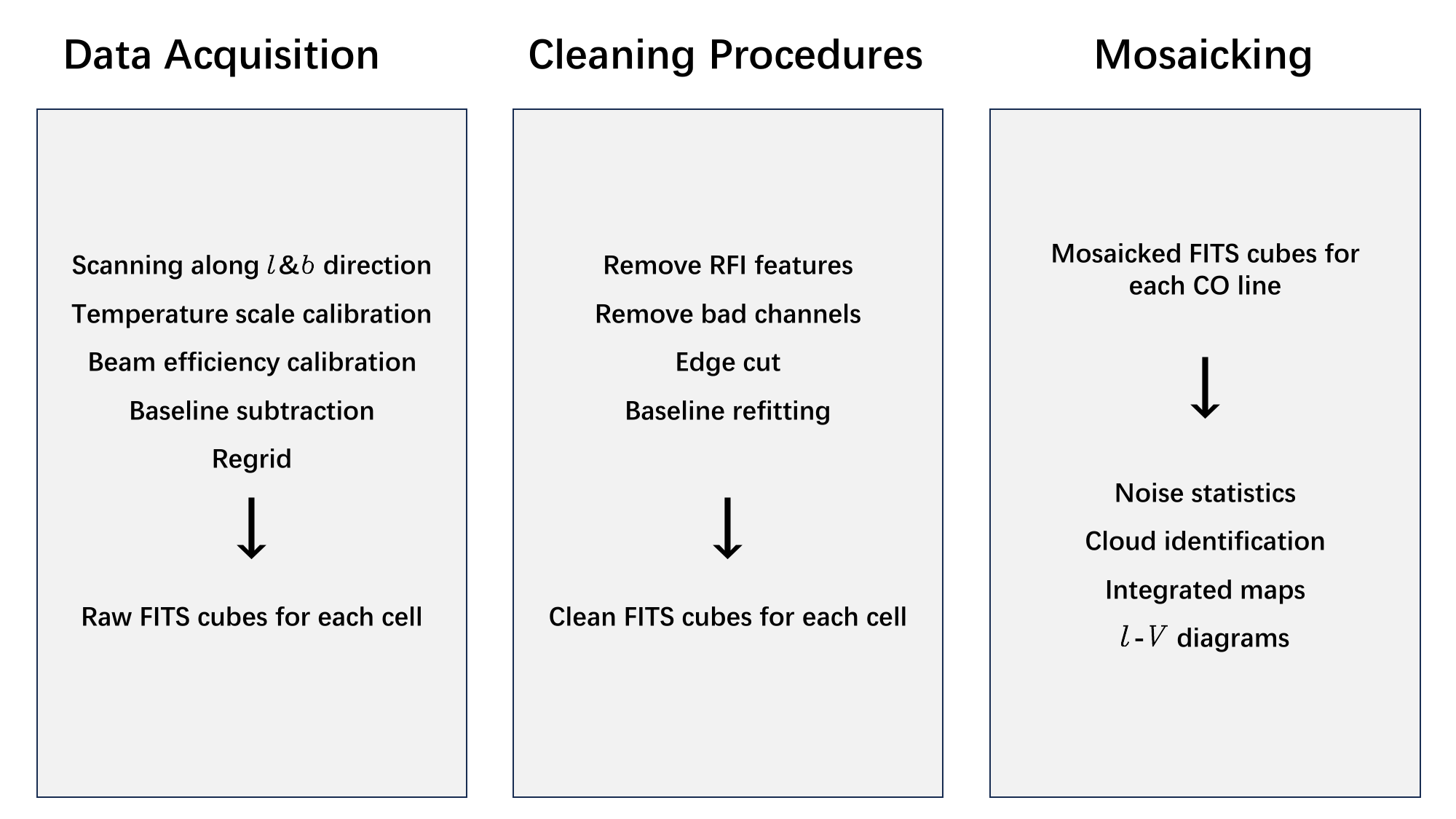} 
\caption{An overview of the data reduction process of the MWISP survey. \label{fig:sketch}}  
\end{figure}

To achieve uniform and full sampling with high mapping efficiency, we developed the On-The-Fly (OTF) mapping scheme for the telescope \citep{2018AcASn..59....3S} and applied it to the MWISP survey. In order to complete the mapping within a reasonably short period of time, the target area of the survey was divided into 9240 sky cell units, each covering an area of 30\arcmin$\times$30\arcmin. The cell unit is also called tile \citep[e.g.,][]{2013ApJS..209....8D} or segment \citep[e.g.,][]{2018PASA...35...29B} in other CO surveys. To minimize scanning artifacts, both $l$ and $b$ directions were scanned. Reference positions of blank sky were selected and confirmed absent of CO emission. The reference positions for all cells are included as part of the DR1.

The scanning data of each unit were processed and regridded into a 3D data cube using the GILDAS software package \citep{2005sf2a.conf..721P}. The size of the Gaussian kernel used in regridding is one-third of the beam size. The regridded datacube was arranged to be 45 \arcmin$\times$45\arcmin\ to host the emissions out-of-bounds footprint of the receiver, taking into account the varying orientation \& field rotation in the sky, and leave 7.5\arcmin\ extension at each side for  diminishing overlap fringe between sky units during mosaicking.  The size of a single voxel is 30\arcsec$\times$30\arcsec\ in the $l$-$b$ plane and about 0.16 \kms\ along the velocity (frequency) axis. The total velocity coverage is about 2600 \kms, exceedingly large for the Galactic CO emission. The target noise level for the survey was $\sim$0.5~K for the \cofs\ line and $\sim$0.3~K for the \coss\ and \cots\ lines. More details on the observing procedures were explained in \citet{2019ApJS..240....9S}.

 Observations of DR1 spanned approximately one decade, from 2011 to 2022. In total, DR1 covers an angular area of 2310 deg$^2$ in the northern sky, including the Galactic range of $9.75\deg\leqslant l \leqslant229.75\deg$ and $|b|\leqslant5.25\deg$. In this section, we detail the MWISP data processing.  A schematic overview of the complete data collection and processing workflow is presented in Figure \ref{fig:sketch}. The pipeline comprises three parts: data acquisition, cleaning procedures, and mosaicking. 

\subsection{Intensity calibration and temperature scale}

For the MWISP survey, we adopted the standard chopper wheel calibration method \citep{1973ARA&A..11...51P,1976ApJS...30..247U}. The obtained temperature scale is the "antenna temperature" after correcting for atmospheric absorption and ohmic loss, which is referred to as $T_A^*$ in the literature. At both the beginning and the end of each scan, an ambient temperature chopper wheel with real-time temperature sensor was switched to the receiver beams to provide a blackbody intensity scale.  In the sideband separation receiving scheme, a calibration error of less than 10\% can be achieved under the conditions that, with the atmospheric opacity at the specific site, the image rejection ratios are $>$10~dB and the gain compression was tuned reasonably small. For further explanation of  calibrations we refer readers to  \citet{1999AcApS..19...55Y} and \citet{2012ITTST...2..593S}.

For the cases of extended objects like molecular clouds, this temperature scale needs to be further corrected for the main-beam efficiency of the telescope, $\eta_{\rm MB}$, to obtain the main-beam brightness temperature, $T_{\rm MB}$.  $T_{\rm MB}$ is  comparable among telescopes of similar beam sizes, according to the relation $T_{\rm MB}=T_A^*/\eta_{\rm MB}$ \citep[see also][]{2019ApJS..240....9S}, where $\eta_{\rm MB}$ represents the convolution efficiency of the source brightness  temperature distribution and the main beam of the telescope. $\eta_{\rm MB}$ was measured for each beam by observing Jupiter, whose brightness temperatures and angular size are known. The annual status report of the PMO 13.7~m telescope provides measured results of $\eta_{\rm MB}$ for the corresponding season. The conversion from $T_A^*$ to $T_{\rm MB}$ was applied to individual OTF scans and all the calibrated MWISP data are denoted in the main-beam temperature scale, $T_{\rm MB}$.

As an attempt to verify the intensity scale, \citet{2021ApJS..256...32S} performed an elaborate comparison of the main beam brightness temperatures between MWISP and those of the Boston University-Five College Radio Astronomy Observatory Galactic Ring Survey  \citep[GRS,][]{2006ApJS..163..145J}. After aligning the spatial and angular resolutions, they found that the temperatures of the two surveys exhibit a consistent linear relation, $T_{\rm MB}(\rm MWISP)  =(0.94\pm0.001)\mathit{T}_{\rm MB}(\rm GRS)+(0.08\pm0.001)$  \citep[see][for plots and detailed analyses]{2021ApJS..256...32S}. The slope is close to 1.0 and the intercept is close to zero, indicating that the calibration procedure of   MWISP is appropriate.

\begin{figure} 
     \centering
     \includegraphics[width = 0.8\linewidth]{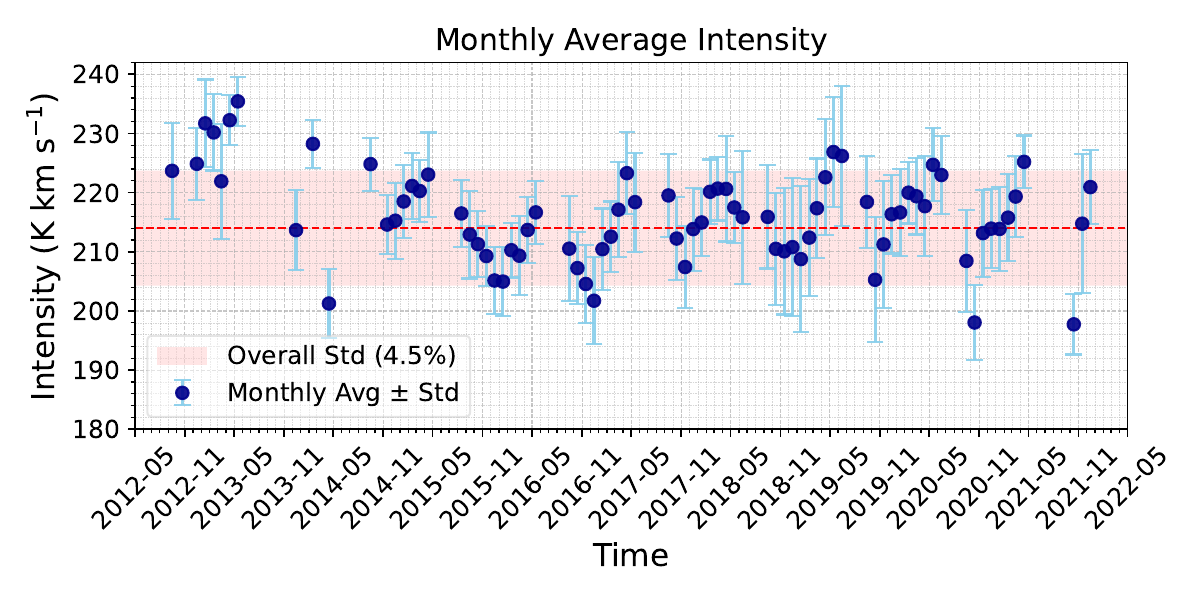}
     \caption{Monthly-averaged integrated intensity of $^{12}$CO ($J=1\rightarrow0$) emission of NGC 2264 during the MWISP survey. The error bars represent the standard deviation of the integrated intensity for each month. The red dashed line and shaded area stand for the overall mean and standard deviation of the integrated intensity, respectively.} 
     \label{fig_bzy}
\end{figure}

\begin{deluxetable*}{cccccccccc}
\setlength{\tabcolsep}{1pt}

\tablecaption{Summary of line monitoring for spectral-line reference sources
\label{tab_bzy}}
\tablewidth{0pt}
\tablehead{
\colhead{Sources} & \colhead{$l$} & \colhead{$b$} & \colhead{$N_{\rm spectra}$} & \colhead{V-Range} & \colhead{W$_{\rm CO}$ Range} &  \colhead{ $\overline{\rm W(\rm CO)}$ }  &  \colhead{ Dispersion(percentage) } & \colhead{$\sigma_{\rm (V-shift)}$} \\
\colhead{} & \colhead{(\arcdeg)} & \colhead{(\arcdeg)}  & \colhead{}   & \colhead{(km s$^{-1}$)} &
\colhead{(K km s$^{-1}$)} & \colhead{(K km s$^{-1}$)}  & \colhead{K km s$^{-1}$} &   \colhead{(km s$^{-1}$)} \\
\colhead{(1)} & \colhead{(2)} & \colhead{(3)} & \colhead{(4)} & 
\colhead{(5)} & \colhead{(6)}  & \colhead{(7)} & \colhead{(8)} & \colhead{(9)}
}
\startdata 
L134N & 4.229 & 35.782 & 11197 & [-3, 7] & [5.26, 15.68] & 10.62 & 1.20 (11.3\%)  & 0.03 \\
W51D & 49.491 & $-$0.369 & 20782 & [40, 80] & [479.09, 707.50] & 620.27 & 29.04(4.7\%) & 0.09 \\
DR21 & 81.682 & 0.540 & 184 & [$-$40, 25] & [407.80, 513.28] & 448.08 & 17.90(4.0\%) & -- \\  
Sh2-140 & 106.799 & 5.312 & 17097 & [$-$40, 10] & [246.57, 401.18] & 324.32 & 16.25(5.0\%) & 0.11 \\
W3(OH) & 133.948 & 1.064 & 9496 & [$-$60, $-$35] & [175.90, 238.28] & 211.23 & 7.62(3.6\%) & 0.23 \\
NGC2264 & 203.316 & 2.055 & 21691 & [$-$10, 25] & [169.84, 251.70] & 214.01 & 9.73(4.5\%) & 0.07 \\
\enddata
\tablecomments{Column 1: identifier of reference sources.  Column 4: number of spectra used to calculate the average integrated intensity.  Column 5: velocity  (kinematic $V_{\rm LSR}$) range calculating integrated intensity. Column 6: minimum and maximum value of integrated intensity of all the spectra. Column 7: averaged integrated intensity.  Column 8: standard deviation and percentage of integrated intensity. Column 9: the dispersion of velocity shift. Note that statistics in this table were based on the $^{12}$CO ($J=1\rightarrow0$) line data.}
\end{deluxetable*}

\subsection{The long-term stability of temperature scale and radial velocity}

In millimeter-wavelength observations, spectral line sources with known intensity and line profiles are commonly used to monitor the correctness and stability of the whole observing system \citep{1981ApJ...250..341K}. For the MWISP survey, six reference sources were routinely monitored during the survey and their coordinates and other relevant information are provided in Table \ref{tab_bzy}.  At the beginning and end of each mapping unit, a nearby reference source was observed in single-point mode using beams 1-9 sequentially, resulting in a check period of about every 3 hours. 

Line intensity is the primary concern in estimating the long-term stability. Among all monitored sources, L134N exhibits a larger relative intensity dispersion compared to the other sources. This is due to the extremely low source elevation frequently occurring at the beginning of observation each day. The rest of the sources all show an intensity dispersion within 5\%. As an example, Figure \ref{fig_bzy} illustrates the time variation in the integrated intensity of NGC 2264 on a monthly basis during the survey period. Overall, during the 2012-2013 observational season, the monthly-averaged  integrated intensity of NGC 2264 was relatively higher, exceeding the 10-year dispersion in most months. Apart from this period, the monthly-averaged intensity of this source remained stable, exhibiting deviations of $\leq$5\% from the average value. The integrated intensity of NGC 2264 also exhibits some periodic fluctuations. The integrated intensity variations of other reference sources are similar. 

\begin{figure} 
     \centering
     \includegraphics[width = 0.9\linewidth]{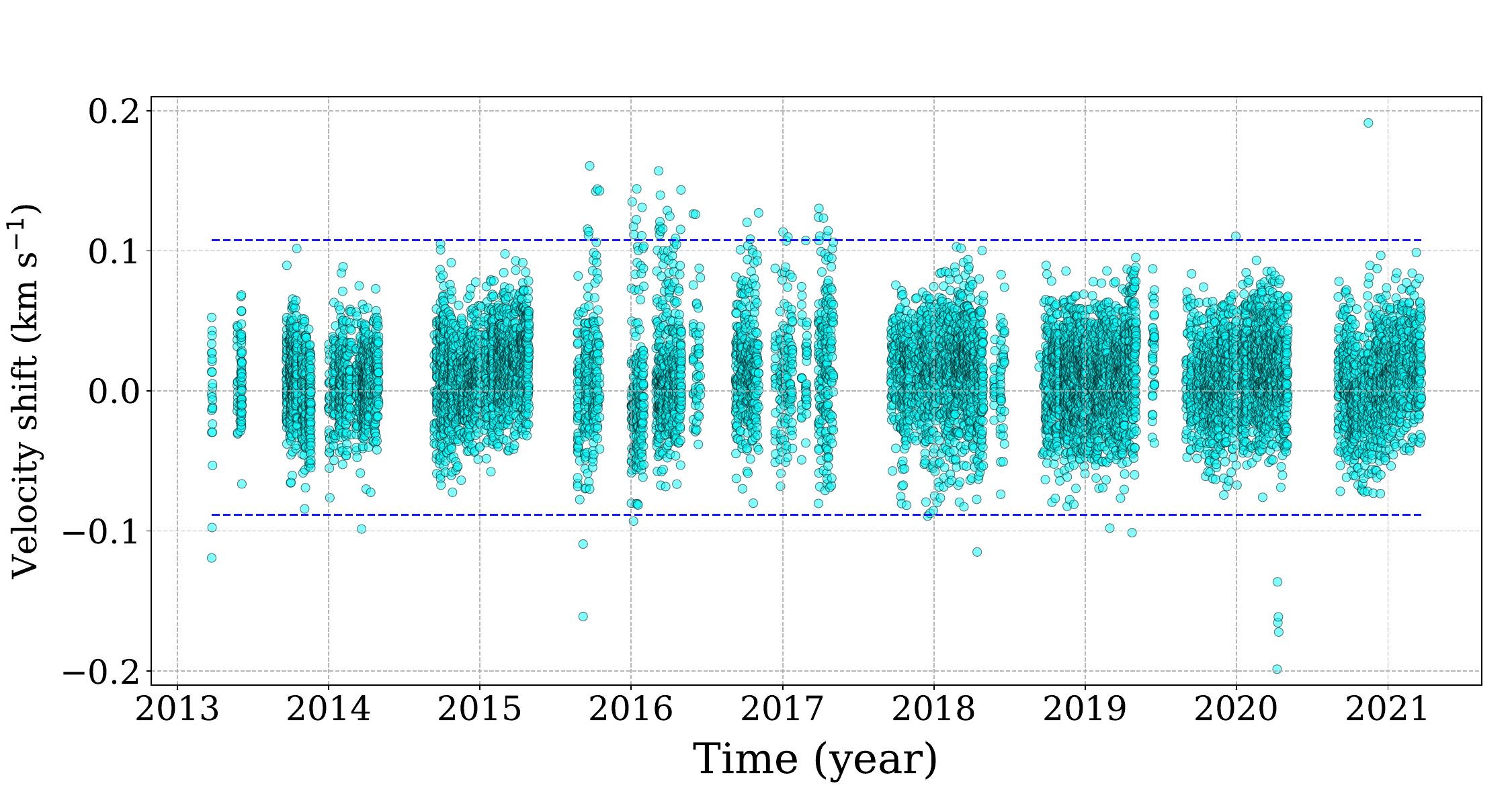}
     \caption{The measured velocity shifts of the reference source L134N from 2013 to 2021. Each cyan dot represents the velocity shift of a $^{12}$CO spectral line of L134N. The blue dashed lines indicate the thresholds of -0.09 km s$^{-1}$ and 0.11 km s$^{-1}$, which are established as the boundaries for identifying the extreme outliers.} 
     \label{L134}
\end{figure}

The long-term stability of radial velocities (kinematic $V_{\rm LSR}$) is another metric for the MWISP data quality. We employ the cross-correlation method to examine the velocity shifts among the $^{12}$CO spectral lines of the standard sources. The standard deviations ($\sigma$) for these measured velocity shifts are listed in column (9) of Table \ref{tab_bzy}.  
According to the statistics, L134N shows the lowest 1$\sigma$ velocity variation of 0.03 km s$^{-1}$, due largely to its narrow intrinsic line width. W3(OH) shows the highest $\sigma$ of 0.23 km s$^{-1}$, followed by S140 with 0.11 km s$^{-1}$, while the remaining sources have values below 0.1 km s$^{-1}$.
Figure \ref{L134} illustrates the measured velocity shifts of L134N from 2013 to 2021, where the values are predominantly below 0.1 km s$^{-1}$.  
We note that, over the survey period,  radial velocities of W51D, Sh2-140, W3(OH), and NGC~2264 exhibit annual fluctuations with small amplitudes of less than 0.2 km s$^{-1}$. 

\section{Data Processing and Cleaning}
\label{sec:clean}

Large surveys are occasionally affected by the radio environments, instrument flaws, or data reduction imperfections. To mitigate these effects and enhance the data quality, we further processed the cell cubes, including removing radio frequency interference (RFI), correcting baselines, and cleaning bad channels. Separate papers will elaborate on the details of those processes; here, we summarize the key steps.  
\subsection{Removing RFI}

RFI is characterized by its transient nature, appearing as sharp, localized jumps in a single scan rather than remaining fixed on the sky across repeating observations. As a result, the spectra affected by RFI are few in number and spatially random, and can be identified based on their distinctive features. RFI can be categorized as broad-band or narrow-band based on its velocity width, and dedicated procedures were developed to detect each type. Regarding the radial velocity, RFI emission in the upper sideband is far from the Galactic \cofs\ emission, so it is simply ignored. However, those in the lower sideband are close to the signal range of \coss\ and \cots, and therefore need to be dealt with carefully. In total, 5139 \coss\ and 73805 \cots\ spectra were identified as contaminated. The affected spectral channels were corrected by replacing them with white noise for isolated RFI, or with the average of uncontaminated scans when the RFI overlapped with the  astronomical signal. 

 \begin{deluxetable}{ccc}
\tablecaption{Signal and baseline velocity ranges.  \label{Tab:vrange}}
 
\tablehead{
\colhead{$l$} & \colhead{Signal range} & \colhead{Baseline range}    \\
\colhead{ (deg)  } & \colhead{(\kms)} & \colhead{(\kms)}   
} 
\startdata
 9.75 $\leq l \leq$ 13  & [-90, +220]  & [-150, 280]     \\ 
13 $< l \leq$ 20  &   [-90, +190]  & [-150, 250] \\ 
20 $< l \leq$ 30  &    [-100, +150] & [-160, 210]  \\ 
30 $< l \leq$ 40  &    [-110, +140]   & [-170, 200] \\ 
40 $< l \leq$ 50  &     [-120, +110] & [-180, 170] \\ 
50 $< l \leq$ 60  &   [-130, +90] &  [-190, 150]  \\ 
60 $< l \leq$ 70  &     [-140, +60] &[-200, 120] \\ 
70 $< l \leq$ 80  &  [-140, +60] &   [-200, 120] \\ 
80 $< l \leq$ 90  & [-140, +50]  &  [-200, 110]  \\ 
90 $< l \leq$ 100  &  [-140, +50]  &  [-200, 110]  \\ 
100 $< l \leq$ 110  & [-140, +50]  & [-200, 110]   \\ 
110 $< l \leq$ 120  &  [-140, +50]  & [-200, 110]   \\ 
120 $< l \leq$ 130  &  [-140, +50]  &  [-200, 110]  \\ 
130 $< l \leq$ 140  &   [-130, +40] &  [-190, 100]  \\ 
140 $< l \leq$ 150  & [-120, +40]  &   [-180, 100]  \\ 
150 $< l \leq$ 160  & [-100, +40]  &   [-160, 100]  \\ 
160 $< l \leq$ 170  &  [-80, +40]  &  [-140, 100]  \\ 
170 $< l \leq$ 180  &  [-70, +40] & [-130, 100]   \\ 
180 $< l \leq$ 190  &   [-60, +50] &  [-120, 110]  \\ 
190 $< l \leq$ 200  &   [-50, +60] &  [-110, 120]  \\ 
200 $< l \leq$ 210  & [-40, +70]  &  [-100, 130]   \\ 
210 $< l \leq$ 220  &   [-40, +90] &  [-100, 150]  \\ 
220 $< l \leq$ 229.75  & [-30, +100]  &  [-90, 160]   \\ 
\enddata
 
\end{deluxetable}

\subsection{Correcting baselines} 
\label{sec:baseline}

After processing RFI effects, we corrected the baselines of CO spectra. To avoid any non-linear distortion, we only removed linear gradients. The velocity range of raw spectra is exceedingly large, which is unfavorable for achieving flat baselines. To address this, we defined the signal and baseline velocity ranges for the Galactic CO emission, and cropped the data cube accordingly. The signal velocity range was empirically determined based on the velocity of Galactic \HI\ emission, and the corresponding baseline range was defined by extending the signal velocity range by 60 \kms\ at both the negative and positive velocity ends. Due to Galactic rotation, this radial velocity range varies with Galactic longitude, as demonstrated in Table \ref{Tab:vrange}. 

The linear gradient was fitted with the 120 \kms\ spectral segments in the baseline range, 60 \kms\ at each of the negative and positive velocity ends for each spectrum. To exclude unexpected contamination or bad channels within the spectra, we removed data points deviating by more than 2.5$\sigma$ from the baseline. Consequently, this baseline correction is an iterative operation, terminating when the slopes and intercepts converge, i.e., when there is a change of less than 1\% relative to the previous iteration. All voxels outside the baseline velocity range were masked to NaN (Not a Number).

\begin{figure}[ht]
\plotone{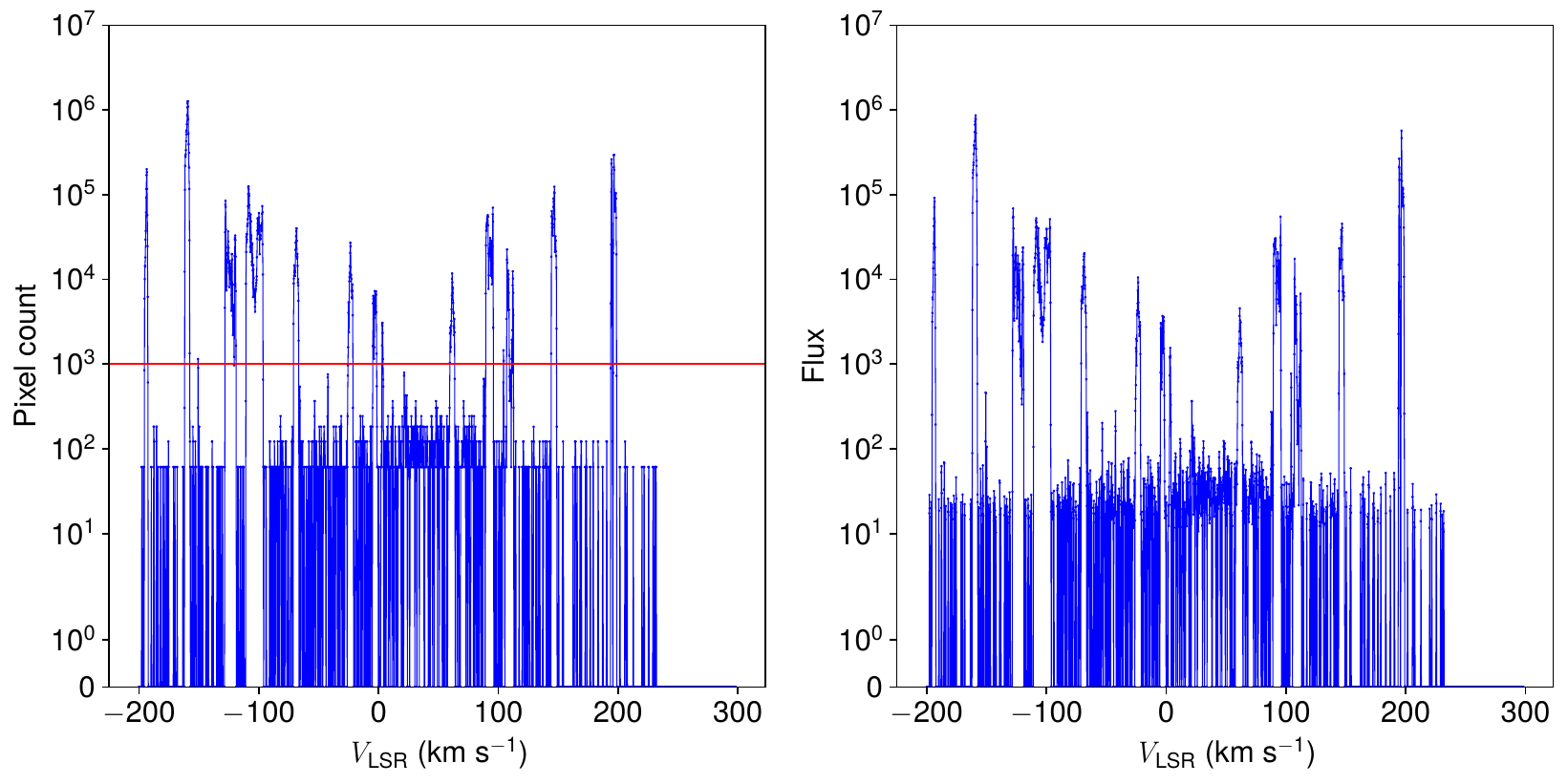} 
\caption{Counts (left) and flux (right) distributions of bad channels for \cots\ over the whole 2310 deg$^2$.  Bad channels are identified by searching stripes along $l$ or $b$ directions. A threshold of 1000 pixels is used to identify velocity channels that contain flawed spectra.  \label{fig:badchannel}}  
\end{figure}

\subsection{Removing bad channels}

The third step involves identifying and removing bad channels caused by receiver malfunctions. A small number of channels did not work properly, producing narrow, channel-fixed,  and intensity-varying spikes in the spectra. To mitigate this effect, we located those channels and replaced their values using linear spline interpolation.   

 Strong bad channels were removed at the beginning of raw data processing, and this section addresses the weak bad channels. Although correcting for Doppler effects may shift the positions of the bad channels, their patterns remain evident due to the short scanning period, during which the channels remain fixed. 

Bad channels are pinpointed by using a spike detection algorithm. In data cubes, these artifacts manifest as stripes along $l$ or $b$ directions, identifiable by detecting spikes in the spectra averaged along the $l$ or $b$ directions. 

To avoid misidentifying narrow CO emission as bad channels, we apply an additional criterion that counts the number of adjacent pixels above a specific threshold in vertical, horizontal, and diagonal directions. Statistically, bad channels are characterized by an excess in the vertical or horizontal directions, distinguishing them from astrophysical signals. 

The bad channel removal procedure works well with \cofs\ due to its relatively wide velocity dispersion. However, caution is required for the \coss\ and \cots\ data, as their narrow radial velocity dispersion makes their emission resemble spikes, which can easily be smeared by interpolation. Consequently, to preserve the integrity of \coss\ and \cots\ data, we use \cofs\ and \coss\ as their references, respectively. Specifically,  bad channels of \coss\ outside the cloud regions (see Section \ref{sec:dbscan} for the definition of molecular clouds with DBSCAN) were processed with the same procedure applied to \cofs.  Within the cloud region, only bad channels that are clearly identified were processed. For \cots, the same procedure was applied but uses \coss\ as a reference.

Bad channels in \cofs\ and \coss\ have been efficiently removed, and no obvious noise structures (typically horizontal lines at specific velocities) are present in the $l$-$V$ plot of DBSCAN-identified clouds or clumps. As for \cots, many bad channels remain in the data cube, but we have chosen not to pursue additional processing. First, most bad channels are beyond the \coss\ emission region. Secondly, they could potentially serve as training datasets for supervised machine learning applications, such as noise pattern recognition or artifact correction.

We show the distribution of \cots\ bad channels in Figure \ref{fig:badchannel}.  After applying the three steps described above, we identified a few minor issues through visual inspection.  For example, one spectrum was incomplete and non-linear baselines were present.  Not all issues were fixed, and due to limited computational resources and time, we stopped processing when the data became acceptable.

After these reductions, we mosaicked all the cells into three large data cubes, corresponding to three CO lines. The DR1 contains both the individual cell cubes and the mosaicked data cubes.

\begin{figure} 
\plotone{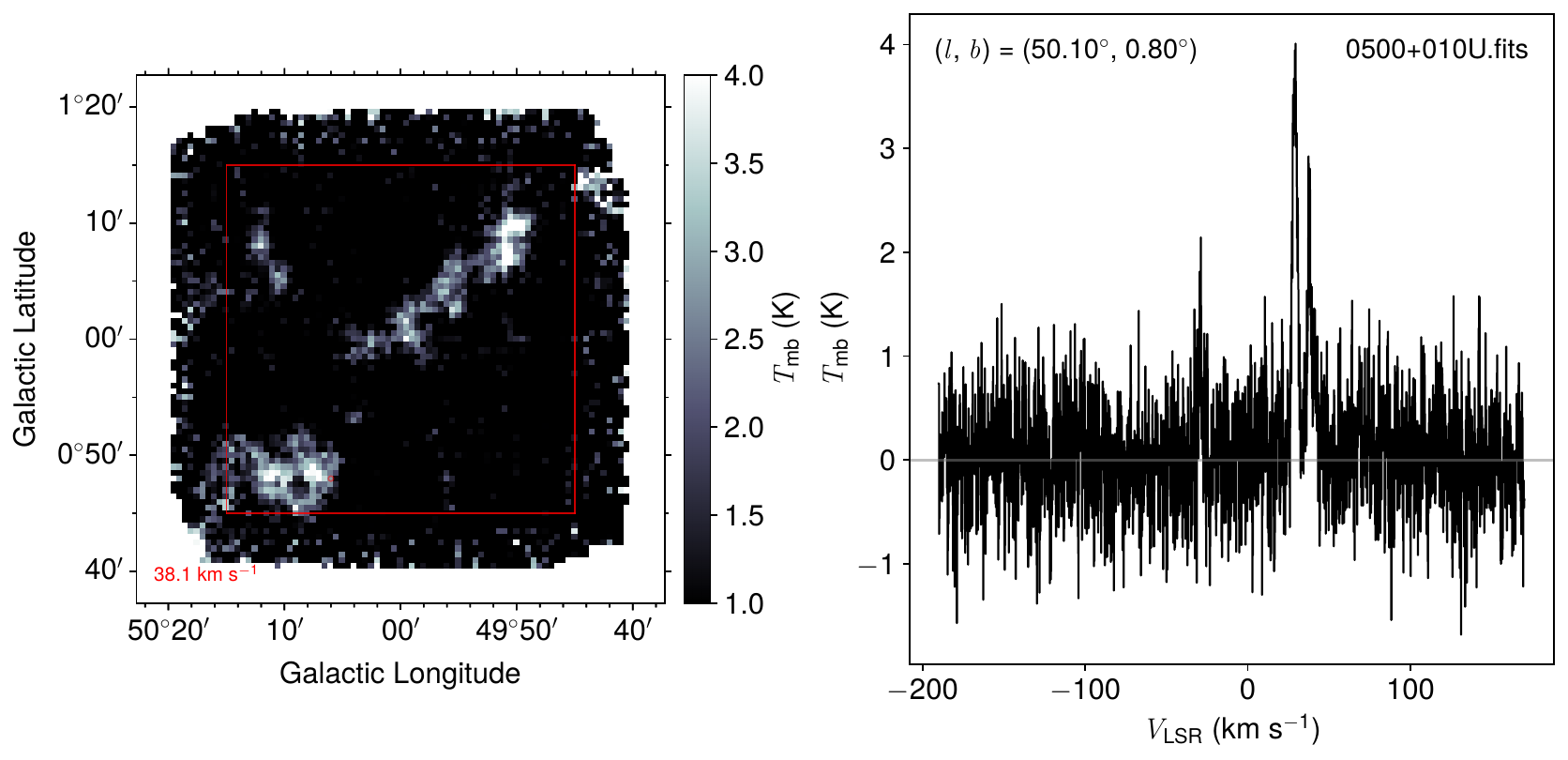} 
\caption{Demonstration of a cell cube named 0500+010U.fits. The red square delineates the center 30\arcmin$\times$30\arcmin\ region of the cell cube. A sample spectrum is displayed in the right panel. \label{fig:cubeexample}}  
\end{figure}

\section{Single Cell Data Cubes}

\label{sec:single}

The DR1 data contain 9,240 cell FITS cubes for each CO line, fully covering a sky area of $9.75\deg\leqslant l \leqslant229.75\deg$ and $|b|\leqslant5.25\deg$. Collectively, DR1 contains 27,720 cell FITS cubes. 

The centers of cell cubes form a regular grid in the $l$-$b$ plane, with a spacing of 0.5\deg. Consequently, the $l$-$b$ coordinates of the cell centers are integers or half-integers in degrees. The \cof, \cos, and \cot\ lines are denoted by U, L, and L2, respectively. For instance, the file name 0500+010U.fits indicates a \cofs\ cell cube at the location of $(l, b)=(50.0\deg, 1.0\deg)$. Each cell cube has a mapping size of 45\arcmin$\times$45\arcmin\  (91$\times$91 pixels), resulting in an overlap of 7.5\arcmin\ between adjacent cell cubes. This overlap diminishes edge effects during scans. As an example, a cell cube is demonstrated in Figure \ref{fig:cubeexample}, where the 30\arcmin$\times$30\arcmin\  (60$\times$60 pixels) center region is delineated with red lines.

According to the minimum and maximum velocity listed in Table \ref{Tab:vrange}, we cropped all cell cubes to a velocity range within [-200, 280] km s$^{-1}$. For \cofs, the $l\times b \times V_{\rm LSR}$ dimension is 91$\times$91$\times$3151, while for \coss\ and \cots, these dimensions are 91$\times$91$\times$3013  and 91$\times$91$\times$3002, respectively. As an example, we display a spectrum of \cofs\ in Figure \ref{fig:cubeexample}.

The size of each cell FITS cube is approximately 100 MB, amounting to about 900 GB for each CO line. In total, the DR1 data from the MWISP survey contain 2.6 TB of cell FITS cube data.

\section{Creating the Mosaic Data Cubes}
\label{sec:mosaic}

\subsection{Mosaic Procedure}
The final step of the data reduction is to mosaic individual cell cubes into large single datacubes. Here, we describe the procedure used to perform the mosaic operation. Due to the significantly increased $b$ range and the overlapping area between adjacent cell cubes, the primary issues of mosaicking are: (1) the stereographic projection effect and (2) merging the overlapping areas.

\begin{figure} 
\centering
\includegraphics[width=0.6\textwidth]{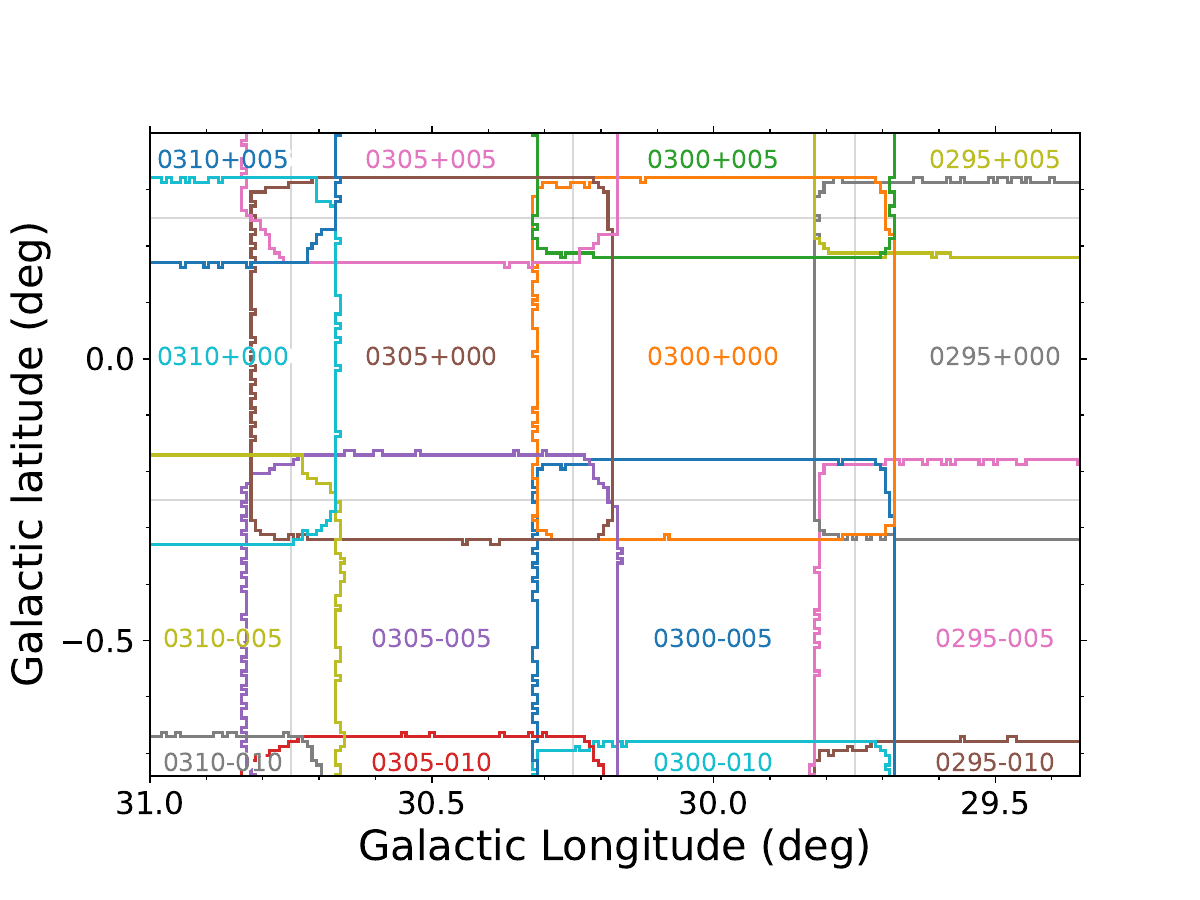}
\caption{An example of cell coverage and overlapping within a field. Boundaries and names of cells are drawn in different colors. Four cells at most contribute to the spectra in the overlapping regions.}
\label{fig:mosaic}
\end{figure}

Data of cells need to be combined to form a large-field data cube with a uniform noise distribution. When mosaicking a specific spatial and spectral region, not only are the cells available within the field included, but also adjacent cells outside the field boundaries. This is because peripheral cells can contribute to the noise level at the edges. Figure~\ref{fig:mosaic} illustrates an example of overlapping among the cells within a field. Data from overlapping cells are combined with weighted averaging, where the weights are derived from the corresponding noise. The data in the mosaic data cube in unit of  main beam temperature, consistent with individual cells, is derived from
\begin{equation}
T_{\rm mb} = \sum_{c} w_{c}\ T_{\mathrm{mb}, c} \Bigg/ \sum_{c} w_{c},
\end{equation}
where $T_{mb, c}$ is the main beam temperature in cell $c$, and $w_{c}$ denotes the weights of cells based on their noise $\sigma_{c}$, with $w_{c} = 1 / \sigma_{c}^2$. The noise level in the resulting mosaic data cube can be estimated with
\begin{equation}\label{equ:noise}
\sigma = \left[ \sum_{c} w_{c} \right]^{-1/2}.
\end{equation}

A mosaicking routine called \verb|mosaic| in \verb|mwispy| \footnote{\url{http://pypi.org/project/mwispy}} package has been developed to combine data cubes of a large number of cells, producing a mosaic data cube in Galactic coordinate system of the plate carr\'{e}e projection (CAR) and Local Standard of Rest (LSR) velocity frame. The routine automatically locates and gathers data cubes and noise images of all available cells based on their cell names and sideband from a file path or a list of paths, in case the data are stored in different directories. Given that the input cells are in Sanson-Flamsteed projection (GLS), which may not exactly coincide with the pixels of the output data cube in the  galactic longitude direction, especially at high latitude (offset by less than 3.42\arcsec\ at $b=5$\arcdeg), spectra falling within each pixel of the output cube are weight-averaged. The mosaic procedure requires input datacubes to be aligned in velocity.
 
In addition to creating a data cube, \verb|mosaic| also produces supplementary outputs. A coverage image is generated to indicate whether a cell is successfully combined to the output cube. A weight image gives the total weights of the mosaicked cube, equal to the sum of weights of cells, and a noise image is calculated with Equation ~\ref{equ:noise}. A log file documents the timeline of the program's execution along with the cells that were involved in the computations.

The computational resources required for \verb|mosaic| to operate correctly scale with the field size and the velocity span. Memory requirements can be approximated as 
\begin{equation}
{\mathrm{Memory} \over \rm GB} = 0.000340 \left( {A \over {\mathrm{deg}^2}} \right) \left( {\Delta v \over {\rm km~s^{-1}}} \right) + 0.863,
\end{equation}
where $A$ is the area of field in square degrees, and $\Delta v$ is the velocity range in \kms. Therefore, it is recommended to allocate over 400~GB of memory for mosaicking all survey data spanning 500 \kms. The processing time is primarily determined by the field size, but it is also influenced by various hardware factors, particularly the read and write speeds of the disk.

\begin{deluxetable}{c|c|ccc}
\tablecaption{Dimensions of mosaicked CO data cubes.   \label{Tab:cube}}
\tablehead{
   \multicolumn{2}{c|}{ }   & \colhead{\cofs} & \colhead{\coss} & \colhead{\cots}   
} 
\startdata
    \multicolumn{2}{c|}{Frequency}   & 115.271 GHz  &  110.201 GHz    &  109.782 GHz   \\
        \multicolumn{2}{c|}{$l$ range}   &  [9.75\deg, 229.75\deg] &   [9.75\deg, 229.75\deg]  &  [9.75\deg, 229.75\deg]  \\
        \multicolumn{2}{c|}{$b$ range}   &  [-5.25\deg, 5.25\deg] &   [-5.25\deg, 5.25\deg]  &   [-5.25\deg, 5.25\deg]   \\
                \multicolumn{2}{c|}{$V_{\rm LSR}$ range}   &  [-200, 300] \kms &   [-200, 300] \kms &   [-200, 300] \kms   \\
   \multicolumn{2}{c|}{$l\times b \times V_{\rm LSR}$ voxel size}   &  $0.5\arcmin\times0.5\arcmin\times0.159$ \kms   & $0.5\arcmin\times0.5\arcmin\times0.166$ \kms &  $0.5\arcmin\times0.5\arcmin\times0.167$ \kms  \\
 \multicolumn{2}{c|}{$l\times b \times V_{\rm LSR}$ dimension}   &  $26401\times1261\times3151$ &  $26401\times1261\times3013$  &    $26401\times1261\times3002$  \\
   \multicolumn{2}{c|}{Total voxel number\tablenotemark{a}}   & $6.22\times$10$^{10}$ &   $5.95\times$10$^{10}$ &    $5.93\times$10$^{10}$  \\
      \multicolumn{2}{c|}{Total pixel number}   &  $3.33\times$10$^{7}$  &  $3.33\times$10$^{7}$  &   $3.33\times$10$^{7}$  \\
\hline 
\enddata 
\tablenotetext{a}{Voxels with NaN were not taken into account.}
\end{deluxetable}

\begin{figure} 
\centering
\includegraphics[width=0.6\textwidth]{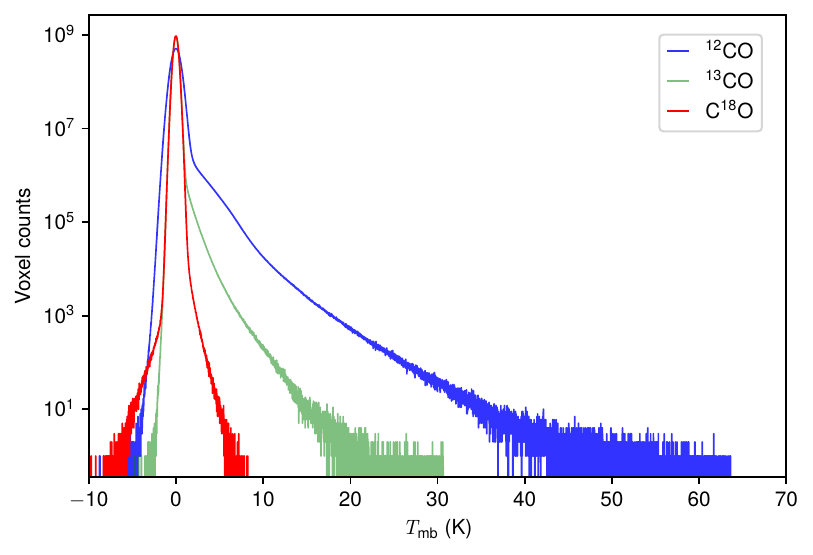} 
\caption{Distribution of main-beam brightness temperature for CO and its isotopologue line emission. For each line, there are two dominant components, the signal component in approximate power-law form and the noise component in Gaussian form.  The excess of voxels with negative temperatures,  more conspicuous in \cots, is due to the residual from bad channel removing.  \label{fig:voxel}}  
\end{figure}
 
\begin{figure} 
\plotone{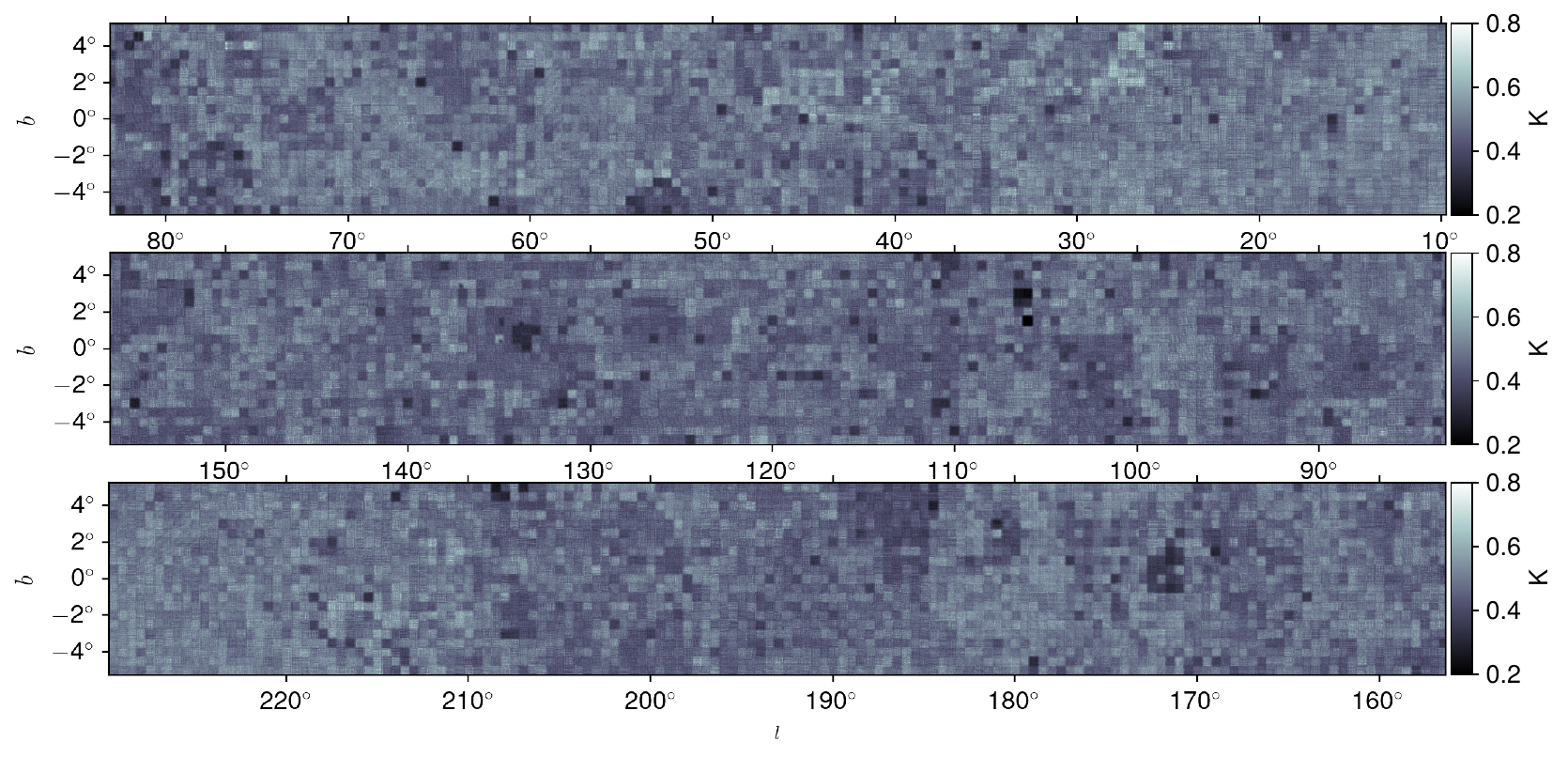} 
\caption{Spectral rms noise of \cofs. \label{fig:noiseimg12}}  
\end{figure}

\begin{figure} 
\plotone{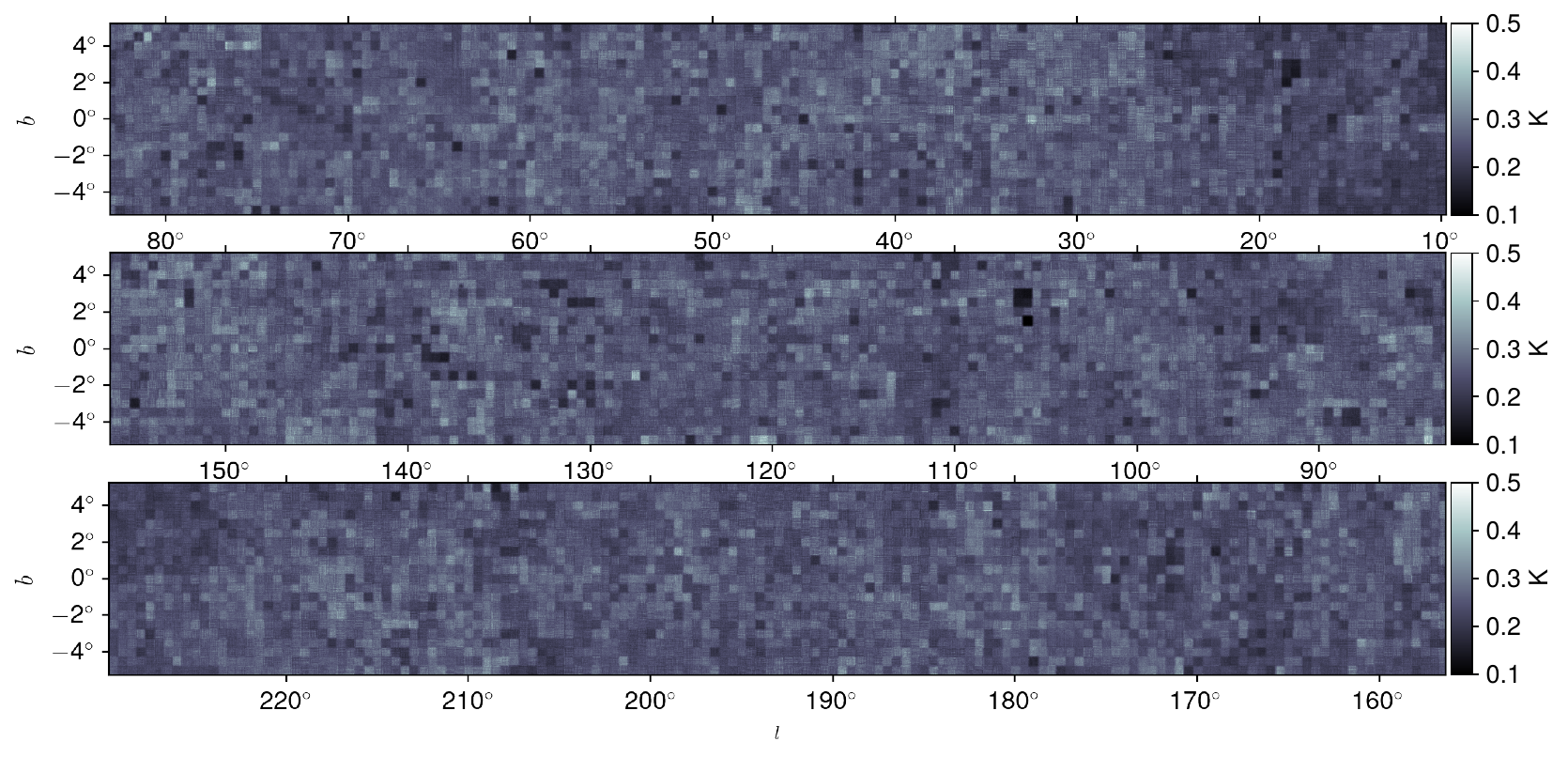} 
\caption{Spectral rms noise of \coss.  \label{fig:noiseimg13}}  
\end{figure}

 \begin{figure} 
\plotone{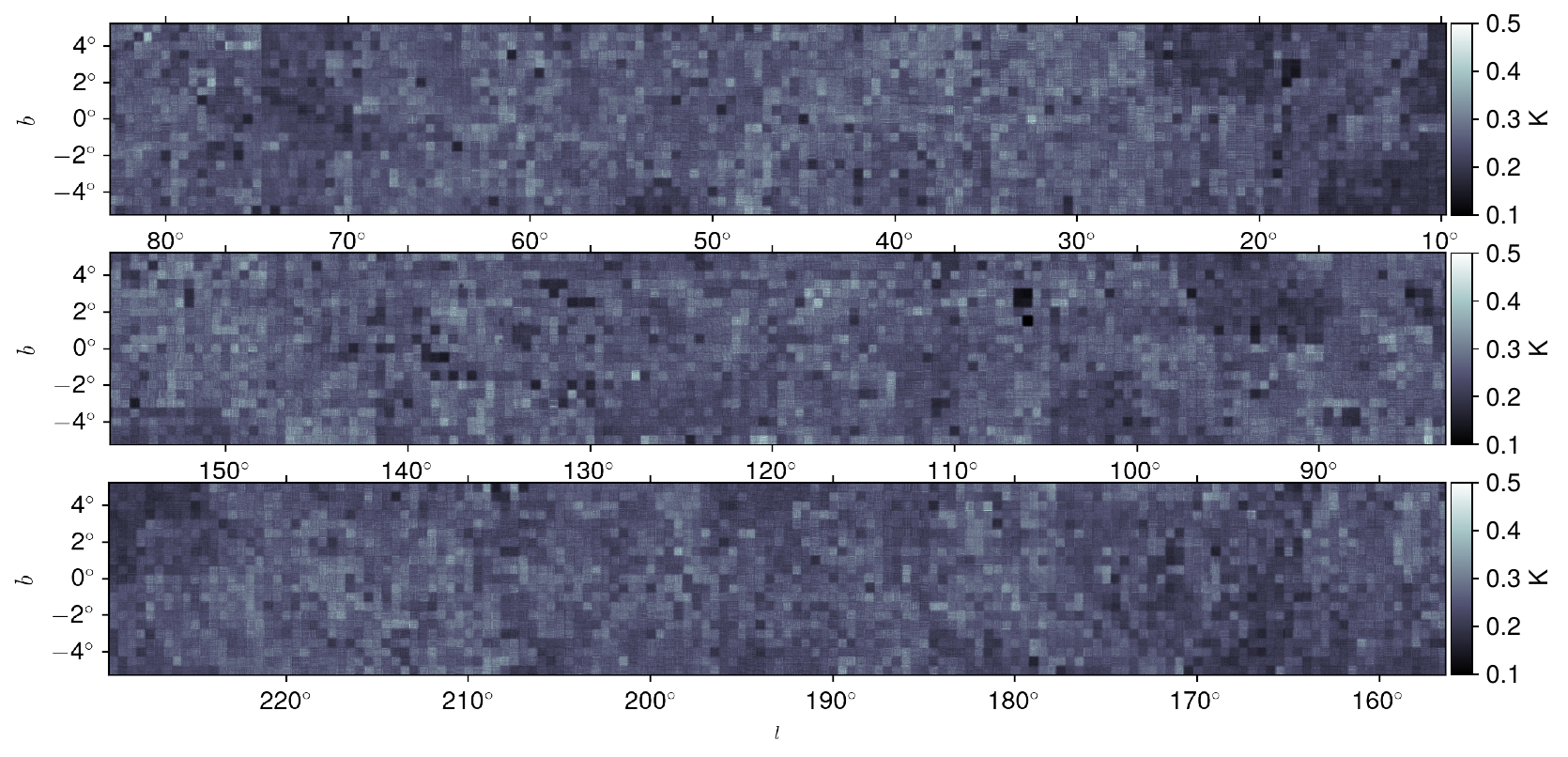} 
\caption{Spectral rms noise of \cots. \label{fig:noiseimg18}}  
\end{figure}

\begin{figure} 
\centering
\includegraphics[width=0.6\textwidth]{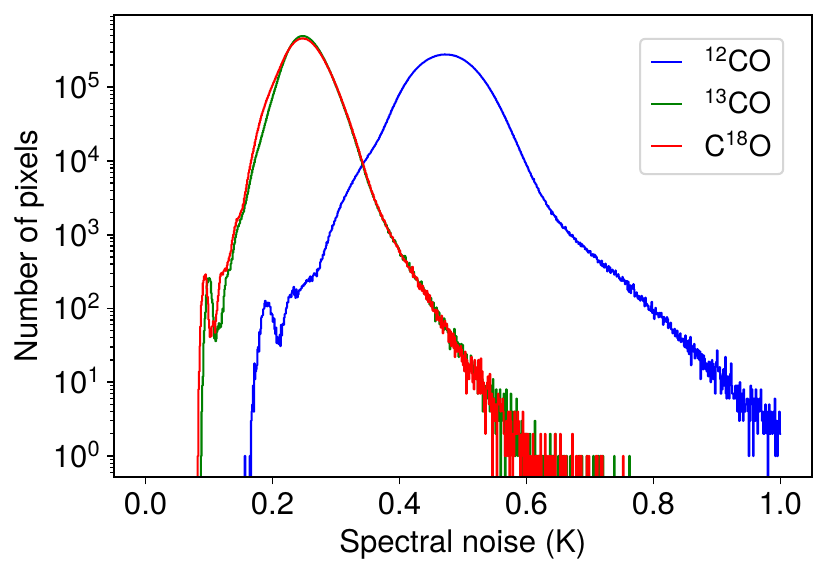} 
\caption{Histogram of spectral noise of the MWISP CO survey. The median noise of \cofs, \coss, and \cots\ is, respectively, 0.47 K, 0.25 K, and 0.25 K. The peaks around 0.2 K for \cofs\ and approximately 0.1 K for \coss\ and \cots\ are due to the low noise at cell ($l$, $b$) = (106.0\deg, 1.5\deg).  \label{fig:noisedis}} 
\end{figure}

\subsection{The Mosaicked Cubes}

Each CO line corresponds to a mosaicked data cube  and an rms map, both saved as FITS files. The file sizes of mosaicked FITS cubes for \cofs, \coss, and \cots\ are 391 GB,  374 GB, and  373 GB, respectively. The differences in file sizes are due to slight changes in the velocity resolution. Users should be cautious when processing the entire data cube, as it requires considerable memory and CPU resources.   

Detailed dimensions of mosaicked data cubes are summarized in Table \ref{Tab:cube}. For all data cubes of three CO isotopologue lines, the $l$-$b$-$V$ coverage spans $9.75\deg\leqslant l \leqslant229.75\deg$, $|b|\leqslant5.25\deg$, and $-200\leqslant V_{\rm LSR} \leqslant300 $ \kms. Although the mosaicked cubes span a velocity range of [-200, 300] \kms, voxels outside the baseline range defined in Table \ref{Tab:vrange} are masked with NaN values. Each of the mosaicked datacubes  contains $3.33\times$10$^{7}$ spectra, which is large in scale compared with similar surveys.

A histogram of main-beam brightness temperature for the three data cubes is demonstrated in Figure \ref{fig:voxel}. Each line contains signal   and noise components.  On average, the brightness temperature decreases successively from \cofs\ to \coss\ and then to \cots.  A brief decomposition of the distribution histogram is described in the following sections.

\subsection{Noise Characterization}
  
In this section, we present the noise characterization of the MWISP DR1 data. The inspection is based on the rms noise maps produced using spectrum segments within the baseline velocity range (see Section \ref{sec:baseline}). For spectral cubes, the noise is essentially a three-dimensional quantity, but for simplicity, we examine the rms of spectra, i.e., two-dimensional rms noise maps. 

Figure \ref{fig:noiseimg12} displays the distribution of traditional 1D spectral (rms) noise in \cofs. The noise is generally uniform, exhibiting small-amplitude fluctuations caused by observing strategies or weather conditions. Due to specific deep investigations,  some low-noise regions are present. See Figure \ref{fig:noiseimg13} and  \ref{fig:noiseimg18} for rms noise maps of \coss\ and  \cots, respectively.

 A close examination reveals faint stripes in the rms noise maps. These stripes are due to the slight radial velocity shifts across the 30\arcmin$\times$30\arcmin\ cells. Along the stripes, the rms noise is approximately 0.03 K higher than the surrounding areas, though this has negligible effects on the spectra. 

Figure \ref{fig:noisedis} shows the distribution of rms  noise for three CO lines. The median values of \cofs, \coss, and \cots\ are 0.47 K, 0.25 K, and 0.25 K, respectively. These values are below the target noise levels.

An overview of physical parameters based on \cofs\ clouds defined by DBSCAN (see Section \ref{sec:co12cat}) is summarized in Table \ref{Tab:prop}. From \cofs\ to \coss\ and then to \cots, the area of emission regions shrinks, from   1225.4~deg$^2$ in \cofs\ to 501.6~deg$^2$ in \coss\ and  35.0~deg$^2$ in \cots.

\begin{deluxetable}{c|c|ccc}
\tablecaption{Properties\tablenotemark{a} of three CO lines based on images. \label{Tab:prop}}
\tablehead{
   \multicolumn{2}{c|}{ }   & \colhead{\cofs} & \colhead{\coss} & \colhead{\cots}   
} 
\startdata
  \multicolumn{2}{c|}{Median noise}   & 0.47 K &   0.25 K &    0.25 K   \\
  \multicolumn{2}{c|}{Emission voxel number }   & 5.86$\times$10$^8$ & 1.21$\times$10$^8$ &  3.71$\times$10$^6$  \\
 \multicolumn{2}{c|}{Total flux (K \kms arcmin$^2$)}   & 6.34$\times10^7$ & 6.12$\times10^6$  & 1.36$\times10^5$      \\
  \multicolumn{2}{c|}{Total angular area}   & 1225.4  deg$^2$ &  501.6  deg$^2$ &   35.0 deg$^2$  \\
 \multicolumn{2}{c|}{Projection pixel number}  &  1.76$\times10^7$  &   7.17$\times10^6$ &   5.03$\times10^5$ \\
 \hline 
\multirow{3}{*}{Brightness temperature}  & Minimum &0.34 K  & 0.18 K &   0.18 K \\
\cline{2-5}  & Median &   2.16 K & 0.95 K & 0.77 K  \\
\cline{2-5}  & Maximum &  63.63 K & 30.66 K  & 6.71 K  \\ 
\hline 
\multirow{2}{*}{$V_{\rm LSR}$ (\kms)}  & Minimum & -122.5     & -115.2    & -64.3     \\
\cline{2-5}  & Median &   10.3 & 19.4 & 38.2  \\
\cline{2-5}  & Maximum &  195.9    &  175.8     &  152.4     \\
\hline  
\enddata 
\tablenotetext{a}{This only accounts for voxels or pixels within envelops determined with DBSCAN.}
\end{deluxetable}

\section{Generated images}
\label{sec:images}
In this section, we demonstrate the DR1 data of the MWISP survey from various perspectives, including integrated images, $l$-$V$  diagrams, and sampling spectra. The color composite images are generated based on the integrated intensities of the three isotopologue  CO lines, i.e., images of each cloud identified with DBSCAN  (see Section \ref{sec:co12cat}).

\begin{figure} 
\plotone{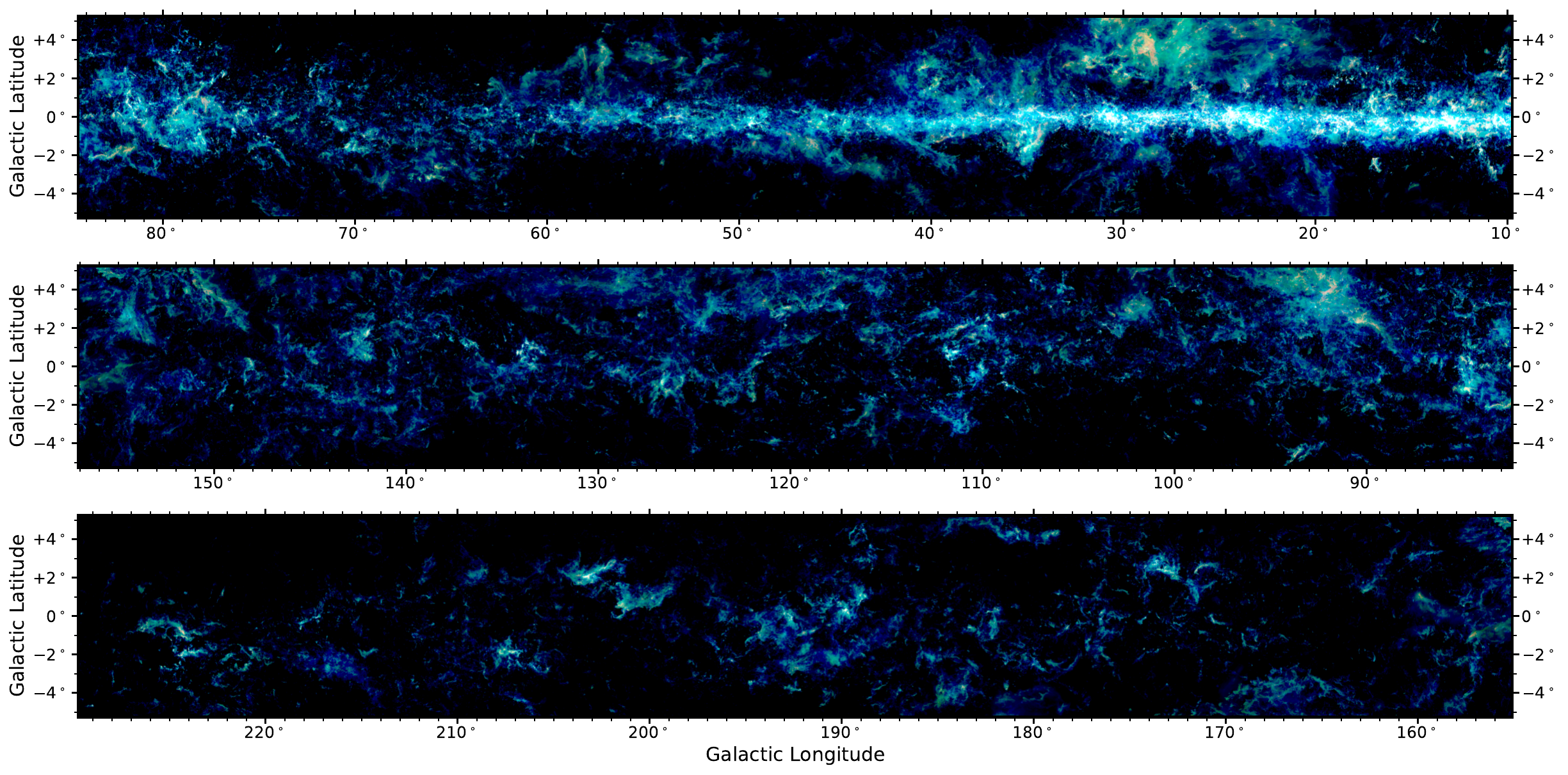} 
\caption{Composite RGB image of of \cots\ (Red), \coss\ (Green), and \cofs\ (Blue). To make the isotopologue emission comparable, the \cots\ and \coss\ emissions are multiplied by factors of 53.5 and 6, respectively \citep[see the flux ratio derived by][]{2023AJ....166..121W}. The intensity was scaled using a square-root stretch to enhance faint emission features. \label{fig:rgb}}  
\end{figure}

 \begin{figure} 
 \plotone{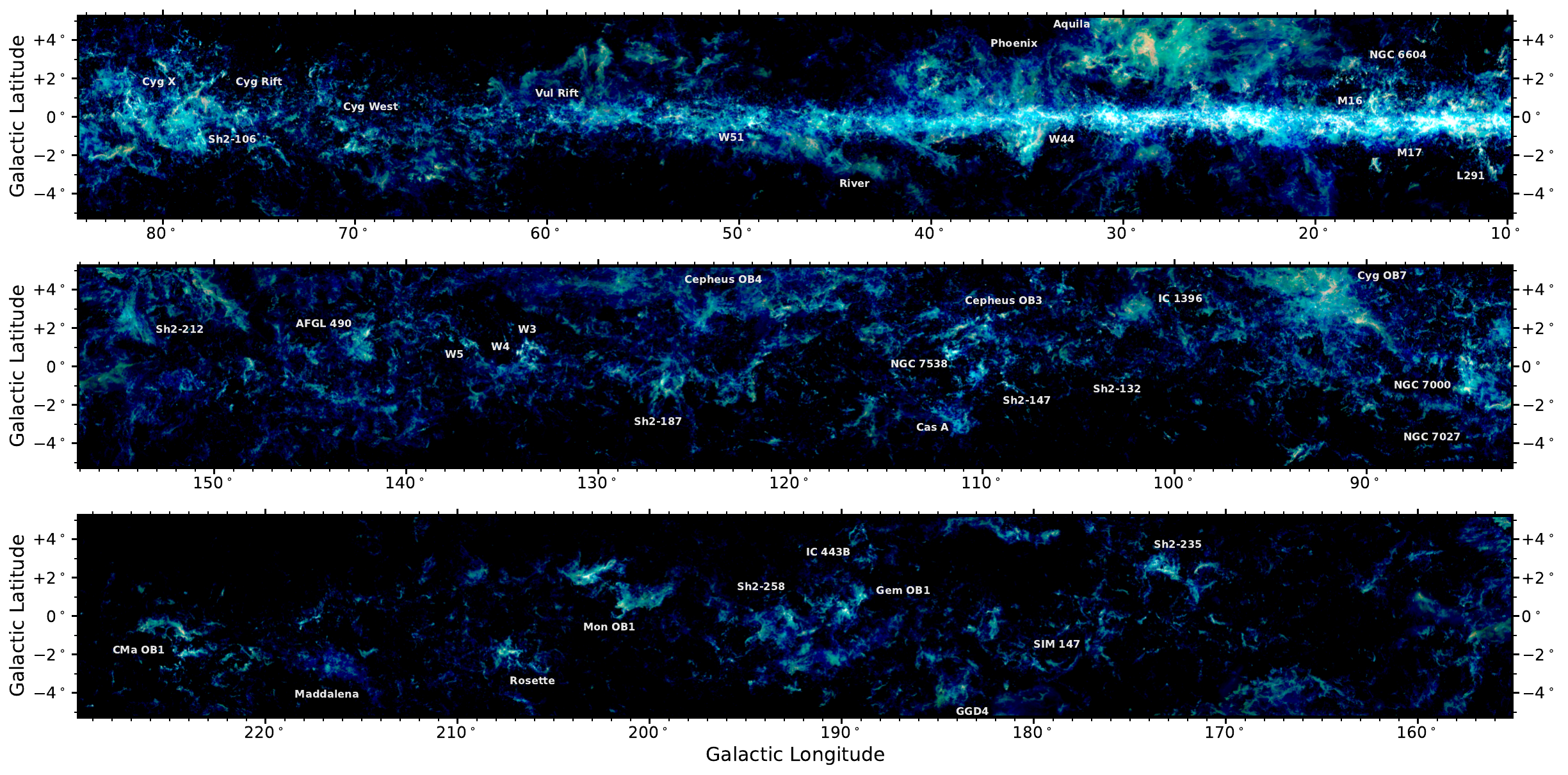} 
\caption{Same as Figure \ref{fig:rgb} but with prominent regions marked.  \label{fig:chart}}  
\end{figure}

\subsection{Individual Regions}

For each CO line, an integrated intensity image is produced by integrating the whole data cube along the radial velocity. We combine the integrated intensity of three CO lines into a composite RGB map, with \cofs\ in blue, \coss\ in green, and \cots\ in red, as illustrated in Figure \ref{fig:rgb}.  Benefiting from the multi-line receiving with a broad dynamic range, the high spatial resolution and the complete sampling of the MWISP \cofs/\coss/\cot\ survey, this composite image demonstrates the overall emission from the surveyed area at an unprecedented quality and exhibits a wealth of information on molecular clouds, such as the morphology, distribution, and internal structures.  For convenient comparison with sky images at other wavelengths, Figure \ref{fig:chart} presents a finder chart for prominent regions in the Galactic plane. 

 The composite RGB image exhibits a high dynamic range, which is 10,557 (about 40 dB) for \cofs, revealing a diversity of cloud structures, including the Galactic disk, Galactic warp and flare, giant molecular clouds (GMCs), dense gas regions, filament networks, and bubbles. Additionally, CO emissions from \HII\ regions, supernova remnants, and high-mass star-forming regions, can also be easily identified. Due to observational effects, molecular clouds at high Galactic latitudes are more likely to be located near the Solar System.  
 
\subsubsection{G10-40}
The first Galactic quadrant encompasses notable large-scale structures such as the Galactic bar, the 3-kpc molecular ring, and multiple spiral arms, harboring substantial reservoirs of molecular gas in the Milky Way. Over the past four decades, this region has been a key target for CO surveys, from early pioneering efforts \citep[e.g., see the summary of CfA survey from][]{2001ApJ...547..792D} to some recent mappings, e.g., the GRS survey \citep{2006ApJS..163..145J}, the COHRS survey, the CHIMPS survey, and the FUGIN survey.  For the most extensive local MC in our map spanning Galactic longitude $l = 10^\circ$ deg to$l = 50^\circ$”, MWISP data have revealed significant diversity in the distribution, structure, and physical properties of molecular gas within the Aquila Rift in the solar neighborhood \citep[e.g., the River and Phoenix MCs in][]{2020ApJ...893...91S}. At larger distances, prominent CO emission from massive star-forming regions along the Galactic plane is clearly delineated in the color maps (e.g., M16, M17, W41, W43, W44, W49A, and W51). Based on the statistical results of the spatial distribution of MCs at Galactic tangent points, we have characterized the structures of the inner Galactic molecular disk. The high-sensitivity MWISP survey has, for the first time, enabled the quantitative identification of distinct thin and thick disk components \citep[with thicknesses of ~85 pc and ~280 pc, respectively,][]{2021ApJ...910..131S}, as well as successfully tracing signatures of Galactic nuclear wind-disk interactions \citep{2022ApJ...930..112S}. Furthermore, the survey systematically reveals extended molecular spiral arm structures beyond the Solar circle \citep{2016ApJ...828...59S,2017ApJS..230...17S,2024ApJ...977L..35S}, demonstrating the kinematic and physical complexity of molecular gas across different Galactic environments. The high-quality datasets also provide a unique perspective for investigating interactions between supernova remnants and MCs, e.g., G22.7-0.2 in \citet{2014ApJ...788..122S}, HESS J1912+101 in \citet{2017ApJ...836..211S}, SS 433 in \citet{2018ApJ...863..103S}, and other cases in \citet{2023ApJS..268...61Z}, offering important implications for understanding the origin of high-energy TeV emissions.
 
\subsubsection{G50-60}
Numerous surveys were conducted toward the G50 region, in the Galactic longitude range of 45 to 60 deg. \citet{1977ApJ...217L.155C} and  \citet{1980ApJ...239L..53C} surveyed the first Galactic quadrant by sparsely sampling and studied the spiral arm and molecular ring structures;   \citet{1986ApJS...60..695C} expanded the observations to a beamwidth-sampled survey of the Galactic plane. \citet{1986ApJS...60....1S} performed a superbeam survey of this region using the 14 m FCRAO telescope. A superbeam survey of \cos\ emission was also conducted by \citet{1988A&A...195...93J}. Subsequent surveys began full sampling observations (beamwidth- or Nyquist-sampling), e.g., the well-known \cos\ GRS Survey   and the \cof\ CfA survey \citep{2001ApJ...547..792D}. Recently, high-sensitivity surveys of the inner Galaxy were conducted that partly covered this region, e.g., the FOREST Unbiased Galactic plane Imaging survey \citep[FUGIN,][]{2017PASJ...69...78U}. The previous surveys only covered the Galactic latitude of $|b|$$\sim<$1 deg, except the CfA \cofs\ survey. The brightest part in this region is the W51 region, which is an active star-forming region, with giant molecular clouds, HII regions, and supernova remnants distributed in it \citep[e.g.,][and references therein]{1998AJ....116.1856C}. Most of the molecular gas is distributed along the galactic plane, as well as different kinds of objects. There is also molecular gas located at relatively high Galactic latitude, e.g., in the Vulpecula Rift. Most high latitude molecular gases are widely distributed in the region, which are located in our neighborhood \citep{2025arXiv250814547Z}.

\subsubsection{G60-75}
The region from $l$ = 60$^\circ$–75$^\circ$ lies between the Cygnus complex and the Vulpecula Rift \citep[$\sim$400 pc,][]{1985ApJ...297..751D}, and includes the Cygnus West area. It forms a transition zone with pervasive tenuous networks of CO emission. Molecular clouds in this region show clumpy and fragmented morphology, unlike the coherent structures in other inner Galactic regions. This area has not been well studied. It was observed by the CfA CO survey \citep{2001ApJ...547..792D} at lower resolution. The Exeter–FCRAO survey \citep{2016ApJ...818..144R} covered a small part of this region. Based on the MWISP survey, the molecular clouds in this region have been systematically studied for the first time \citep{2023ApJS..267...30L}.

\subsubsection{G75-105}
A substantial concentration of molecular gas is distributed across the Galactic longitude range from 75$^\circ$ to 105$^\circ$.
The Cygnus X region, located toward the western end of this range, stands out as the most prominent feature, hosting the most massive molecular complex in this longitude sector. To the southwest lies Sh2-106, a compact bipolar H{\sc ii} region embedded in molecular gas. Further along the longitude, notable objects include NGC~7000 (the North America Nebula) and the  planetary nebula NGC~7027. At higher Galactic latitude, the large cloud Kh 141 = TGU 541 is associated with the young OB association Cygnus OB7. The bright emission nebula IC~1396 is part of a large H{\sc ii} region associated with CO clouds, and further east, the bubble-like structure of Sh2-132 marks the eastern extent of this longitude range, shaped by stellar winds and radiation from embedded massive stars.
Following the initial mapping by \citet{1985ApJ...297..751D}, several surveys over subsequent decades have promoted our understanding of molecular clouds in this region by improving spatial coverage and sampling. \citet{1992ApJS...81..267L} conducted a CO survey toward the Cygnus complex, which was later incorporated into the CfA survey \citep{1985ApJ...297..751D}. Local $^{13}$CO emission was observed with similar resolution by \citet{1994ApJS...95..419D}. More recent surveys with $\sim1^{\prime}$ resolution have focused on the dense gas within the Cygnus X region, though they cover only a few selected areas, approximately $\sim10~\mathrm{deg}^2$ in the $J=2\rightarrow1$ and $3\rightarrow2$ transitions \citep{2006A&A...458..855S}, and $\sim35~\mathrm{deg}^2$ in the $J=1\rightarrow0$ transition \citep{2011A&A...529A...1S}.
A more recent effort targeting the diffuse ISM component has been carried out with slightly higher resolution than our survey. This survey, known as Exeter-FCRAO (EXFC) survey \citep{2016ApJ...818..144R}, maps the Galactic plane over the longitude range from 55$^\circ$ to 100$^\circ$ and latitude range from $-1.4^{\circ\!}$ to $+1.9^{\circ\!}$ in both $^{12}$CO and $^{13}$CO.

\subsubsection{G105-120}
The region from $l$ = 105$^\circ$–120$^\circ$ was previously observed by the CfA survey \citep{2001ApJ...547..792D} and the FCRAO Outer Galaxy Survey \citep{1998ApJS..115..241H}. Detailed statistical properties of molecular clouds in this region have been studied based on MWISP data \citep{2019ApJ...878...44M}. Several well-known star-forming complexes are located in this region, including Cep OB3 \citep[700 pc;][]{1996ApJ...463..630H,2005A&A...438.1163K,2009ApJ...693..406M} in the Local Arm, and NGC 7538 \citep[2.65 kpc;][]{2009ApJ...693..406M}, the Cas GMC \citep[3.4 kpc;][]{1995ApJ...440..706R,2019ApJ...878...44M}, and molecular clouds associated with S147 and S152 \citep[2.63–2.92 kpc;][]{2019ApJ...885..131R,2022RAA....22i5004L} in the Perseus Arm. Together, these sub-regions constitute a representative sample spanning different evolutionary stages and offer valuable laboratories for studying the lifecycle of molecular clouds and star formation. 

\subsubsection{G120-130}
In the G120-130 region, there is relatively less molecular gas compared to the inner Galaxy. Fewer surveys were conducted toward this region than toward the inner Galaxy. \citet{1980ApJ...239L..53C} surveyed this region by sparsely sampling  and studied the spiral arm structure. A beamwidth-sampled survey was  performed by \citet{1998ApJS..115..241H} using the FCRAO telescope. \citet{2010PASJ...62.1277Y} conducted a Nyquist-sampled \cofs\ and \coss\ ($J=2\rightarrow1$) survey along $b=0^\circ$ deg and compared the results with the \cof\ data from the CfA CO survey. Molecular gas is highly structured in the G120 region, at both low and high Galactic latitudes. The brightest part is molecular cloud SH 2-187 at a low Galactic latitude, which is in a shell-like structure. The high-latitude gas is mainly distributed in the Cepheus Flare Shell around the Cepheus OB4, which was mapped via superbeam observations by \citet{1989ApJ...347..231G}. A molecular filament of 390 pc was discovered and identified as a spur blown from the Local Arm \citep{2023AJ....165...16C}.

\subsubsection{G130-140}
Gas emission from G130-140 in Figures 12-13 is conspicuous at two of the most active sites of ongoing massive star formation, W3 and W5, which are located in the Perseus arm. The 
intervening region between them, corresponding to W4, shows a notable gap in molecular gas emission and is known  as one of the nearest examples of a Galactic super bubble. At higher latitudes, the extended gas emission primarily  originates from nearby local dark clouds. Molecular gas content in this region has been extensively studied via emission from the main \cofs\ molecular line using the CfA 1.2 m telescope \citep{1978ApJ...226L..39L,1996ApJ...458..561D} and the FCRAO 14 m telescope \citep{1998ApJS..115..241H}, although areas with  $b$ $<$ $-$3$^{\circ}$ were not covered by the FCRAO survey. The new data, with improved sensitivity and combined observations of three CO isotopologues, enable several advances: the discovery of a new segment of a distant spiral arm \citep{2015ApJ...798L..27S}, the identification of hundreds of outflow candidates \citep{2019ApJS..242...19L}, and a more accurate inventory of gas physical properties in both arm and inter-arm regions \citep{2020ApJS..246....7S}.

\subsubsection{G140-170}

The Galactic G140-170 region was observed in the CfA  CO survey \citep{1987ApJ...322..706D,2001ApJ...547..792D}. Apart from this survey, no other systematic molecular line surveys were conducted in this region. Compared to the inner Galactic plane, CO molecular gas in this region appears diffuse and faint, mainly distributed in the Local Arm and Perseus Arm \citep{2024ApJ...977L..35S}. Within the Local Arm, the molecular gas is further divided into the Gould’s Belt layer (~150-300 pc) and the Camelopardalis layer (~700-900 pc). A few star-forming regions, such as AFGL 490 and Sh-212 \citep{2022ApJ...938...44G}, are located in this area. Based on the MWISP survey, several large-scale filamentary molecular clouds have been observed \citep{2017ApJS..229...24D}.

\subsubsection{G190-220}
 In the third Galactic quadrant, the molecular gas is distributed in a clustered pattern, mainly concentrating in four well-known giant molecular cloud complexes: Gem OB1, Mon OB1, Rosette, and Maddalena. The Columbia Survey \citep{1987ApJ...322..706D} first achieved complete CO observational coverage of this region, revealing the overall distribution characteristics of the molecular gas. \citet{1985ApJ...294..231M} first discovered and named the Maddalena cloud using the  1.2 m Columbia millimeter-wave telescope. \citet{1996A&A...315..578O} separated the clouds of the Local Arm (d $\approx$ 0.8 kpc) and the Perseus Arm (d $\approx$ 3.5 kpc) based on \cofs\  data in the Mon OB1 region. \citet{1995ApJ...445..246C} investigated the large-scale morphology and properties of the molecular gas toward the Gem OB1 region using \cofs\ and \coss\ data from the FCRAO 14-m telescope.  
 
The MWISP is the second complete CO survey covering this region with high sensitivity and multi-line CO data that reveal detailed structures. Abundant molecular gas emissions have been detected in key regions such as the Gem OB1 MC complex \citep{2017ApJS..230....5W}, Rosette Nebula \citep{2018ApJS..238...10L}, Monoceros OB1 \citep{2024ApJ...966..202Z}, and Sh 2-287 \citep{2016A&A...588A.104G}. Distances and masses for 11 MCs, including the Maddalena cloud and Sh 2-287, were derived, indicating the Perseus Arm is at about 2.4 kpc \citep{2019ApJ...885...19Y}.

\subsubsection{G220-230}

The brightest emission in the G220 region originates from the filamentary CMa OB1 complex in the Local Arm (Figures \ref{fig:rgb} and \ref{fig:chart}). Unlike the first and second Galactic quadrants, this region has not been extensively mapped \citep[see Table 1 of][and references therein]{2023ApJS..268....1D}. A recent $^{12}$CO and $^{13}$CO survey with arcminute-scale resolution has provided more detailed information on the large-scale molecular gas distribution and properties across part of this region, covering approximately 25 deg$^2$ \citep{2020A&A...633A.147B,2021A&A...654A.144B}. The MWISP data enable a more complete census of CO outflow candidates and the identification of relatively denser C$^{18}$O clumps in the CMa OB1 star-forming region  \citep{2021ApJS..252...20L}. Furthermore, we expand the known molecular content of the area by revealing a rich collection of nearby filamentary structures, along with compact, discrete clouds that clearly trace the distant Outer arm \citep{2023ApJS..268....1D}.

\begin{figure} 
\plotone{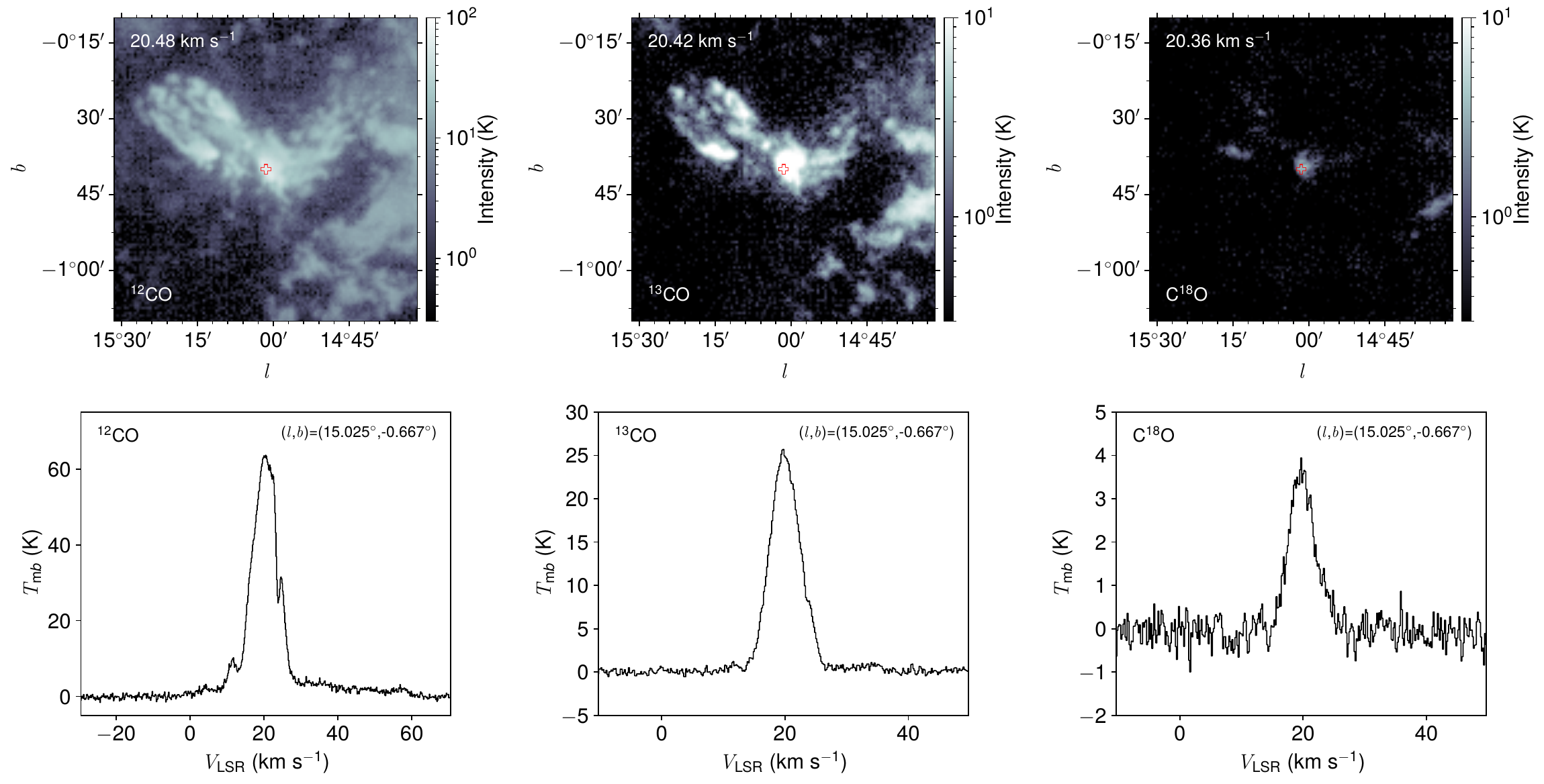} 
\caption{Image (top panels) and spectra (bottom panels) toward positions with the maximum \cofs\ brightness temperature. The images are single channel map at the velocity corresponding to the maximum brightness temperature, and from left to right, the displayed lines are  \cofs, \coss, and \cots. Positions of spectra are  marked with empty red plus symbols in top panels. \label{fig:max}}  
\end{figure}

\begin{figure} 
\plotone{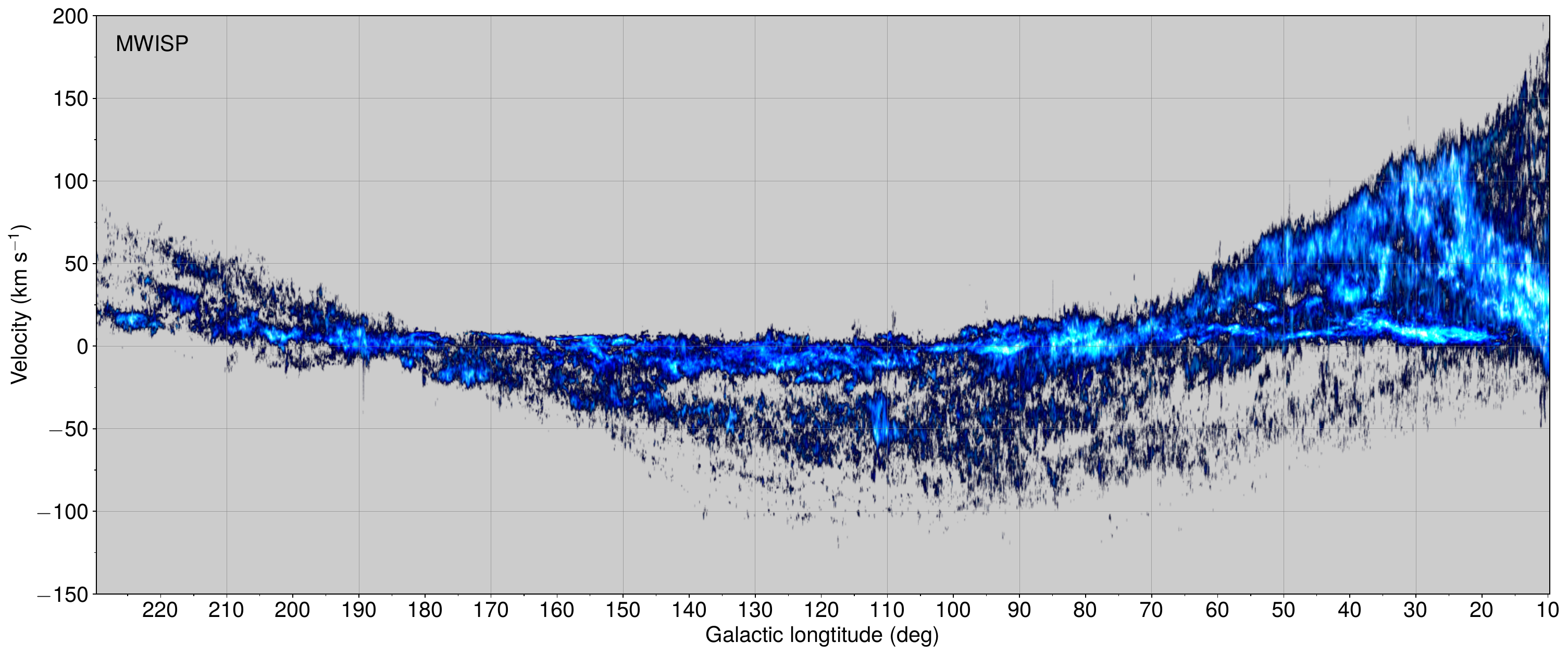} 
\caption{ $l$-$V$ RGB map of three CO emission with \cots\ in Red, \coss\ in Green, and \cofs\ in Blue. The displaying data are the averaged intensity over the $b$ axis. To enhance the visual contrast,  the data within [0, 0.2] K for \cots, [0, 0.1] K for \coss, and [0, 0.02] K for \cofs\ are clipped, square-root transformed, and normalized to [0, 1] for RGB color encoding. \label{fig:lv}}  
\end{figure}

\begin{figure} 
\plotone{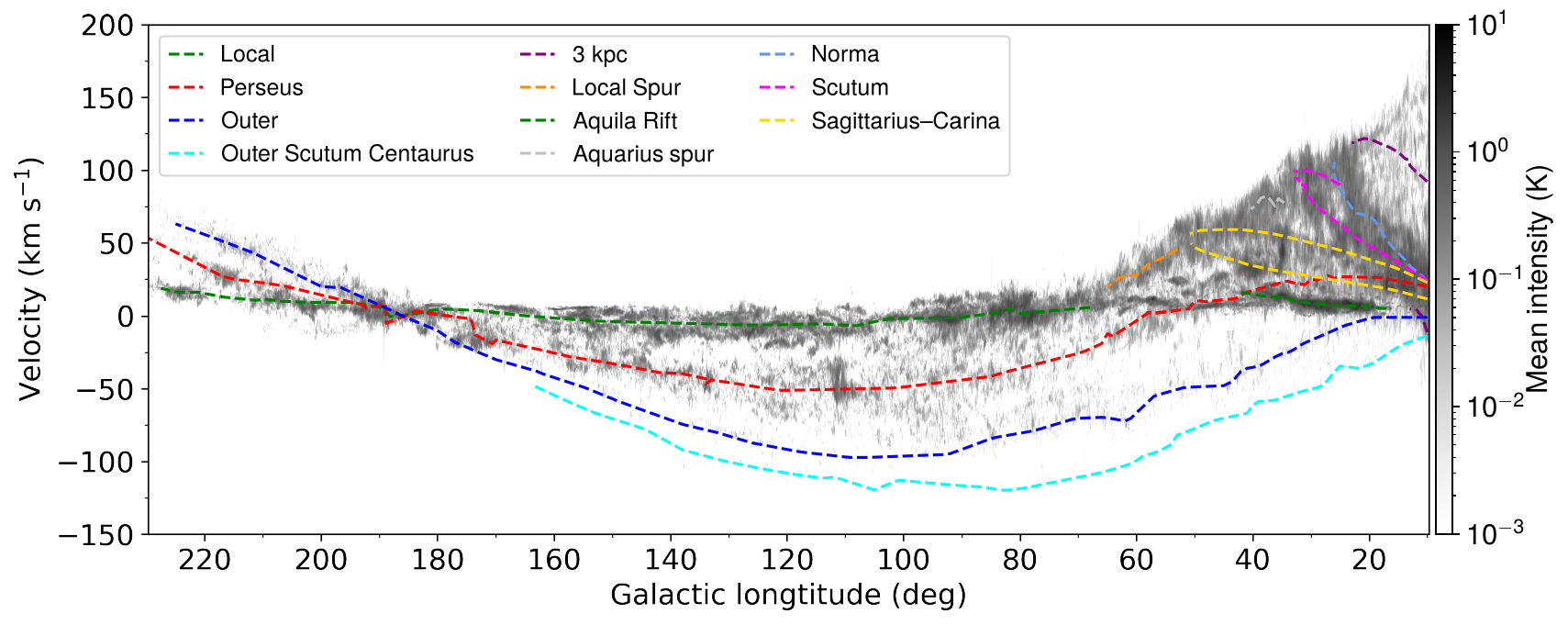} 
\caption{$l$-$V$ diagram of \cofs\ emission overlaid with the arm model of \citet{2019ApJ...885..131R}. \label{fig:arm}}  
\end{figure}

 \begin{figure} 
 \centering
 \includegraphics[width=0.8\textwidth]{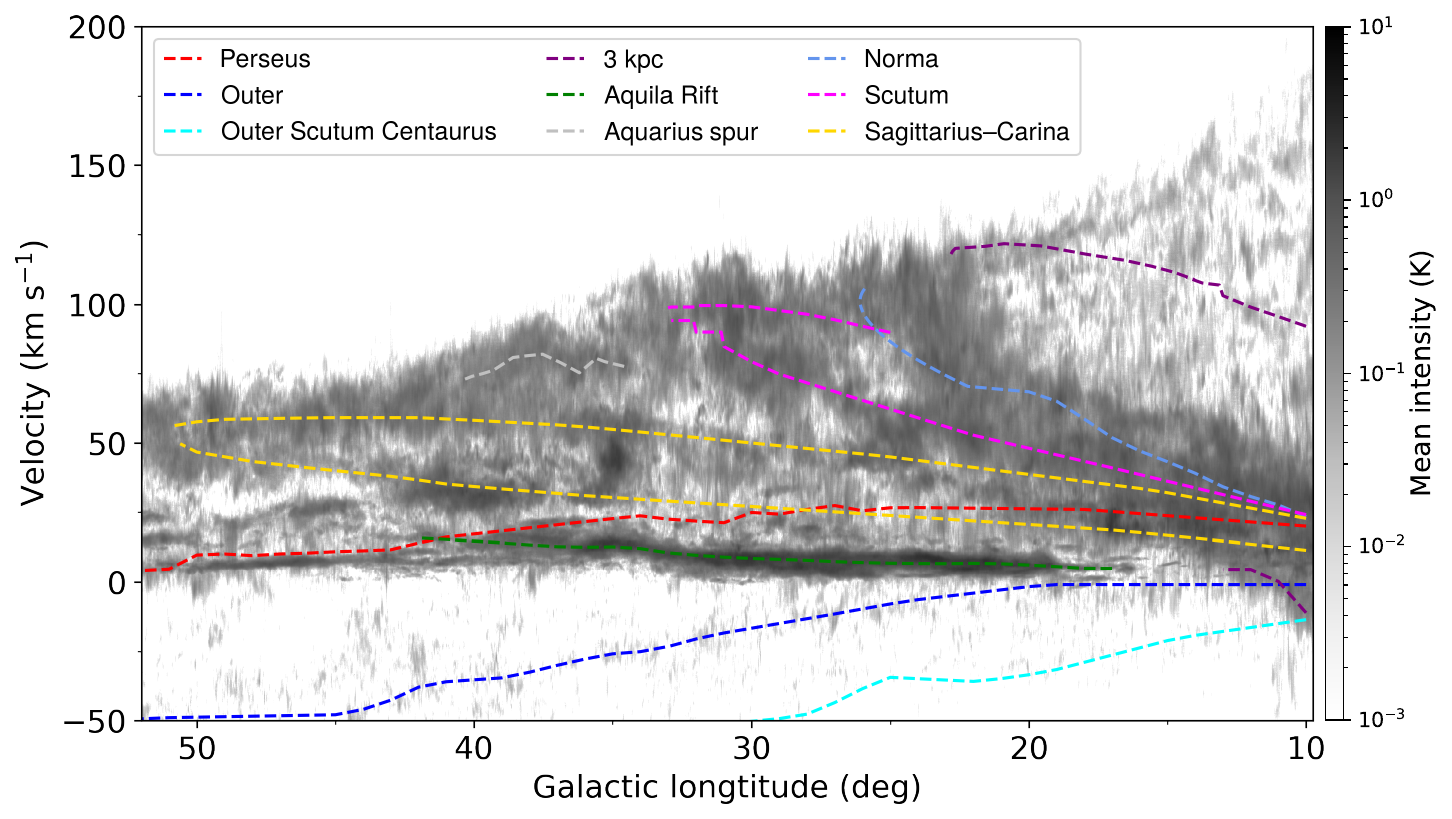} 
\caption{Zoom in of Figure \ref{fig:arm} toward the inner Galaxy.  \label{fig:armQ1}}  
\end{figure}

\subsection{Spectra with the maximum temperature}
As illustrative examples, we show the spectra that correspond to the maximum \cofs\ brightness temperature lines in Figure \ref{fig:max}. The maximum brightness temperature of \cofs\ is 63.6~K. In comparison, at the same position, the brightness temperature observed by the CfA survey \citep{2001ApJ...547..792D} is approximately 8~K,  and it was not the maximum brightness temperature in the CfA survey datacube. The significant increase in brightness temperature is due to the refined angular resolution of the MWISP survey, which is about ten times higher than that of the CfA survey. This difference indicates that resolution is critical for the fidelity of molecular cloud images. 

\subsection{$l$-$V$  diagrams}
Figure \ref{fig:lv} displays the three-color $l$-$V$ map of molecular clouds, using the same color code as in Figure \ref{fig:rgb}. An alternative version of the $l$-$V$ map, with \cofs\  overlaid by an arm model \citep{2019ApJ...885..131R}, is shown in Figure \ref{fig:arm}. The flux of CO emission decreases significantly from the Local Arm to the Perseus Arm, and then to the Outer Arm, indicating that the observed emission at far distances is incomplete \citep[see][for a detailed investigation]{2021ApJS..256...32S}. 

 A zoomed-in view of molecular clouds in the first Galactic quadrant is shown in Figure \ref{fig:armQ1}. Many molecular clouds deviate from the arm model of \citet{2019ApJ...885..131R}. Interpreting all the bridges between arm segments as spurs would yield a significantly different Galactic structure. A detailed analysis of the large-scale structure of molecular clouds is provided by  \citet{2024ApJ...977L..35S}. 

\subsection{Comparison of MWISP with Other surveys}
\begin{figure} 
    \plotone{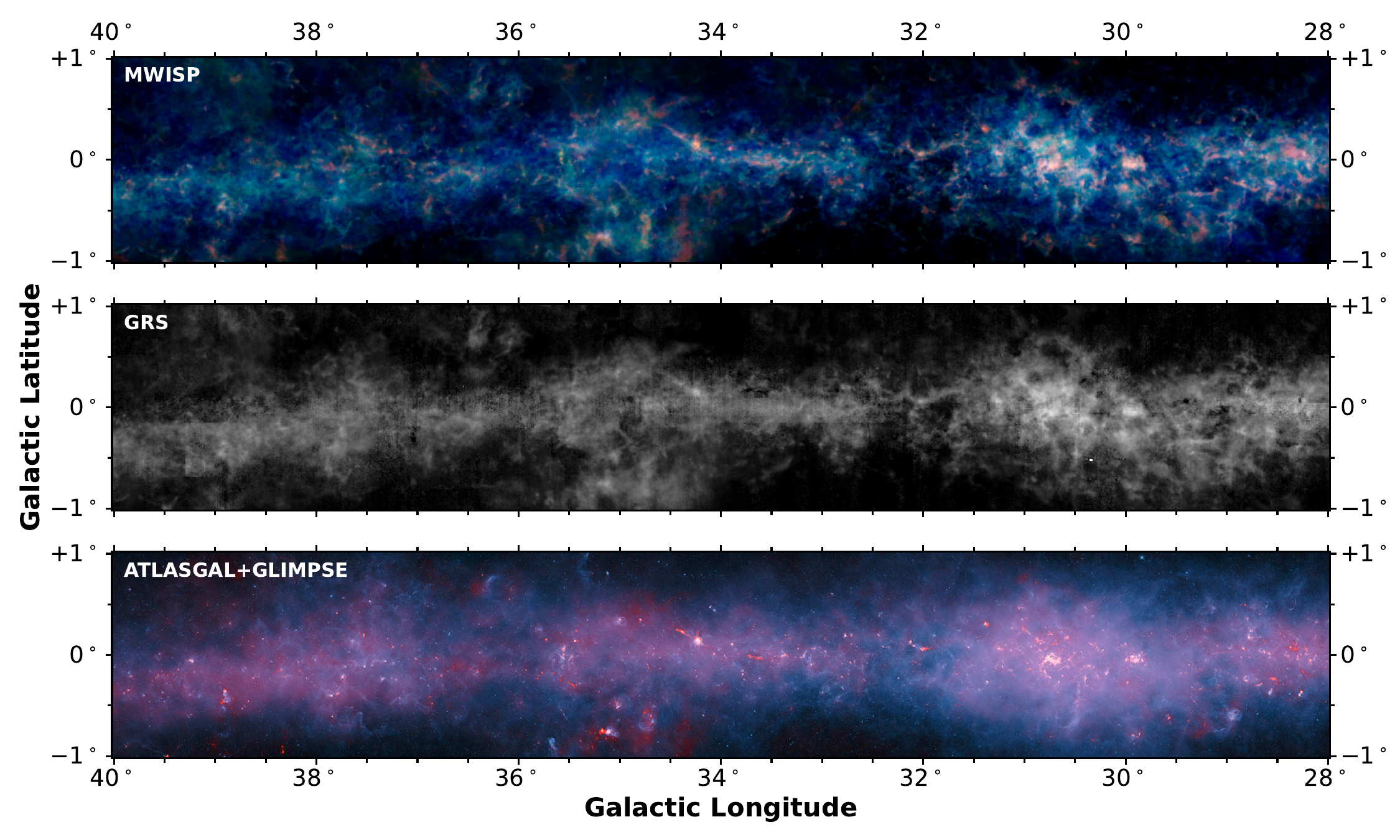} 
    \caption{ Comparison of the MWISP (upper), the GRS (middle), the ATLASGAL--GLIMPSE (bottom) survey. The MWISP RGB image is composed of \cots\ (red), \coss\ (green), and \cofs\ (blue) emissions. The Galactic Ring Survey (GRS) \coss\ data are obtained from the Boston University–FCRAO Galactic Ring Survey archive (\url{https://www.bu.edu/galacticring/new_index.htm}).  The ATLASGAL--GLIMPSE composite image is adapted from the ESO press release (\url{https://www.eso.org/public/news/eso1606/}).  \label{fig:comparesurvey}}  
\end{figure}

 In this section, we compare the MWISP survey with the GRS \citep{2006ApJS..163..145J}, the ATLASGAL \citep{2009A&A...504..415S}, and the GLIMPSE \citep{2009PASP..121..213C} surveys, and the purpose is to demonstrate the improved image quality and multi-line capability of the MWISP survey, as well as its ability to bridge molecular-line and far-infrared continuum observations of the Galactic interstellar medium. The results are shown in Figure~\ref{fig:comparesurvey}.  

 In Figure~\ref{fig:comparesurvey}, the MWISP survey is shown with representative existing datasets over a region of the Galactic plane. The upper panel shows the MWISP composite RGB image combining C$^{18}$O, $^{13}$CO, and $^{12}$CO, tracing molecular gas of different densities. The middle panel presents the corresponding \cos\  integrated intensity map from the GRS. Most of the strong emission features in both panels appear  similar between GRS and MWISP, but the latter exhibits more comprehensive features in the faint parts. This is consistent with the flux statistics reported in  \citet{2021ApJS..256...32S}. The bottom panel shows a composite image from the ATLASGAL (870~$\mu$m, red) and Spitzer GLIMPSE (mid-IR, blue) surveys, highlighting the complementary distribution of dust continuum emission associated with the CO structures. 
\begin{deluxetable}{c|c|cc}
\tablecaption{Properties of \cofs\ cloud samples\tablenotemark{a}.  \label{Tab:overcat}}
\tablehead{
   \multicolumn{2}{c|}{ }  &  \colhead{Clouds complete in PPV}   & \colhead{All molecular clouds} 
} 
\startdata
  \multicolumn{2}{c|}{Total cloud number }    &  103517  &  104054 \\  
  \multicolumn{2}{c|}{Clouds with \coss\ emission}     & 10790  & 10935  \\
    \multicolumn{2}{c|}{Total \coss\ clumps associated with clouds}   &  23332   &  49920  \\
        \multicolumn{2}{c|}{Clouds with \cots\ emission}  & 304  &  325  \\
            \multicolumn{2}{c|}{Total \cots\ clumps associated with clouds}    & 731 &  5875 \\
\hline  
\multirow{3}{*}{Angular area}  & Minimum & 1.00 arcmin$^2$ & 1.00 arcmin$^2$   \\
\cline{2-4}  & Median &  7.75 arcmin$^2$  & 7.75 arcmin$^2$    \\
 \cline{2-4}  & Maximum   &   15.52 deg$^2$   &   466.92 deg$^2$  \\
  \hline
\multirow{3}{*}{Flux (K \kms arcmin$^2$)}  & Minimum & 0.44      &   0.44   \\
\cline{2-4}  & Median  &     8.04  &    8.07\\
\cline{2-4}  & Maximum& 4.60$\times 10^5$   &  4.41$\times 10^7$  \\ 
\hline
\multirow{3}{*}{Peak intensity}  & Minimum &  0.91 K&  0.91 K     \\
\cline{2-4}  & Median &     3.02 K &     3.02 K   \\
 \cline{2-4}  & Maximum  &      46.95 K &      63.63 K  \\
 \hline
 \multirow{3}{*}{Equivalent width (\kms)}  & Minimum &   0.20   &   0.20    \\
\cline{2-4}  & Median &       1.00    &       1.00     \\
 \cline{2-4}  & Maximum &     42.00   &     42.00    \\
 \hline
  \multirow{3}{*}{$V_{\rm LSR}$ (\kms) }  & Minimum &   -121.14   &   -121.14   \\
\cline{2-4}  & Median   &      0.34  &      0.36      \\
 \cline{2-4}  & Maximum &   195.26  &   195.26    \\ 
\enddata
\tablenotetext{a}{For comparison, we list cloud properties based on both PPV-complete and all molecular clouds. A complete cloud  in PPV datacube means its outer envelope traced by DBSCAN is completely within the mosaicked data cube, not truncated by any edge of the cube.}
\end{deluxetable}

\begin{figure} 
 
\gridline{\fig{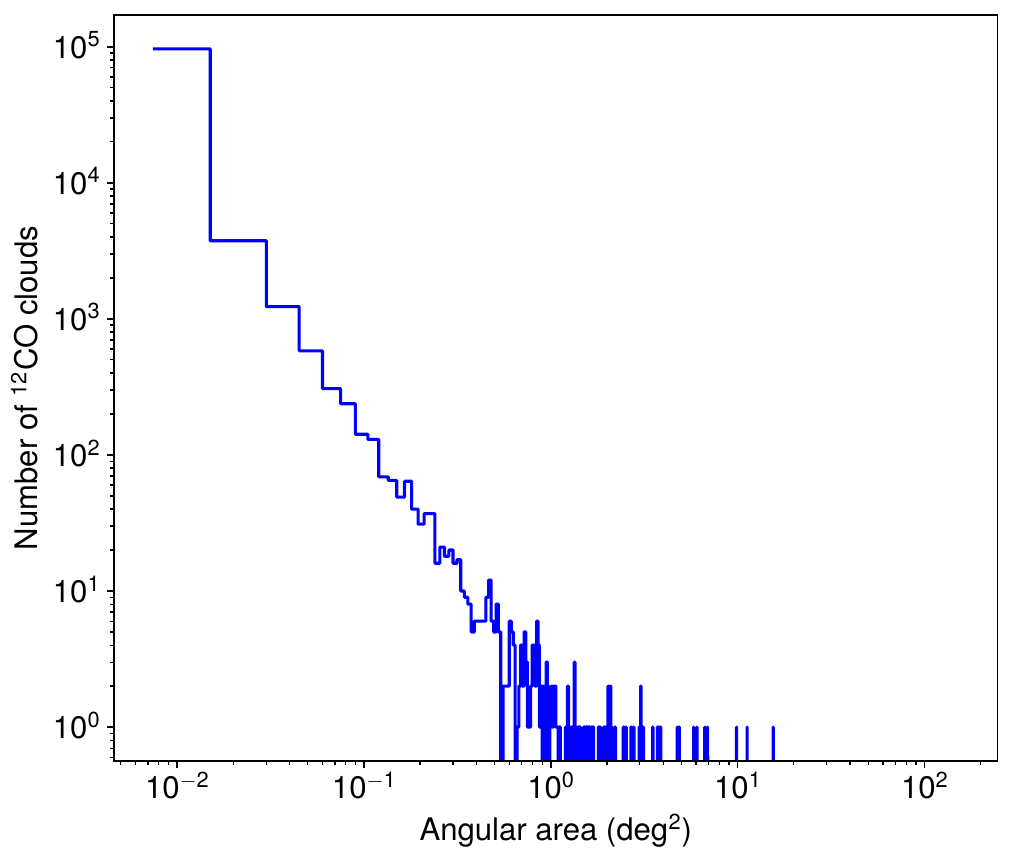}{0.45\textwidth}{} \fig{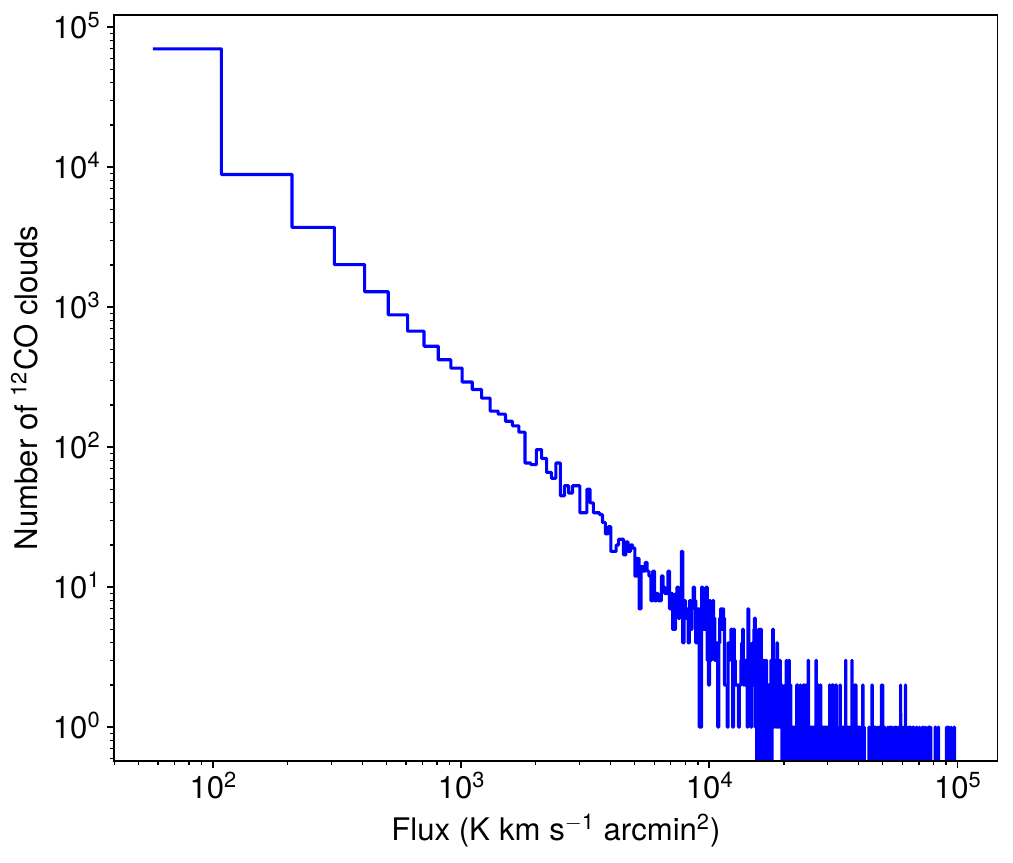}{0.45\textwidth}{}  }
\gridline{\fig{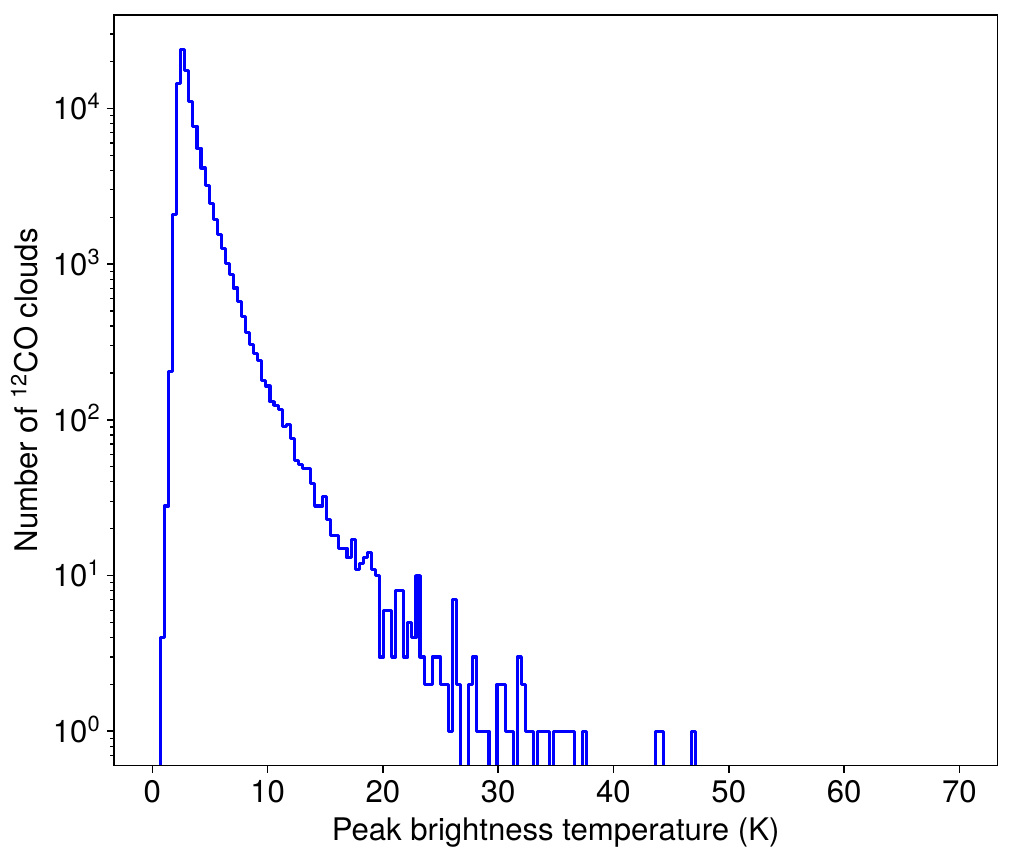}{0.45\textwidth}{} \fig{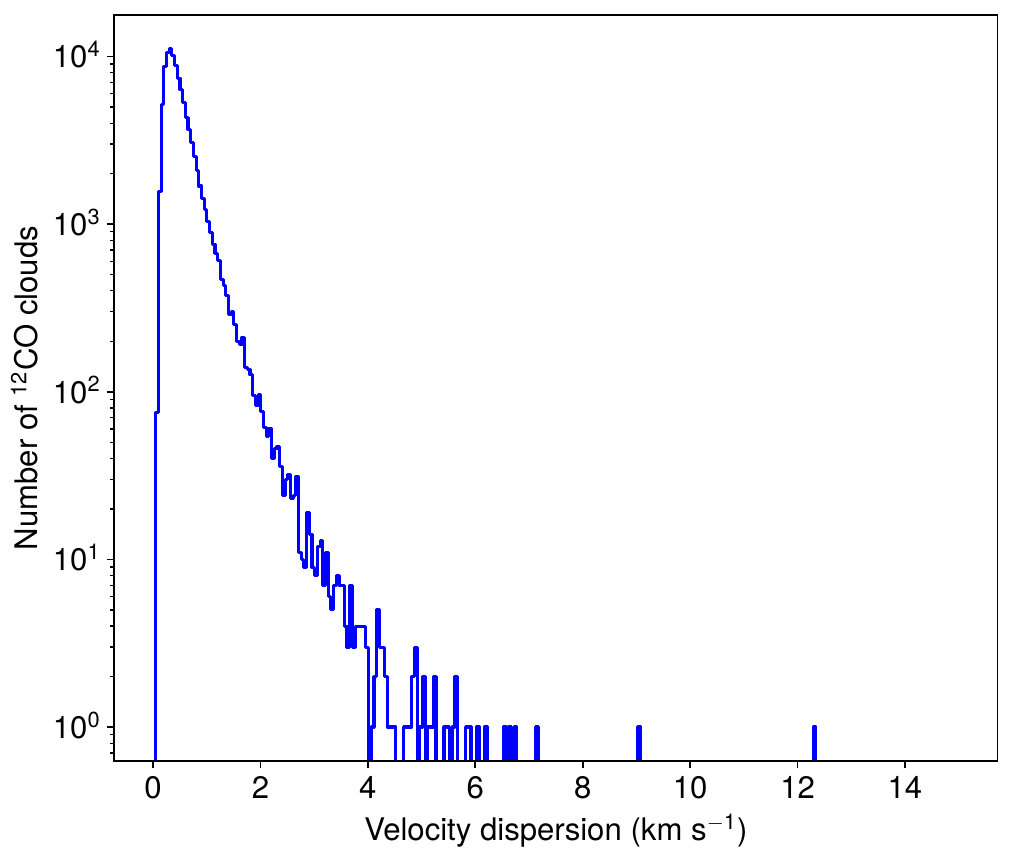}{0.45\textwidth}{}  }
\caption{ Statistics of 103,517 complete molecular clouds in PPV data cube. \label{fig:statcat} }
\end{figure}

 \begin{figure*} 
 
\gridline{\fig{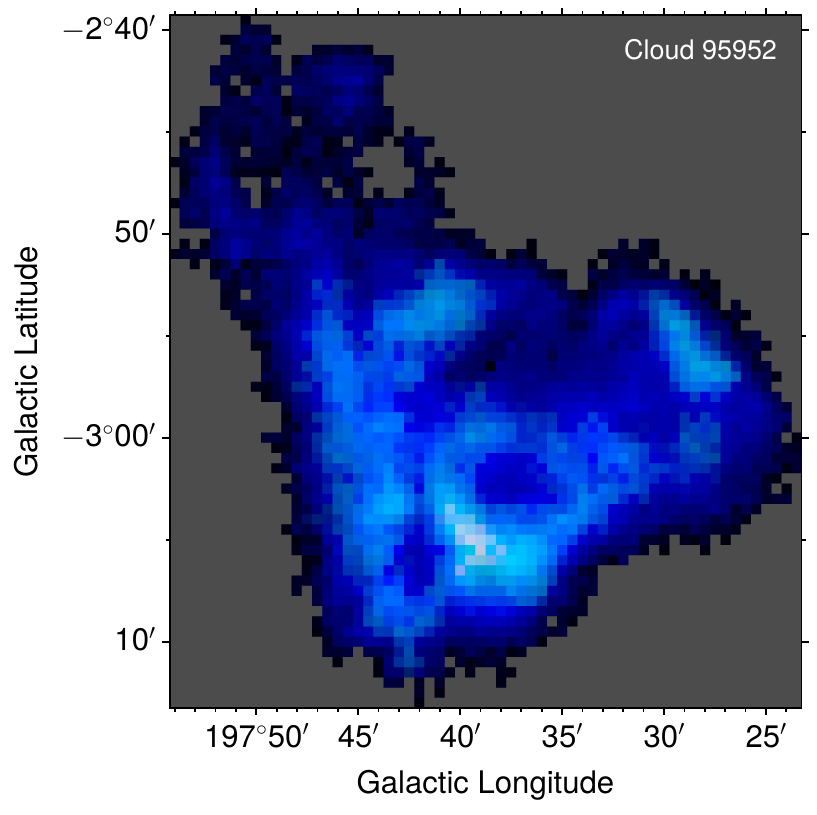}{0.27\textwidth}{} \fig{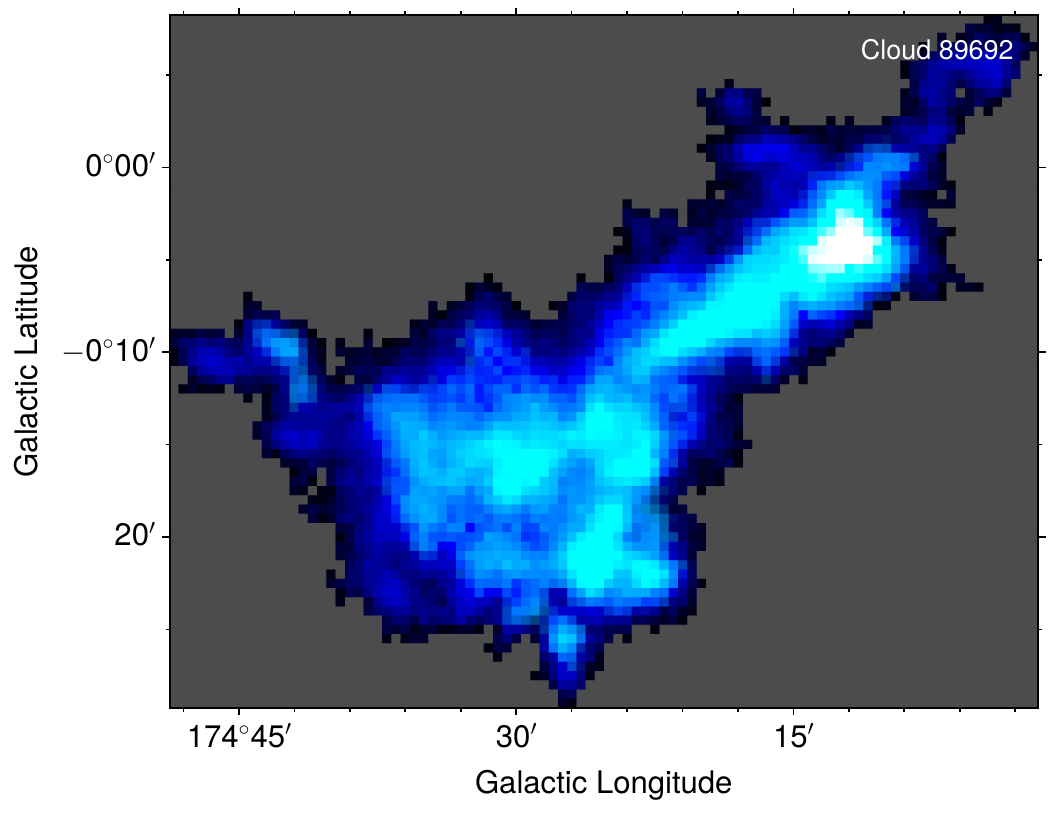}{0.27\textwidth}{} \fig{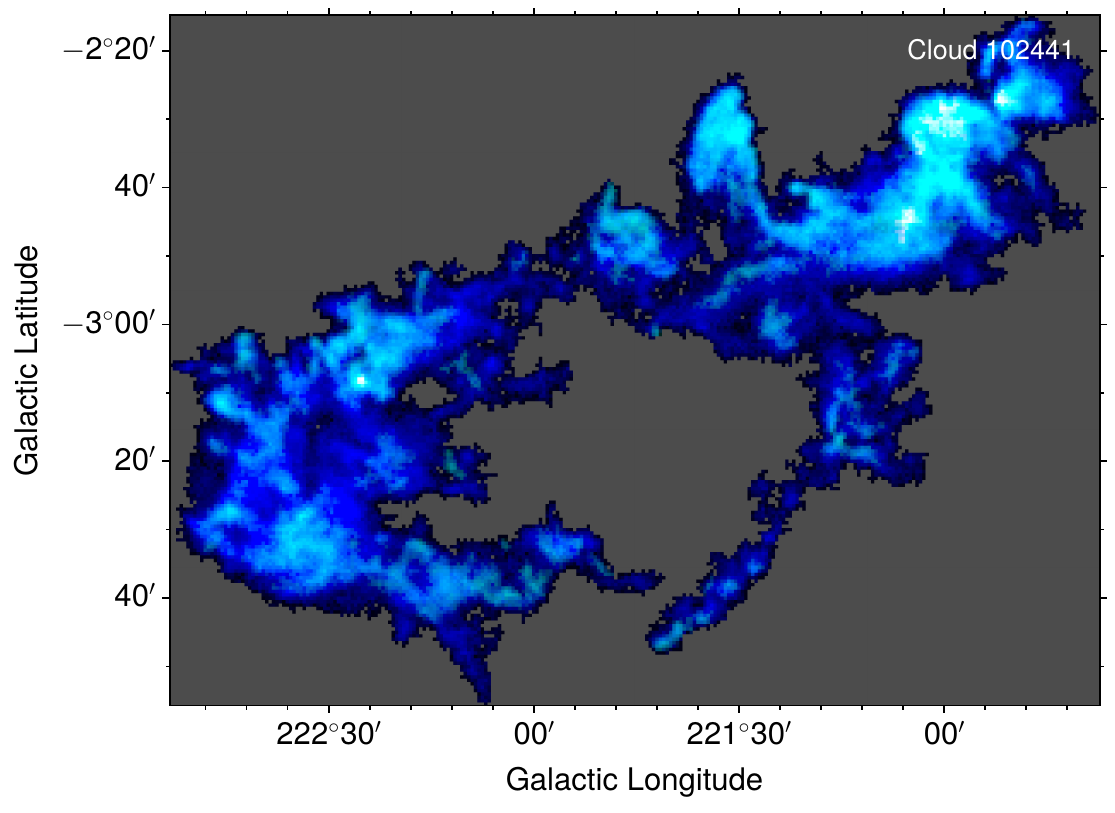}{0.27\textwidth}{}  }

\gridline{\fig{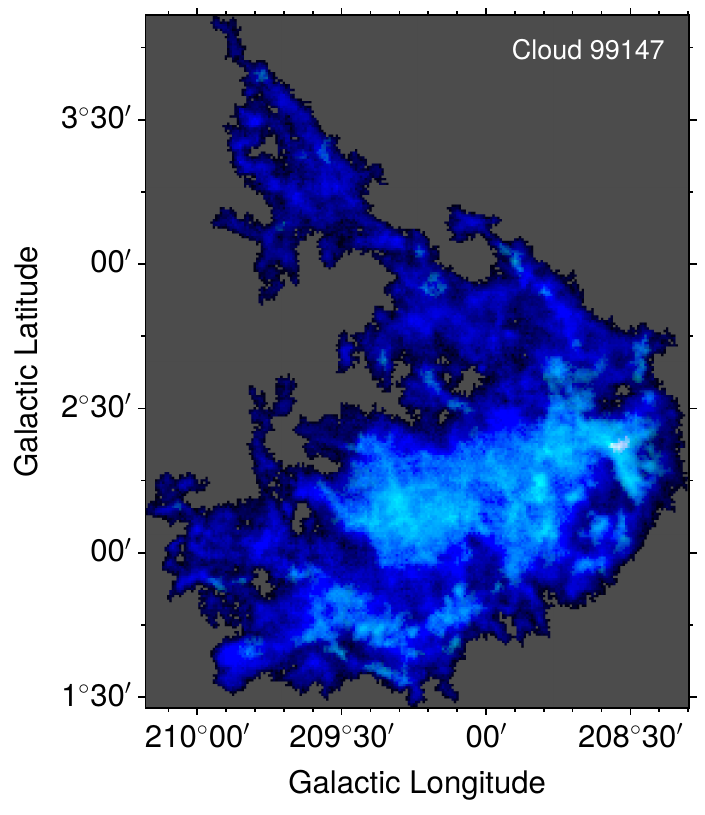}{0.27\textwidth}{} \fig{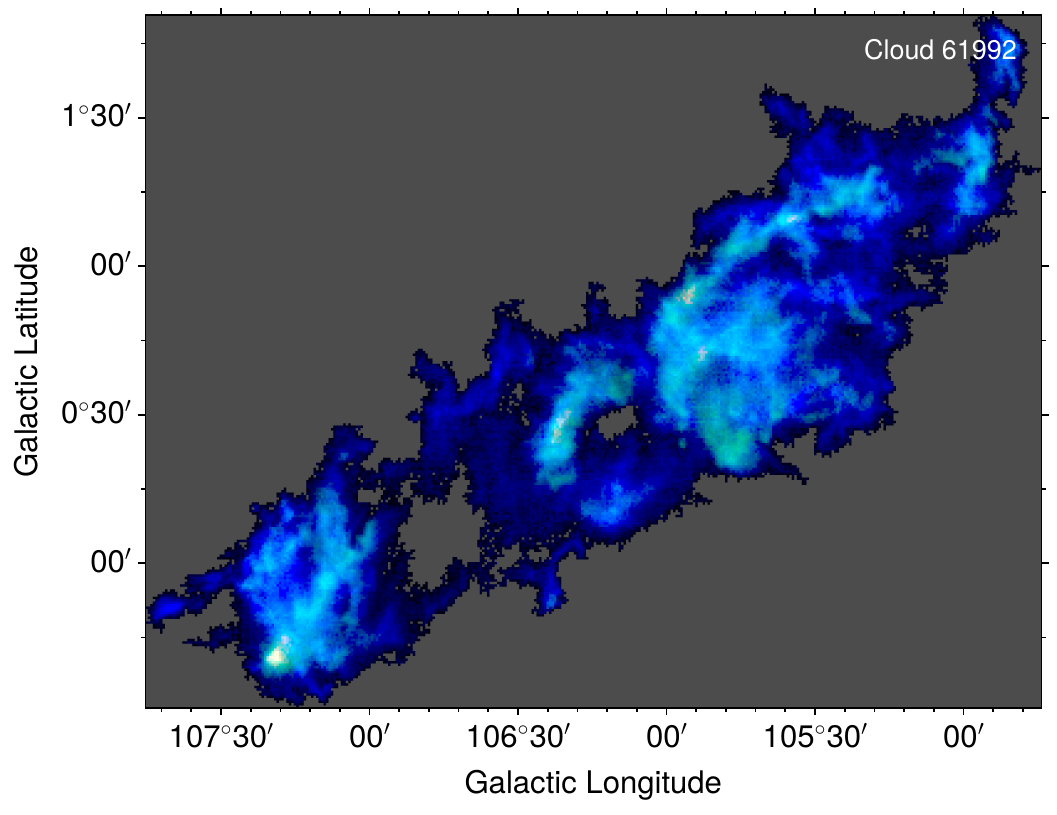}{0.27\textwidth}{} \fig{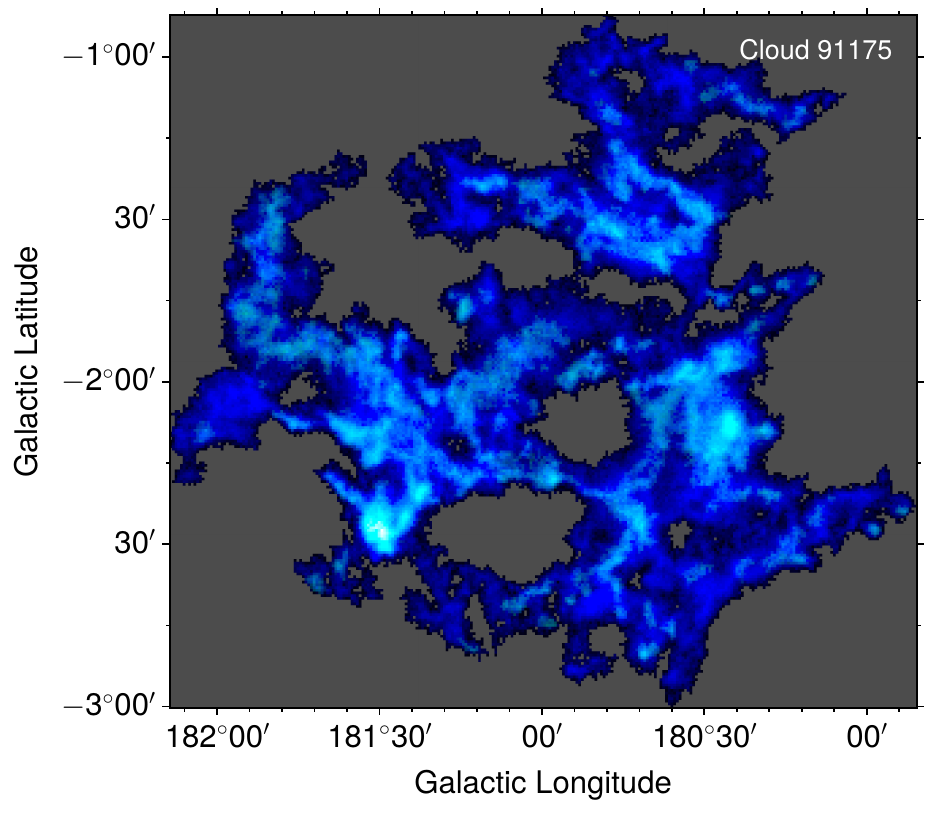}{0.27\textwidth}{}  }

\gridline{\fig{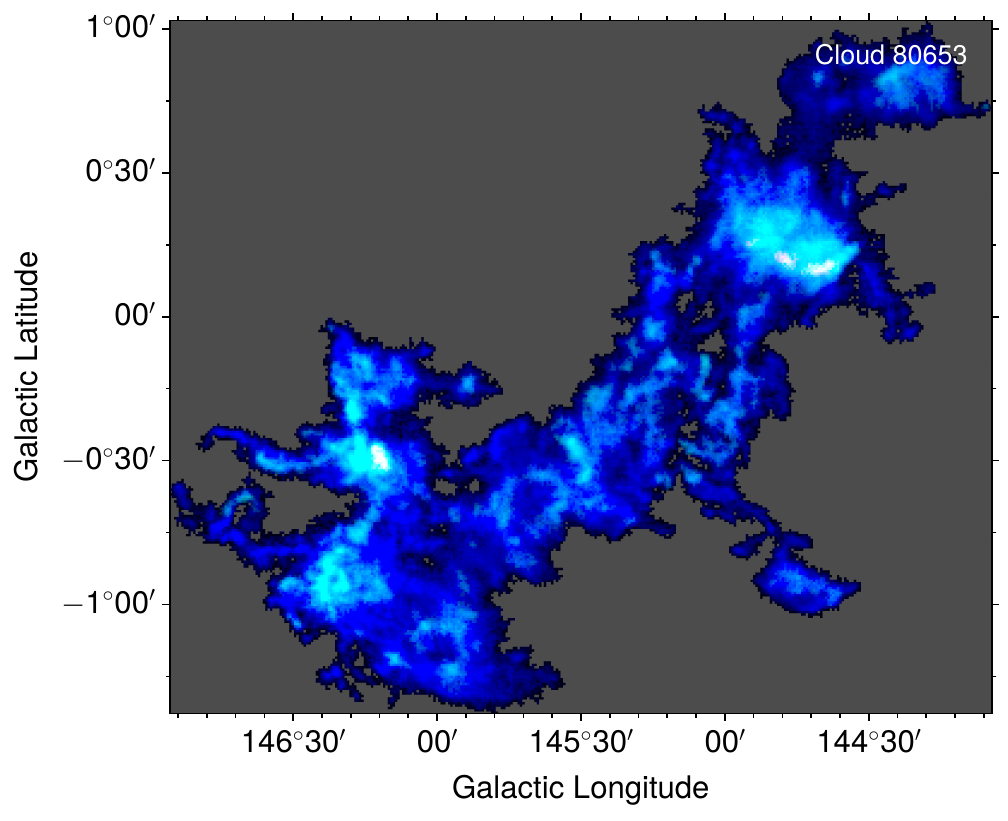}{0.27\textwidth}{} \fig{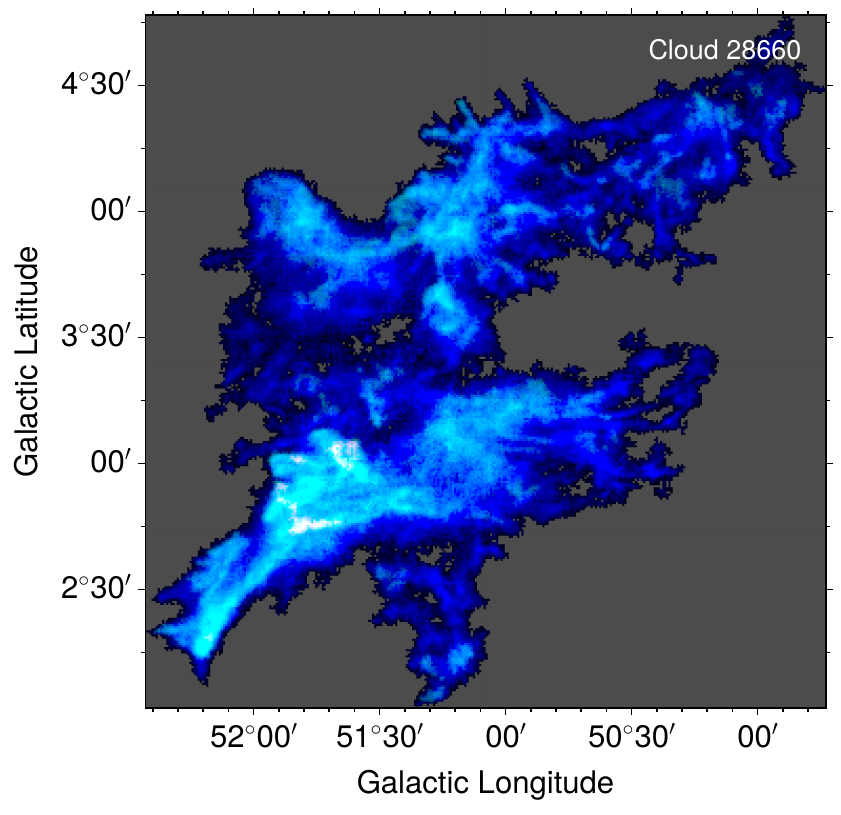}{0.27\textwidth}{} \fig{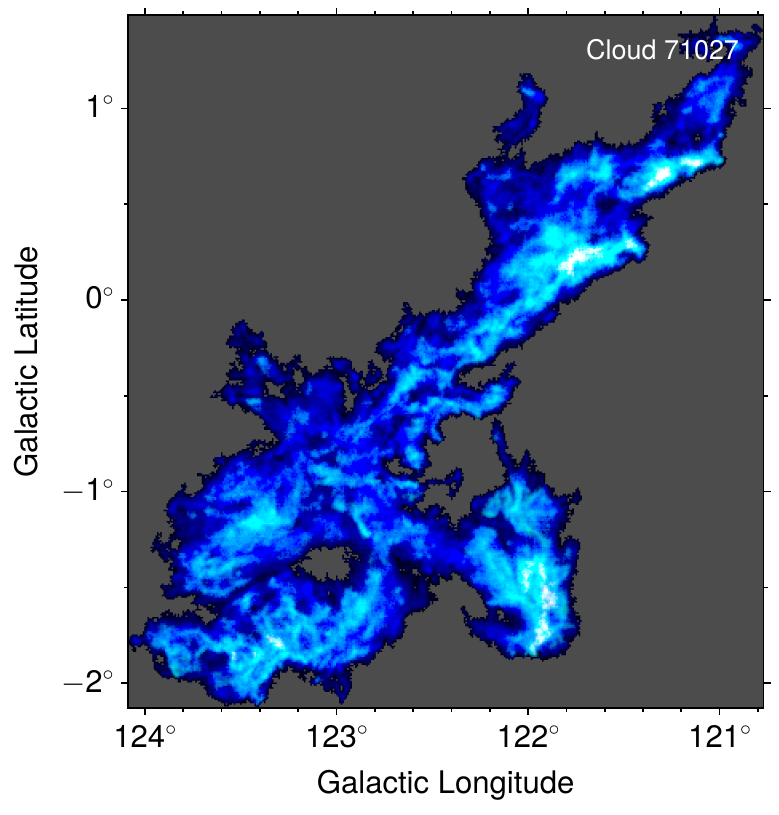}{0.27\textwidth}{}  }

\gridline{\fig{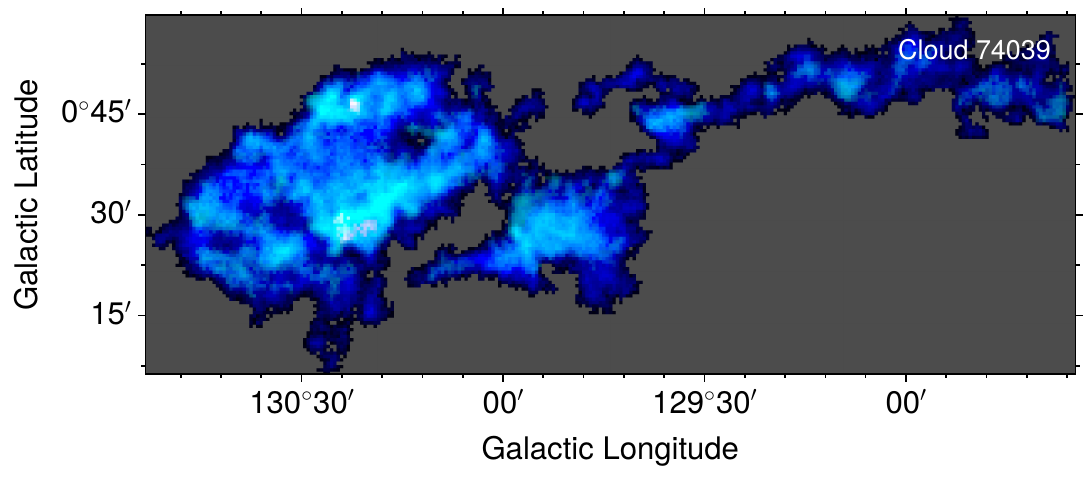}{0.27\textwidth}{} \fig{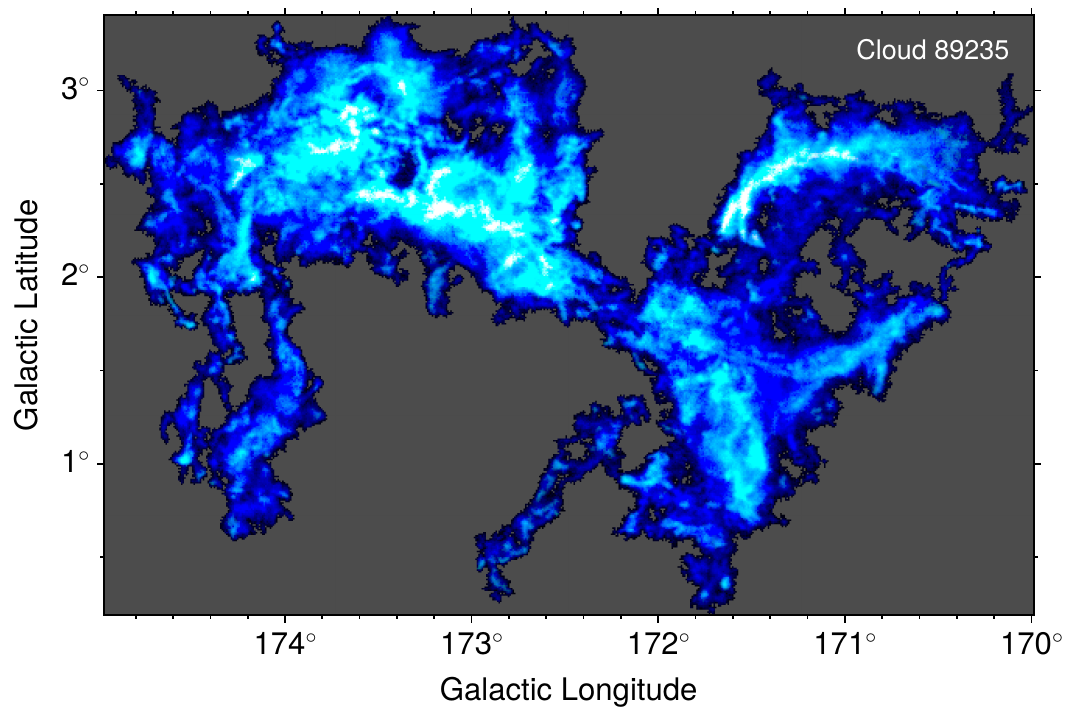}{0.27\textwidth}{} \fig{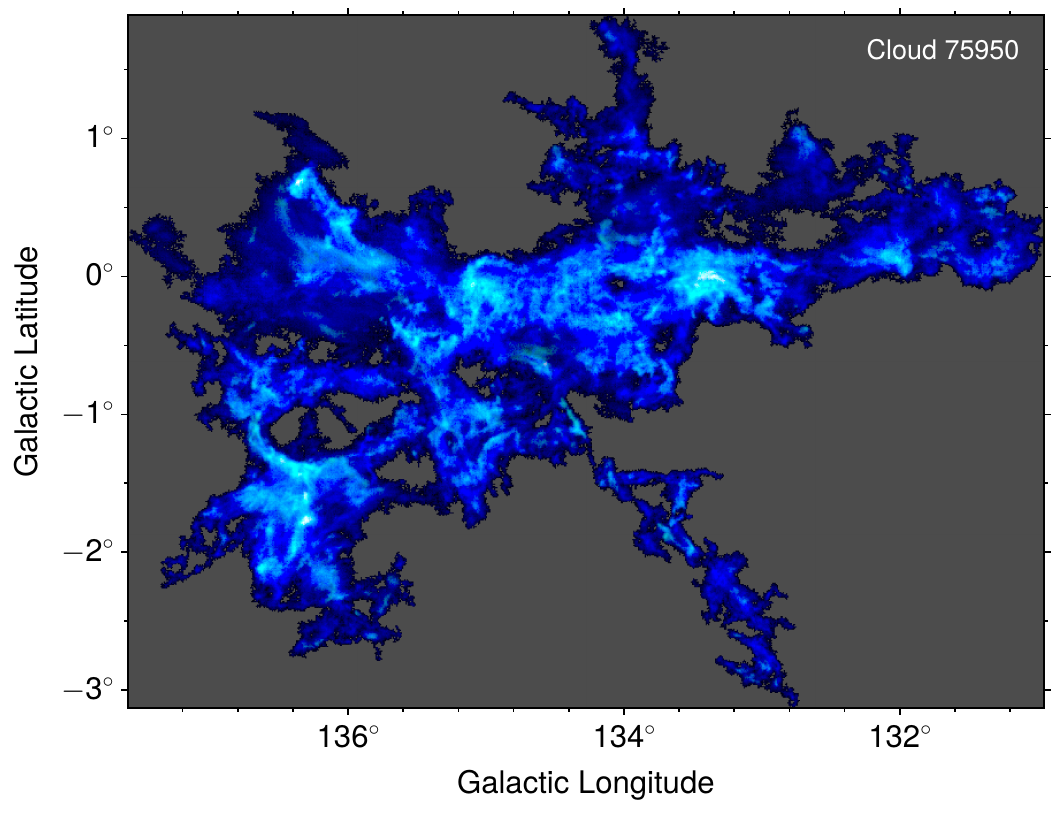}{0.27\textwidth}{}  }

\caption{A gallary of RGB images of molecular clouds. \cofs, \coss, and \cots\ emission are color-coded with blue, green, and red, respectively. \label{fig:cloudgallary} }
\end{figure*}

\section{the cloud catalog}
\label{sec:co12cat}

 In this section, we describe the molecular cloud samples constructed using the DR1 data of the MWISP survey. In principle, the edges of molecular clouds are defined with \cofs\ emission, while \coss\ and \cots\ emissions are considered as internal structures associated with the clouds.  

In order to identify molecular clouds, the first attempt was made to decompose contiguous structures in PPV data cube. On the one hand, the main purpose at this step is to search for the largest isolated structures \citep[e.g.,][]{2015ARA&A..53..583H}. No further decomposition was made in \cofs\ to look for substructures. We leave substructure identification to \coss\ and \cots\ which trace densities higher than \cofs.  On the other hand, when two physically independent molecular clouds are observed as a connected structure in PPV data cube, it is difficult to reconstruct their individual PPV images. This uncertainty of PPV data is discussed in detail by \citet{2013ApJ...777..173B}. Molecular clouds at different distances along the line of sight may possess close radial velocities, leading to superposition in PPV data \citep[e.g.,][]{1992ApJ...384...95A,2001ApJ...546..980O}, while components within a single molecular cloud may appear as separate structures in PPV data cube due to relatively large velocity differences \citep[e.g.,][]{2000ApJ...532..353P}.  Both cases pose a difficulty in identifying a cloud uniquely within the PPV data cube and this drawback should be given particular attention in the crowded regions such as the first and fourth quadrants. Consequently, we have made no attempt to further decompose \cofs\ substructures within a cloud. Instead, we only attempt to identify \coss\ and \cots\ substructures that are more localized and therefore the possibility of overlap is much lower. We will arrive at statistics of the surface filling factor in Section \ref{sec:sff}.

Several popular algorithms, such as SCIMES \citep{2015MNRAS.454.2067C} and Dendrograms \citep{2008ApJ...679.1338R}, have been proposed.  These algorithms are helpful in identifying individual clouds from a 3D data cube. However, some of these algorithms try to split large structures in PPV data cube into hierarchical substructures.  Such algorithms may consume a large amount of computing resources and thus are not practical for a large datacube. Given the large size of the MWISP DR1 data, the algorithm that we seek needs to be both computationally efficient and capable of clearly defining molecular clouds. Based on these requirements, DBSCAN was examined as a suitable algorithm for constructing molecular cloud samples from the MWISP survey data \citep{2020ApJ...898...80Y}.  
 
\subsection{The DBSCAN algorithm} 
\label{sec:dbscan}
DBSCAN \citep{Ester1996} is an unsupervised machine learning clustering algorithm that isolates  contiguous structures from PPV data cubes.  Due to the irregular morphology of molecular clouds, the DBSCAN algorithm runs on mosaicked datacbues of three CO lines. Before applying the DBSCAN algorithm, a cutoff of 2$\sigma$ is imposed on the entire data cube. To reject noise clusters, four additional criteria are further applied: (1) an area at least 1 beam size in the $l$-$b$ plane projection, which requires an area of $\geqslant$ 2$\times$2 pixels  (one-beam size); (2)  a minimum span of three velocity channels; (3) a minimum PPV voxel number of 16;  (4) a minimum peak intensity of 5$\sigma$. Detailed descriptions and comparisons with other algorithms for the application of DBSCAN to MWISP data are presented in \citet{2020ApJ...898...80Y}.

 The cloud catalog is incomplete for small-sized or faint molecular clouds. For instance, the requirement of a minimum peak of 5$\sigma$ is a strong filtering criterion, primarily influenced by sensitivity.  In addition to sensitivity, small molecular clouds are also affected by the angular resolution and the limitations of the algorithms used in detection \citep{2022AJ....164...55Y}. 

As mentioned above, we also apply the DBSCAN algorithm to the  \coss\ and \cots\ data cubes but structures identified in these two lines are considered as clumps or dense cores associated with (or within) the molecular clouds defined by DBSCAN in \cofs.  Before running the DBSCAN algorithm on the datacubes, we imposed extra mask on the data cubes using \cofs\ for \coss\ and \coss\ for \cots,  in addition to the 2$\sigma$ cutoff. A mask is an accompany datacube with voxel values equal to 1 for regions of interest and equal to 0 for unrelated regions. This approach ensures that \coss\ regions are enclosed by \cofs,  and \cots\ regions are enclosed by \coss.

\begin{deluxetable*}{cccccccccc}
\tabletypesize{\tiny}
\tablecaption{\cofs\ cloud catalog\tablenotemark{\rm 1}.\label{Tab:cloudDis}}
\tablehead{ 
   Index &  $l$\tablenotemark{\rm 2}  &   $b$ &\colhead{$V_{\rm LSR}$}  & $\sigma_l$ & $\sigma_b$ & $\sigma_{V_{\rm LSR}}$  & $T_{\rm peak}$ &   Angular area & Flux  \\
   &  (\deg)  &  (\deg) & (\kms)  &  (\deg)  &  (\deg) & (\kms)  & (K)    &arcmin$^2$ &   (K \kms\ arcmin$^2$)   
}
\colnumbers 
\startdata
1 & 9.77188$\pm$0.00044 &-0.97570$\pm$0.00053 & -0.465$\pm$0.026 & 0.00637$\pm$0.00035 & 0.00812$\pm$0.00036 & 0.324$\pm$0.020 & 2.44$\pm$0.48 & 3.00$\pm$0.21 & 1.97$\pm$0.12  \\ 
2 & 9.77637$\pm$0.00065 &-1.99177$\pm$0.00050 & -11.929$\pm$0.027 & 0.00756$\pm$0.00048 & 0.00493$\pm$0.00039 & 0.208$\pm$0.024 & 2.70$\pm$0.48 & 2.25$\pm$0.24 & 1.247$\pm$0.093  \\ 
3 & 9.78032$\pm$0.00036 &0.01172$\pm$0.00037 & -4.606$\pm$0.025 & 0.00825$\pm$0.00027 & 0.00840$\pm$0.00027 & 0.595$\pm$0.017 & 2.45$\pm$0.49 & 4.75$\pm$0.25 & 4.25$\pm$0.16  \\ 
4 & 9.78262$\pm$0.00037 &-1.64957$\pm$0.00045 & 21.769$\pm$0.020 & 0.00741$\pm$0.00028 & 0.00944$\pm$0.00032 & 0.355$\pm$0.014 & 2.60$\pm$0.48 & 5.00$\pm$0.36 & 3.94$\pm$0.16  \\ 
5 & 9.78847$\pm$0.00035 &1.36788$\pm$0.00038 & 30.711$\pm$0.027 & 0.00743$\pm$0.00026 & 0.00838$\pm$0.00026 & 0.628$\pm$0.016 & 2.42$\pm$0.47 & 4.50$\pm$0.26 & 4.54$\pm$0.18  \\ 
6 & 9.79163$\pm$0.00073 &-1.93839$\pm$0.00048 & -0.507$\pm$0.017 & 0.01444$\pm$0.00041 & 0.00877$\pm$0.00037 & 0.216$\pm$0.015 & 2.47$\pm$0.48 & 5.75$\pm$0.34 & 3.17$\pm$0.15  \\ 
7 & 9.79226$\pm$0.00053 &0.33548$\pm$0.00054 & -12.232$\pm$0.030 & 0.00598$\pm$0.00041 & 0.00683$\pm$0.00039 & 0.305$\pm$0.023 & 2.35$\pm$0.45 & 2.25$\pm$0.22 & 1.427$\pm$0.099  \\ 
:& : &: & : & : & : & : & : & :   & :  \\ 
103515 & 229.70241$\pm$0.00037 &-0.07449$\pm$0.00033 & 70.748$\pm$0.015 & 0.01332$\pm$0.00025 & 0.01158$\pm$0.00024 & 0.478$\pm$0.011 & 3.09$\pm$0.53 & 9.75$\pm$0.39 & 10.04$\pm$0.25  \\ 
103516 & 229.71299$\pm$0.00031 &-1.06747$\pm$0.00030 & 38.702$\pm$0.017 & 0.00962$\pm$0.00021 & 0.00920$\pm$0.00023 & 0.504$\pm$0.011 & 2.95$\pm$0.47 & 6.75$\pm$0.32 & 8.45$\pm$0.23  \\ 
103517 & 229.72255$\pm$0.00048 &-2.79442$\pm$0.00048 & 26.662$\pm$0.023 & 0.00669$\pm$0.00034 & 0.00661$\pm$0.00035 & 0.223$\pm$0.019 & 2.69$\pm$0.48 & 3.25$\pm$0.27 & 1.81$\pm$0.11  \\ 
\enddata
\tablenotetext{1}{This table lists 10 example clouds, and a complete machine readable version of the cloud catalog can be obtained at \url{https://doi.org/10.57760/sciencedb.27351}.} 
\tablenotetext{2}{Galactic coordinates ($l$, $b$, and $V_{\rm LSR}$) and their sizes ($\sigma_l$, $\sigma_b$, and $\sigma_{V_{\rm LSR}}$) are the mean and standard derivation of voxels weighted by temperatures, and see \citet{2006PASP..118..590R} for detailed definitions and formula.  The uncertainties of positions and sizes are estimated with error propergation. The uncertainty of $T_{\rm peak}$ is obtained based on the 2-D rms map shown in Figure \ref{fig:noiseimg12}. The uncertainty in the angular area is estimated by treating each pixel as an independent Bernoulli process, where a value of 1 indicates  that at least one voxel in the pixel spectrum exceeds 2$\sigma$. The flux uncertainty is obtained by propagating the temperature uncertainties of the individual voxels. Measurements are accurate to two significant digits of uncertainties.}  
\end{deluxetable*} 

\tabletypesize{\scriptsize}

\subsection{Catalogs of \cofs\ clouds and its associated dense-gas clumps in \coss\ and \cots\ } 

In total, we identified 104,054 molecular clouds using the \cofs\ PPV data cube, and among those clouds, 103,517 are complete in PPV data cube.  Basic statistics of these molecular clouds are summarized in Table \ref{Tab:overcat}. Parameters of 103,517 clouds are demonstrated in Table \ref{Tab:cloudDis}. 

Distributions of the \cofs\ clouds, including angular area, flux, peak intensity, and equivalent linewidth, are plotted in Figure \ref{fig:statcat} for 103,517 clouds that are complete in PPV data cube. Both the angular area and flux show power-law distributions, consistent with results reported by \citet{2020ApJ...898...80Y}. The decline in the number of molecular clouds at low peak intensities and small equivalent linewidths is due to observational limits. A gallery of three-color images of molecular clouds is displayed in Figure \ref{fig:cloudgallary}. This catalog may serve as a working catalog of molecular clouds for different purposes of study. 

The numbers of \coss\ and \cots\ clumps are summarized in Table \ref{Tab:overcat}. For a comparison, we display statistics based on both PPV-complete and all molecular clouds. For PPV-complete molecular clouds not touching the PPV edges, the fraction exhibiting \coss\ emission is about 10.4\%, while those with \cots\ account for 0.3\%. However, these fractions may vary significantly with detection algorithms and should be considered lower limits. The largest molecular cloud is incomplete in the PPV cube, but it contains most of the \cos\ and \cot\ clumps.

It is worth reminding that the construction of catalogs for clouds and clumps from a datacube is not unique. In addition to the algorithm, the boundary—and therefore the derived cloud parameters—show strong dependence on the applied noise levels \citep{2022AJ....164...55Y}. In  the analysis of the MWISP data, there have been several attempts to identify clouds and/or substructures using different algorithms \citep{2021A&A...645A.129Y, 2021ApJ...910..131S, 2023ApJS..267...32J,2024ApJ...977L..35S,2025arXiv250901955J}. Improving the algorithm and noise assessment may lead to an increased number of clouds with detectable isotopologue emission. Therefore, the catalogs provided in this release are recommended to be regarded as interim products. Besides, further comparison among different catalogs is needed to understand plausible bias involved in producing various catalogs.

\begin{figure} 
    \plotone{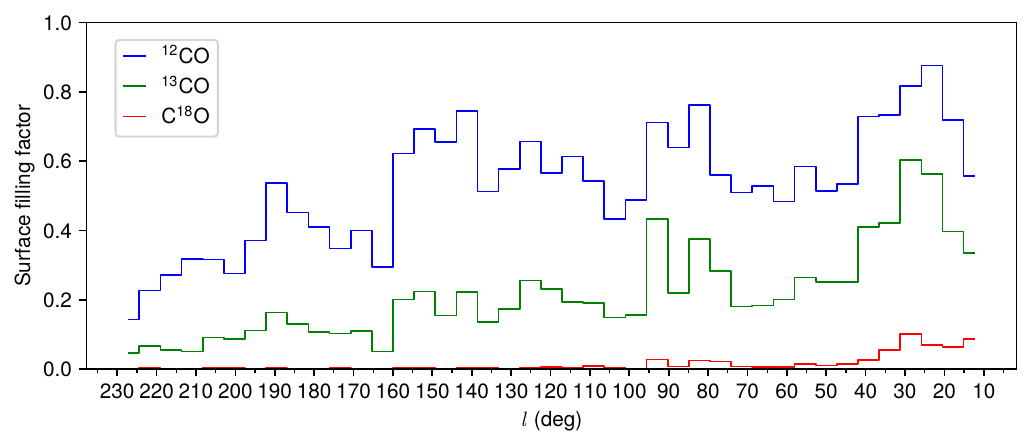} 
    \caption{Surface filling factor of three CO line emission. \label{fig:sff}}  
\end{figure}
 

\section{The released Data Files}
\label{sec:release}

The MWISP DR1 data are released via the Science Data Bank (ScienceDB) platform, an open and trustworthy platform  dedicated to global open scientific data sharing. The data can be openly accessed at \url{https://doi.org/10.57760/sciencedb.27351} \citep{scienceDBMWISP}. Table \ref{Tab:releasefiles} summarizes the released data files.

The file system for the released data consists of single-cell data cubes in FITS, single-cell rms noise images in FITS, mosaicked data cubes in FITS, mosaicked rms noise images in FITS, catalogs of \cofs\ clouds and \coss\ \& \cots\ clumps, tables of statistics, and ancillary files. Collectively, there are about 56 thousand files with a total size of about 3.7 TB. 

In this data release, we provide 9240$\times$3 single-cell FITS, accompanied by an equal number of rms noise images for these single cells in FITS format. These individual data files are provided for maximum convenience across a variety of applications. In addition, the observing log files and information on the reference positions used in observations are also included.

Three mosaicked FITS cubes, each for one CO isotopologue, are also released, along with the corresponding rms noise images. The mosaicked datacubes are the largest files within the release.
Color images generated from the mosaicked data cubes are also provided, including both labeled and unlabeled versions. The images are produced at both high and low resolutions.

The catalogs of \cofs\ clouds and their associated \coss\ and \cots\ clumps are released in CSV format. To make these catalogs easier to use, we also provide the data for each cloud and clump as individual FITS files, packaged in three ZIP files. RGB images of 304 molecular clouds that are complete in PPV cubes and have \cots\ clumps detected are also provided.

Along with the data, we include the pipeline tools developed for processing the MWISP data. Some of these pipelines have been distributed in public domains, such as on GitHub.

\begin{deluxetable}{c|c|cccc}
\tablecaption{Summary of released data files \label{Tab:releasefiles}} 
\tablehead{
   \multicolumn{2}{c|}{File category}   & Number of files   & \colhead{File format}    & \colhead{File size} 
} 
\startdata
    \multicolumn{2}{c|}{Single cell FITS}   &  9240$\times$3   & FITS & 2.6 TB \\
    \multicolumn{2}{c|}{Single cell rms FITS}   &  9240$\times$3    & FITS & $<$0.1 TB  \\
  \multicolumn{2}{c|}{Log files}   & 1   & ZIP & $<$0.1 TB  \\
    \multicolumn{2}{c|}{Reference positions}   & 1   & CSV & $<$0.1 TB \\
    \hline
  \multicolumn{2}{c|}{Mosaicked FITS}   & 1$\times$3   & FITS & 1.1 TB \\
 \multicolumn{2}{c|}{Mosaicked rms FITS}     &  1$\times$3 & FITS &  $<$0.1 TB \\
   \multicolumn{2}{c|}{Integrated FITS}     &  1$\times$3 & FITS & $<$0.1 TB \\
  \multicolumn{2}{c|}{Images}     &  5  & PNG, ZIP & $<$0.1 TB \\
  \hline 
 \multicolumn{2}{c|}{Catalog}    &  1$\times$3 & CSV & $<$0.1 TB \\
  \multicolumn{2}{c|}{Cloud/clump FITS}     &  1$\times$3 & ZIP & $<$0.1 TB \\
  \hline
  \multicolumn{2}{c|}{Code}     &  15  & Python,FITS & $<$0.1 TB \\
    \hline
    \multicolumn{2}{c|}{Tables and figures}     &  32  &  PDF, PNG, ZIP & $<$0.1 TB \\
  \hline
  \multicolumn{2}{c|}{Sum}     &  55,519\tablenotemark{a}    & &3.7 TB \\
\hline  
\enddata 
\tablenotetext{a}{This includes README and auxiliary files used to demonstrate the code usage. This number may change due to  updates after the data release.}
\end{deluxetable}

\section{Discussion}

\label{sec:discuss}

\subsection{Surface filling factors}
\label{sec:sff}
The surface filling factors (SFFs) derived from the MWISP survey provide a general impression of the beam filling factors of CO lines. The SFF variations are consistent with previous studies on the beam filling factor \citep{2021ApJ...910..109Y}. 
 Based on integrated images, we derived SFFs of  53.0\%, 21.6\%, and 1.5\% for \cofs, \coss, and \cots, respectively. The significant difference in SFFs among the three lines calls for caution when comparing intensities among CO isotopologue lines, particularly for small-angular-size structures. 

The SFF is expected to vary across the Galactic plane. Figure \ref{fig:sff} demonstrates the variation of SFFs with respect to the Galactic longitude, derived from the integrated intensity maps (see Figure \ref{fig:rgb}). Notably, the SFFs of all three CO lines tend to decrease toward the outer Galaxy, and the SFF of \coss\ is significantly lower than that of  \cofs, with \cots\ even less. In the outer Galaxy, the SFF of \cots\ drops to approximately zero. 

\begin{figure} 
\centering
\plottwo{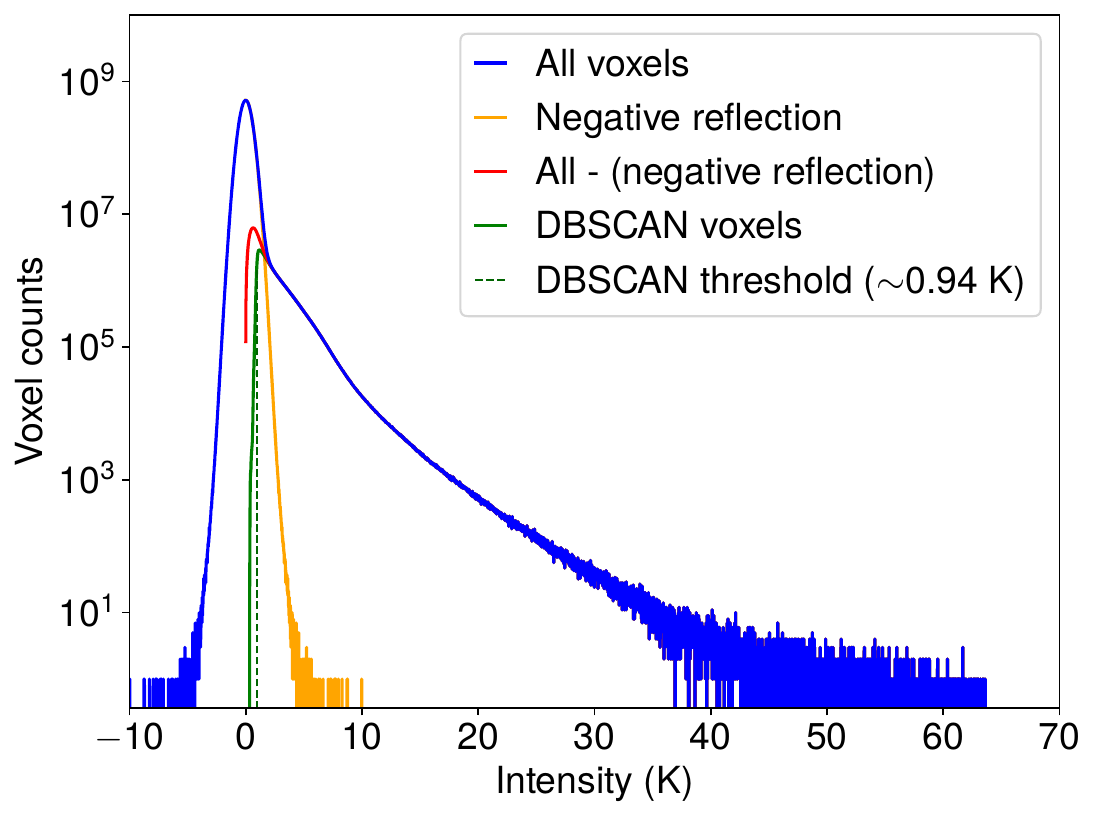}{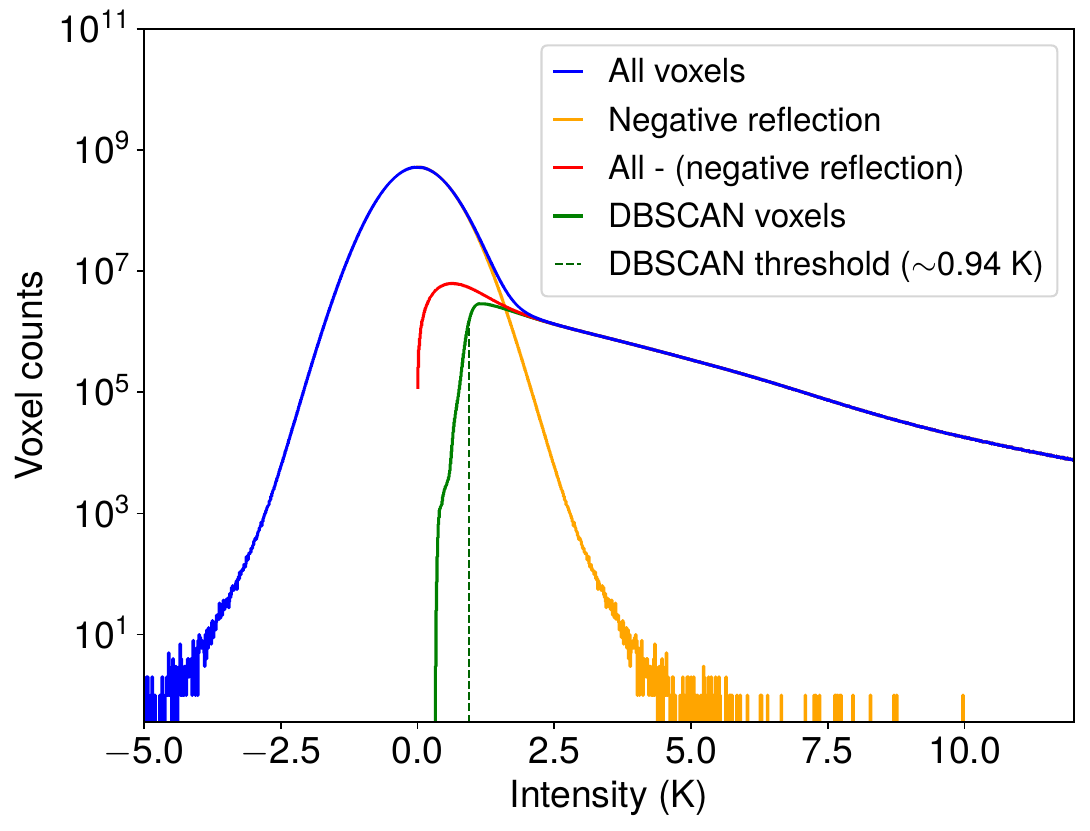}

\caption{Intensity histograms of \cofs\ voxels. The solid orange line is the reflection negative part of the blue lines. The right panel is a zoomed view of the left panel at low intensities. \label{fig:totalflux}}  
\end{figure}

\begin{figure} 
    \plotone{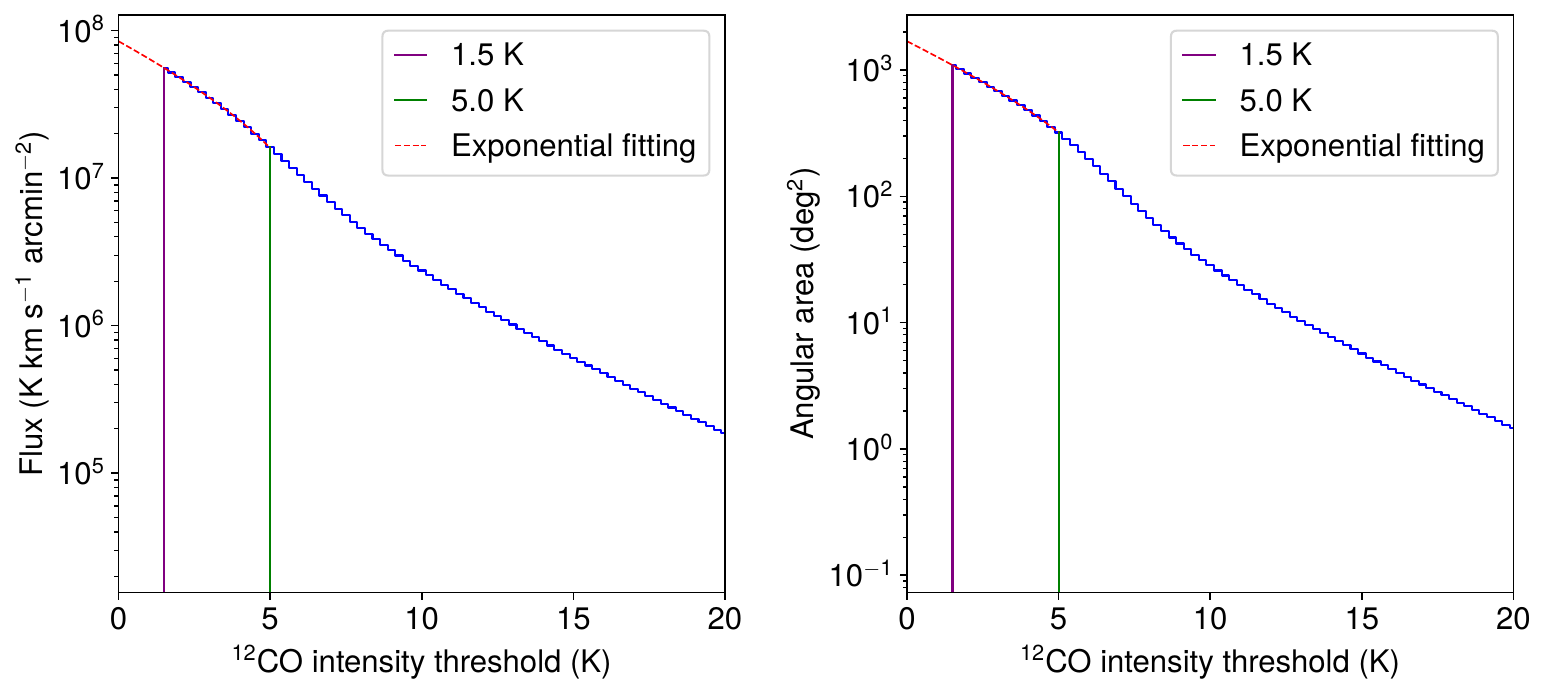} 
    \caption{ Extrapolations of flux and the angular area for the cloud regions.  The blue lines represent the total flux and angular area above certain \cofs\ intensity thresholds across the whole mosaicked PPV data cube. To minimize the noise  effect and fitting uncertainties, the exponential function is fitted with  the data segment between 1.5 and 5.0 K. The fitting exponential function for flux is $\mathit y=(1.04\pm0.01)\times10^8\mathit e^{(-0.2158\pm0.006)\mathit x}-(1.9\pm0.2)\times10^7$, and is $\mathit y=(1933\pm7)\mathit e^{(-0.2463\pm0.003)\mathit x}-(243\pm11) $ for the angular area.  \label{fig:fluxextrapolate}}  
\end{figure}

\begin{deluxetable}{c|cc}
\tablecaption{Flux statistics based on the MWISP DR1 data.  \label{Tab:fluxmeasure}}
\tablehead{
   & \colhead{Flux} &  \colhead{Completeness}   \\
      & \colhead{( K \kms arcmin$^2$)}   
} 
\startdata
  cloud catalog &   6.34$\pm$0.32$\times$$10^7$\tablenotemark{a}  &     82\%/75\%\tablenotemark{b}  \\
   Observed total flux in  \cofs\ PPV cube &   7.69$\pm$0.38 $\times$$10^7$       &91\%\tablenotemark{c}  \\  
      Flux extrapolated to 0 K &    8.47$\pm$0.19$\times$$10^7$  &  $--$  \\  
\enddata
\tablenotetext{a}{ The flux errors for total flux in the cloud catalog and observed in PPV cube is estimated to be 5\%, and the error of estimated flux is derived with exponential fitting parameter errors.} 
\tablenotetext{b}{The completeness of flux in the DBSCAN catalog is approximately 82\% with respect to observed flux in the \cofs\ PPV cube, and is about 75\% with respect to the estimated flux at 0 K.}
\tablenotetext{c}{The completeness of flux observed in PPV cubes is estimated with respect to the estimated flux at 0 K.}
 
\end{deluxetable}

 \subsection{The completeness of flux in present survey and its CO catalog}
 
 In this section, we try to estimate the completeness of flux captured in the cloud catalog constructed using the DBSCAN algorithm and the MWISP DR1 data. From the cloud catalog, the measured flux is approximately 6.34$\pm$0.32$\times$10$^7$ K \kms arcmin$^2$, encompassing both complete and incomplete clouds. The flux error is about 5\%, estimated based on the dispersion of integrated intensities of reference sources listed in Table \ref{tab_bzy}.  Since the DBSCAN cloud-finding algorithm applied here picked up CO emissions above 2$\sigma$ (approximately 0.94~K) and 60$\arcsec\times60\arcsec$, it excluded clouds with peak intensities below 5$\sigma$ and with small angular sizes, and some flux may be lost due to the application of these criteria. 

The total observed flux in a PPV data cube can be estimated by analyzing the histogram of all voxels. As shown in Figure \ref{fig:totalflux}, since the histogram of Gaussian noise is symmetrical, the integration of the blue curve, i.e., the area under the curve, is approximately the total flux in the PPV cube. In this way, the total flux within the DR1 \cofs\ datacube is estimated to be 7.69$\pm$0.38$\times10^7$ K \kms\ arcmin$^2$, indicating that the DBSCAN algorithm recovered about 82\% of the total observed flux in the PPV datacube. It should be emphasized that the percentage of flux picked up by a catalog depends highly on the algorithm. This situation is true not only for \cofs\ flux mentioned here, but also for \coss\ and \cots.

However, due to the limited sensitivity, a fraction of CO flux remains undetected by the MWISP survey. As illustrated in Figure \ref{fig:fluxextrapolate}, we can estimate the total CO flux by extrapolating the flux down to 0 K using an exponential function \citep{2021ApJ...910..109Y}. This extrapolation suggests that a hypothetical telescope  with infinite sensitivity would detect a flux of approximately 8.47$\pm$0.19$\times$$10^7$ K \kms\ arcmin$^2$. In such a case, the current DBSCAN catalog would capture about 75\% of the total flux. The flux statistics are summarized in Table \ref{Tab:fluxmeasure}. If we consider the extrapolated flux at 0 K as the total CO flux in the surveyed region, the overall flux completeness of MWISP DR1 is approximately 91\%. However, this completeness decreases significantly with increasing distances \citep{2021ApJS..256...32S}. Because the flux is dominated by local clouds, the derived completeness is biased toward local structures. Detailed investigations are required for more accurate estimates.

\section{Summary}
\label{sec:summary}

 The MWISP survey maps $J=1\rightarrow0$ transitions of three CO isotopologue lines, covering 2310 degree$^2$  ($9.75\deg\leqslant l \leqslant229.75\deg$, $|b|\leqslant5.25\deg$, and -200 \kms\ $\leqslant$ $V_{\rm LSR}$ $\leqslant$ 200 \kms) along the northern Galactic plane with 9240 cells for each CO isotopologue line.  We present the first data release of the MWISP survey, covering a $(l,b,V_{\rm LSR})$ size of $220\deg\times10.5\deg\times500$ \kms. 
 
\begin{enumerate} 
\item For each CO isotopologue line, the 9240 cell FITS cubes and the mosaicked data cubes are hereby released, publicly accessible at https://doi.org/10.57760/sciencedb.27351. As part of the data release, the log files that contain the coordinates of reference positions are also released.
\item To demonstrate the quality of the MWISP DR1 data set, we generated color images for the integrated intensities, $l$-$V$ diagrams, and some pictures of individual clouds. These images, diagrams, and pictures are available along with the data release.
\item A cloud catalog is built based on the \cofs\ data cube and the DBSCAN algorithm, containing 103,517 complete molecular clouds. Further application of the  DBSCAN detection algorithm to this cloud sample shows that 10,790 (10.4\%) clouds possess significant \coss\ emission, and 304 (0.3\%) clouds contain \cots\ emission. Catalogs from \cofs, clumps from  \coss, and cores from \cots\ are all released. 
\end{enumerate}
The dataset released in this work will provide a legacy for the  molecular gas distribution along the northern Galactic plane. This dataset, together with the existing database, is expected to be valuable for a wide range of applications in future multi-wavelength studies.

\section{Facilities and Software tools}


%

\vspace{5mm}
\facilities{PMO 13.7~m, Science Data Bank (ScienceDB)}


\software{GILDAS \citep{2005sf2a.conf..721P}, Numpy \citep{harris2020array}, Matplotlib \citep{Hunter:2007}, Astropy \citep{2013A&A...558A..33A}, SciPy \citep{2020SciPy-NMeth}, mwispy (\url{https://github.com/shbzhang/mwispy}), DBSCAN \citep[\url{https://github.com/shbzhang/mwispy/tree/main/DBSCAN},][]{Ester1996}, CubeView (\url{https://github.com/shbzhang/mwispy/tree/main/cubeView})}

\begin{acknowledgments}

This work is resulted from the Milky Way Imaging Scroll Painting (MWISP) project, which is a multi-line survey in  \cofs/\coss/\cots\ along the northern galactic plane with the PMO 13.7~m telescope. We are grateful to the operating staff members at the PMO 13.7~m telescope for their long-term support. 
We appreciate Yuan Wang, Jing Yue, He Zhao, Jiansheng Zhou, and a number of students, in particular Shuaibo Bian, Xinyu Du, Haoran Feng, Jiancheng Feng, Dongqing Ge, Facheng Li, Junyu Li, Lianghao Lin, Zehao Lin, Qiang Liu, Wenshan Long, Xiangjun Shao, Chao Song, Yuan Wang, Fang Xiong, Yang Yang, Shuling Yu, Xin Zeng, Xiaoliang Zhan, Qi Zhao, Wei Zhou,  for their contributions which made the phase I of the MWISP survey a successful one. 
We acknowledge Science Data Bank (ScienceDB) for the support of data archiving.
MWISP was sponsored by National Key R\&D Program of China with grants 2023YFA1608000 \& 2017YFA0402701 and by CAS Key Research Program of Frontier Sciences with grant QYZDJ-SSW-SLH047. This work was also supported by the National Natural Science Foundation of China through grants 12041305 \& 11233007. 
We thank the anonymous referee for a constructive report which helped significantly in improving the paper.

\end{acknowledgments} 
 
\appendix
\setcounter{table}{0}
\renewcommand{\thetable}{A\arabic{table}}
\section{Nomenclature}

 \begin{deluxetable}{ll}[ht]
\tablecaption{Nomenclature used in describing a datacube. \label{Tab:nomenclature}}
 
\tablehead{ Terminology  &  Explanation  }   
\startdata
$l$ &   Galactic longitude  \\
$b$ &   Galactic latitude  \\
$V_{\rm LSR}$ &  The kinematic Local Standard of Rest (LSR)\\ 
Pixel & Unit of two-dimensional images   \\ 
Voxel &   Unit of three-dimensional datacbues  \\ 
Spectrum &  The $V_{\rm LSR}$-$T_{\rm mb}$ distribution at at each pixel position, i.e., the brightness distribution along the $V_{\rm LSR}$ axis at a specific position. \\
Cell &   The angular area (45\arcmin$\times$45\arcmin) produced by the OTF scanning of a 30\arcmin$\times$30\arcmin\ region with respect to the central beam \\ 
PPV data cube & 3D data cubes with axes of $l$-$b$-$V_{\rm LSR}$, also called PPV space \\
\enddata

\end{deluxetable}

 \bibliographystyle{aasjournal}
 \bibliography{refMCsample}

\begin{thebibliography}{}
\expandafter\ifx\csname natexlab\endcsname\relax\def\natexlab#1{#1}\fi
\providecommand{\url}[1]{\href{#1}{#1}}
\providecommand{\dodoi}[1]{doi:~\href{http://doi.org/#1}{\nolinkurl{#1}}}
\providecommand{\doeprint}[1]{\href{http://ascl.net/#1}{\nolinkurl{http://ascl.net/#1}}}
\providecommand{\doarXiv}[1]{\href{https://arxiv.org/abs/#1}{\nolinkurl{https://arxiv.org/abs/#1}}}

\bibitem[{{Adler} \& {Roberts}(1992)}]{1992ApJ...384...95A}
{Adler}, D.~S., \& {Roberts}, Jr., W.~W. 1992, \apj, 384, 95,
  \dodoi{10.1086/170854}

\bibitem[{{Astropy Collaboration} {et~al.}(2013){Astropy Collaboration},
  {Robitaille}, {Tollerud}, {Greenfield}, {Droettboom}, {Bray}, {Aldcroft},
  {Davis}, {Ginsburg}, {Price-Whelan}, {Kerzendorf}, {Conley}, {Crighton},
  {Barbary}, {Muna}, {Ferguson}, {Grollier}, {Parikh}, {Nair}, {Unther},
  {Deil}, {Woillez}, {Conseil}, {Kramer}, {Turner}, {Singer}, {Fox}, {Weaver},
  {Zabalza}, {Edwards}, {Azalee Bostroem}, {Burke}, {Casey}, {Crawford},
  {Dencheva}, {Ely}, {Jenness}, {Labrie}, {Lim}, {Pierfederici}, {Pontzen},
  {Ptak}, {Refsdal}, {Servillat}, \& {Streicher}}]{2013A&A...558A..33A}
{Astropy Collaboration}, {Robitaille}, T.~P., {Tollerud}, E.~J., {et~al.} 2013,
  \aap, 558, A33, \dodoi{10.1051/0004-6361/201322068}

\bibitem[{{Bania}(1977)}]{1977ApJ...216..381B}
{Bania}, T.~M. 1977, \apj, 216, 381, \dodoi{10.1086/155478}

\bibitem[{{Barnes} {et~al.}(2015){Barnes}, {Muller}, {Indermuehle},
  {O'Dougherty}, {Lowe}, {Cunningham}, {Hernandez}, \&
  {Fuller}}]{2015ApJ...812....6B}
{Barnes}, P.~J., {Muller}, E., {Indermuehle}, B., {et~al.} 2015, \apj, 812, 6,
  \dodoi{10.1088/0004-637X/812/1/6}

\bibitem[{{Beaumont} {et~al.}(2013){Beaumont}, {Offner}, {Shetty}, {Glover}, \&
  {Goodman}}]{2013ApJ...777..173B}
{Beaumont}, C.~N., {Offner}, S. S.~R., {Shetty}, R., {Glover}, S. C.~O., \&
  {Goodman}, A.~A. 2013, \apj, 777, 173, \dodoi{10.1088/0004-637X/777/2/173}

\bibitem[{{Benedettini} {et~al.}(2020){Benedettini}, {Molinari}, {Baldeschi},
  {Beltr{\'a}n}, {Brand}, {Cesaroni}, {Elia}, {Fontani}, {Merello}, {Olmi},
  {Pezzuto}, {Rygl}, {Schisano}, {Testi}, \&
  {Traficante}}]{2020A&A...633A.147B}
{Benedettini}, M., {Molinari}, S., {Baldeschi}, A., {et~al.} 2020, \aap, 633,
  A147, \dodoi{10.1051/0004-6361/201936096}

\bibitem[{{Benedettini} {et~al.}(2021){Benedettini}, {Traficante}, {Olmi},
  {Pezzuto}, {Baldeschi}, {Molinari}, {Elia}, {Schisano}, {Merello}, {Fontani},
  {Rygl}, {Brand}, {Beltr{\'a}n}, {Cesaroni}, {Liu}, \&
  {Testi}}]{2021A&A...654A.144B}
{Benedettini}, M., {Traficante}, A., {Olmi}, L., {et~al.} 2021, \aap, 654,
  A144, \dodoi{10.1051/0004-6361/202141433}

\bibitem[{{Braiding} {et~al.}(2015){Braiding}, {Burton}, {Blackwell},
  {Gl{\"u}ck}, {Hawkes}, {Kulesa}, {Maxted}, {Rebolledo}, {Rowell}, {Stark},
  {Tothill}, {Urquhart}, {Voisin}, {Walsh}, {de Wilt}, \&
  {Wong}}]{2015PASA...32...20B}
{Braiding}, C., {Burton}, M.~G., {Blackwell}, R., {et~al.} 2015, \pasa, 32,
  e020, \dodoi{10.1017/pasa.2015.20}

\bibitem[{{Braiding} {et~al.}(2018){Braiding}, {Wong}, {Maxted}, {Romano},
  {Burton}, {Blackwell}, {Filipovi{\'c}}, {Freeman}, {Indermuehle}, {Lau},
  {Rebolledo}, {Rowell}, {Snoswell}, {Tothill}, {Voisin}, \& {de
  Wilt}}]{2018PASA...35...29B}
{Braiding}, C., {Wong}, G.~F., {Maxted}, N.~I., {et~al.} 2018, \pasa, 35, e029,
  \dodoi{10.1017/pasa.2018.18}

\bibitem[{{Carpenter} \& {Sanders}(1998)}]{1998AJ....116.1856C}
{Carpenter}, J.~M., \& {Sanders}, D.~B. 1998, \aj, 116, 1856,
  \dodoi{10.1086/300534}

\bibitem[{{Carpenter} {et~al.}(1995){Carpenter}, {Snell}, \&
  {Schloerb}}]{1995ApJ...445..246C}
{Carpenter}, J.~M., {Snell}, R.~L., \& {Schloerb}, F.~P. 1995, \apj, 445, 246,
  \dodoi{10.1086/175692}

\bibitem[{{Chen} {et~al.}(2023){Chen}, {Sun}, {Feng}, {Zhang}, {Guo}, {Xu},
  {Su}, {Sun}, {Zhang}, {Zhou}, {Chen}, {Yan}, {Zhang}, {Fang}, \&
  {Yang}}]{2023AJ....165...16C}
{Chen}, X., {Sun}, L., {Feng}, J., {et~al.} 2023, \aj, 165, 16,
  \dodoi{10.3847/1538-3881/ac9ea2}

\bibitem[{{Churchwell} {et~al.}(2009){Churchwell}, {Babler}, {Meade},
  {Whitney}, {Benjamin}, {Indebetouw}, {Cyganowski}, {Robitaille}, {Povich},
  {Watson}, \& {Bracker}}]{2009PASP..121..213C}
{Churchwell}, E., {Babler}, B.~L., {Meade}, M.~R., {et~al.} 2009, \pasp, 121,
  213, \dodoi{10.1086/597811}

\bibitem[{{Cohen} {et~al.}(1980){Cohen}, {Cong}, {Dame}, \&
  {Thaddeus}}]{1980ApJ...239L..53C}
{Cohen}, R.~S., {Cong}, H., {Dame}, T.~M., \& {Thaddeus}, P. 1980, \apjl, 239,
  L53, \dodoi{10.1086/183290}

\bibitem[{{Cohen} {et~al.}(1986){Cohen}, {Dame}, \&
  {Thaddeus}}]{1986ApJS...60..695C}
{Cohen}, R.~S., {Dame}, T.~M., \& {Thaddeus}, P. 1986, \apjs, 60, 695,
  \dodoi{10.1086/191101}

\bibitem[{{Cohen} \& {Thaddeus}(1977)}]{1977ApJ...217L.155C}
{Cohen}, R.~S., \& {Thaddeus}, P. 1977, \apjl, 217, L155,
  \dodoi{10.1086/182560}

\bibitem[{{Colombo} {et~al.}(2015){Colombo}, {Rosolowsky}, {Ginsburg},
  {Duarte-Cabral}, \& {Hughes}}]{2015MNRAS.454.2067C}
{Colombo}, D., {Rosolowsky}, E., {Ginsburg}, A., {Duarte-Cabral}, A., \&
  {Hughes}, A. 2015, \mnras, 454, 2067, \dodoi{10.1093/mnras/stv2063}

\bibitem[{{Colombo} {et~al.}(2022){Colombo}, {Duarte-Cabral}, {Pettitt},
  {Urquhart}, {Wyrowski}, {Csengeri}, {Neralwar}, {Schuller}, {Menten},
  {Anderson}, {Barnes}, {Beuther}, {Bronfman}, {Eden}, {Ginsburg}, {Henning},
  {K{\"o}nig}, {Lee}, {Mattern}, {Medina}, {Ragan}, {Rigby},
  {S{\'a}nchez-Monge}, {Traficante}, {Yang}, \& {Wienen}}]{2022A&A...658A..54C}
{Colombo}, D., {Duarte-Cabral}, A., {Pettitt}, A.~R., {et~al.} 2022, \aap, 658,
  A54, \dodoi{10.1051/0004-6361/202141287}

\bibitem[{{Cubuk} {et~al.}(2023){Cubuk}, {Burton}, {Braiding}, {Wong},
  {Rowell}, {Maxted}, {Eden}, {Alsaberi}, {Blackwell}, {Enokiya}, {Feijen},
  {Filipovi{\'c}}, {Freeman}, {Fujita}, {Ghavam}, {Gunay}, {Indermuehle},
  {Hayashi}, {Kohno}, {Nagaya}, {Nishimura}, {Okawa}, {Rebolledo}, {Romano},
  {Sano}, {Snoswell}, {Tothill}, {Tsuge}, {Voisin}, {Yamane}, \&
  {Yoshiike}}]{2023PASA...40...47C}
{Cubuk}, K.~O., {Burton}, M.~G., {Braiding}, C., {et~al.} 2023, \pasa, 40,
  e047, \dodoi{10.1017/pasa.2023.44}

\bibitem[{{Dame} {et~al.}(2001){Dame}, {Hartmann}, \&
  {Thaddeus}}]{2001ApJ...547..792D}
{Dame}, T.~M., {Hartmann}, D., \& {Thaddeus}, P. 2001, \apj, 547, 792,
  \dodoi{10.1086/318388}

\bibitem[{{Dame} \& {Thaddeus}(1985)}]{1985ApJ...297..751D}
{Dame}, T.~M., \& {Thaddeus}, P. 1985, \apj, 297, 751, \dodoi{10.1086/163573}

\bibitem[{{Dame} {et~al.}(1987){Dame}, {Ungerechts}, {Cohen}, {de Geus},
  {Grenier}, {May}, {Murphy}, {Nyman}, \& {Thaddeus}}]{1987ApJ...322..706D}
{Dame}, T.~M., {Ungerechts}, H., {Cohen}, R.~S., {et~al.} 1987, \apj, 322, 706,
  \dodoi{10.1086/165766}

\bibitem[{{Dempsey} {et~al.}(2013){Dempsey}, {Thomas}, \&
  {Currie}}]{2013ApJS..209....8D}
{Dempsey}, J.~T., {Thomas}, H.~S., \& {Currie}, M.~J. 2013, \apjs, 209, 8,
  \dodoi{10.1088/0067-0049/209/1/8}

\bibitem[{{Digel} {et~al.}(1996){Digel}, {Lyder}, {Philbrick}, {Puche}, \&
  {Thaddeus}}]{1996ApJ...458..561D}
{Digel}, S.~W., {Lyder}, D.~A., {Philbrick}, A.~J., {Puche}, D., \& {Thaddeus},
  P. 1996, \apj, 458, 561, \dodoi{10.1086/176839}

\bibitem[{{Dobashi} {et~al.}(1994){Dobashi}, {Bernard}, {Yonekura}, \&
  {Fukui}}]{1994ApJS...95..419D}
{Dobashi}, K., {Bernard}, J.-P., {Yonekura}, Y., \& {Fukui}, Y. 1994, \apjs,
  95, 419, \dodoi{10.1086/192106}

\bibitem[{{Dong} {et~al.}(2023){Dong}, {Sun}, {Xu}, {Lin}, {Bian}, {Hao},
  {Liu}, {Li}, {Yang}, {Su}, {Zhou}, {Zhang}, {Yan}, \&
  {Chen}}]{2023ApJS..268....1D}
{Dong}, Y., {Sun}, Y., {Xu}, Y., {et~al.} 2023, \apjs, 268, 1,
  \dodoi{10.3847/1538-4365/acde81}

\bibitem[{{Du} {et~al.}(2017){Du}, {Xu}, {Yang}, \&
  {Sun}}]{2017ApJS..229...24D}
{Du}, X., {Xu}, Y., {Yang}, J., \& {Sun}, Y. 2017, \apjs, 229, 24,
  \dodoi{10.3847/1538-4365/aa5d9d}

\bibitem[{{Du} {et~al.}(2016){Du}, {Xu}, {Yang}, {Sun}, {Li}, {Zhang}, \&
  {Zhou}}]{2016ApJS..224....7D}
{Du}, X., {Xu}, Y., {Yang}, J., {et~al.} 2016, \apjs, 224, 7,
  \dodoi{10.3847/0067-0049/224/1/7}

\bibitem[{Ester {et~al.}(1996)Ester, Kriegel, Sander, \& Xu}]{Ester1996}
Ester, M., Kriegel, H.-P., Sander, J., \& Xu, X. 1996, in Proceedings of the
  Second International Conference on Knowledge Discovery and Data Mining,
  KDD'96 (AAAI Press), 226--231.
\newblock \url{http://dl.acm.org/citation.cfm?id=3001460.3001507}

\bibitem[{{Gong} {et~al.}(2016){Gong}, {Mao}, {Fang}, {Zhang}, {Su}, {Yang},
  {Jiang}, {Xu}, {Wang}, {Wang}, {Lu}, \& {Sun}}]{2016A&A...588A.104G}
{Gong}, Y., {Mao}, R.~Q., {Fang}, M., {et~al.} 2016, \aap, 588, A104,
  \dodoi{10.1051/0004-6361/201527334}

\bibitem[{{Grenier} {et~al.}(1989){Grenier}, {Lebrun}, {Arnaud}, {Dame}, \&
  {Thaddeus}}]{1989ApJ...347..231G}
{Grenier}, I.~A., {Lebrun}, F., {Arnaud}, M., {Dame}, T.~M., \& {Thaddeus}, P.
  1989, \apj, 347, 231, \dodoi{10.1086/168112}

\bibitem[{{Guo} {et~al.}(2022){Guo}, {Chen}, {Feng}, {Sun}, {Zhang}, {Wang},
  {Su}, {Sun}, {Yan}, {Zhang}, {Zhou}, {Zhang}, {Fang}, \&
  {Yang}}]{2022ApJ...938...44G}
{Guo}, W., {Chen}, X., {Feng}, J., {et~al.} 2022, \apj, 938, 44,
  \dodoi{10.3847/1538-4357/ac8933}

\bibitem[{Harris {et~al.}(2020)Harris, Millman, van~der Walt, Gommers,
  Virtanen, Cournapeau, Wieser, Taylor, Berg, Smith, Kern, Picus, Hoyer, van
  Kerkwijk, Brett, Haldane, del R{\'{i}}o, Wiebe, Peterson,
  G{\'{e}}rard-Marchant, Sheppard, Reddy, Weckesser, Abbasi, Gohlke, \&
  Oliphant}]{harris2020array}
Harris, C.~R., Millman, K.~J., van~der Walt, S.~J., {et~al.} 2020, Nature, 585,
  357, \dodoi{10.1038/s41586-020-2649-2}

\bibitem[{{Heyer} \& {Dame}(2015)}]{2015ARA&A..53..583H}
{Heyer}, M., \& {Dame}, T.~M. 2015, \araa, 53, 583,
  \dodoi{10.1146/annurev-astro-082214-122324}

\bibitem[{{Heyer} {et~al.}(1998){Heyer}, {Brunt}, {Snell}, {Howe}, {Schloerb},
  \& {Carpenter}}]{1998ApJS..115..241H}
{Heyer}, M.~H., {Brunt}, C., {Snell}, R.~L., {et~al.} 1998, \apjs, 115, 241,
  \dodoi{10.1086/313086}

\bibitem[{{Heyer} {et~al.}(1996){Heyer}, {Carpenter}, \&
  {Ladd}}]{1996ApJ...463..630H}
{Heyer}, M.~H., {Carpenter}, J.~M., \& {Ladd}, E.~F. 1996, \apj, 463, 630,
  \dodoi{10.1086/177277}

\bibitem[{Hunter(2007)}]{Hunter:2007}
Hunter, J.~D. 2007, Computing in Science \& Engineering, 9, 90,
  \dodoi{10.1109/MCSE.2007.55}

\bibitem[{{Jackson} {et~al.}(2006){Jackson}, {Rathborne}, {Shah}, {Simon},
  {Bania}, {Clemens}, {Chambers}, {Johnson}, {Dormody}, {Lavoie}, \&
  {Heyer}}]{2006ApJS..163..145J}
{Jackson}, J.~M., {Rathborne}, J.~M., {Shah}, R.~Y., {et~al.} 2006, \apjs, 163,
  145, \dodoi{10.1086/500091}

\bibitem[{{Jacq} {et~al.}(1988){Jacq}, {Despois}, \&
  {Baudry}}]{1988A&A...195...93J}
{Jacq}, T., {Despois}, D., \& {Baudry}, A. 1988, \aap, 195, 93

\bibitem[{{Jiang} {et~al.}(2023){Jiang}, {Chen}, {Zheng}, {Jiang}, {Huang},
  {Zeng}, {Zeng}, \& {Luo}}]{2023ApJS..267...32J}
{Jiang}, Y., {Chen}, Z., {Zheng}, S., {et~al.} 2023, \apjs, 267, 32,
  \dodoi{10.3847/1538-4365/acda89}

\bibitem[{{Jiang} {et~al.}(2025){Jiang}, {Yan}, {Yang}, {Zheng}, {Chen}, {Su},
  {Jiang}, {Chen}, {Zhou}, {Huang}, {Luo}, {Feng}, \&
  {Liu}}]{2025arXiv250901955J}
{Jiang}, Y., {Yan}, Q.-Z., {Yang}, J., {et~al.} 2025, arXiv e-prints,
  arXiv:2509.01955, \dodoi{10.48550/arXiv.2509.01955}

\bibitem[{{Kharchenko} {et~al.}(2005){Kharchenko}, {Piskunov}, {R{\"o}ser},
  {Schilbach}, \& {Scholz}}]{2005A&A...438.1163K}
{Kharchenko}, N.~V., {Piskunov}, A.~E., {R{\"o}ser}, S., {Schilbach}, E., \&
  {Scholz}, R.~D. 2005, \aap, 438, 1163, \dodoi{10.1051/0004-6361:20042523}

\bibitem[{{Kutner} \& {Ulich}(1981)}]{1981ApJ...250..341K}
{Kutner}, M.~L., \& {Ulich}, B.~L. 1981, \apj, 250, 341, \dodoi{10.1086/159380}

\bibitem[{{Lada} {et~al.}(1978){Lada}, {Elmegreen}, {Cong}, \&
  {Thaddeus}}]{1978ApJ...226L..39L}
{Lada}, C.~J., {Elmegreen}, B.~G., {Cong}, H.~I., \& {Thaddeus}, P. 1978,
  \apjl, 226, L39, \dodoi{10.1086/182826}

\bibitem[{{Leung} \& {Thaddeus}(1992)}]{1992ApJS...81..267L}
{Leung}, H.~O., \& {Thaddeus}, P. 1992, \apjs, 81, 267, \dodoi{10.1086/191693}

\bibitem[{{Li} {et~al.}(2018){Li}, {Wang}, {Zhang}, {Ma}, {Fang}, \&
  {Yang}}]{2018ApJS..238...10L}
{Li}, C., {Wang}, H., {Zhang}, M., {et~al.} 2018, \apjs, 238, 10,
  \dodoi{10.3847/1538-4365/aad963}

\bibitem[{{Li} {et~al.}(2022){Li}, {Zheng}, {Li}, {Pang}, {Tang}, {Milone},
  {Wang}, {Wang}, \& {Jiang}}]{2022RAA....22i5004L}
{Li}, C., {Zheng}, Z., {Li}, X., {et~al.} 2022, Research in Astronomy and
  Astrophysics, 22, 095004, \dodoi{10.1088/1674-4527/ac7bf1}

\bibitem[{{Li} {et~al.}(2023){Li}, {Wang}, {Ma}, {Lin}, {Su}, {Li}, {Sun},
  {Zhou}, \& {Yang}}]{2023ApJS..267...30L}
{Li}, C., {Wang}, H., {Ma}, Y., {et~al.} 2023, \apjs, 267, 30,
  \dodoi{10.3847/1538-4365/acd9a7}

\bibitem[{{Li} {et~al.}(2019){Li}, {Xu}, {Sun}, {Yan}, {Ma}, \&
  {Yang}}]{2019ApJS..242...19L}
{Li}, Y., {Xu}, Y., {Sun}, Y., {et~al.} 2019, \apjs, 242, 19,
  \dodoi{10.3847/1538-4365/ab1e55}

\bibitem[{{Lin} {et~al.}(2021){Lin}, {Sun}, {Xu}, {Yang}, \&
  {Li}}]{2021ApJS..252...20L}
{Lin}, Z., {Sun}, Y., {Xu}, Y., {Yang}, J., \& {Li}, Y. 2021, \apjs, 252, 20,
  \dodoi{10.3847/1538-4365/abccd8}

\bibitem[{{Ma} {et~al.}(2021){Ma}, {Wang}, {Li}, {Lin}, {Sun}, \&
  {Yang}}]{2021ApJS..254....3M}
{Ma}, Y., {Wang}, H., {Li}, C., {et~al.} 2021, \apjs, 254, 3,
  \dodoi{10.3847/1538-4365/abe85c}

\bibitem[{{Ma} {et~al.}(2019){Ma}, {Wang}, {Zhang}, {Li}, \&
  {Yang}}]{2019ApJ...878...44M}
{Ma}, Y., {Wang}, H., {Zhang}, M., {Li}, C., \& {Yang}, J. 2019, \apj, 878, 44,
  \dodoi{10.3847/1538-4357/ab1ea7}

\bibitem[{{Ma} {et~al.}(2022){Ma}, {Wang}, {Zhang}, {Wang}, {Zhang}, {Liu},
  {Li}, {Zheng}, {Yuan}, \& {Yang}}]{2022ApJS..262...16M}
{Ma}, Y., {Wang}, H., {Zhang}, M., {et~al.} 2022, \apjs, 262, 16,
  \dodoi{10.3847/1538-4365/ac7797}

\bibitem[{{Maddalena} \& {Thaddeus}(1985)}]{1985ApJ...294..231M}
{Maddalena}, R.~J., \& {Thaddeus}, P. 1985, \apj, 294, 231,
  \dodoi{10.1086/163291}

\bibitem[{{Moscadelli} {et~al.}(2009){Moscadelli}, {Reid}, {Menten},
  {Brunthaler}, {Zheng}, \& {Xu}}]{2009ApJ...693..406M}
{Moscadelli}, L., {Reid}, M.~J., {Menten}, K.~M., {et~al.} 2009, \apj, 693,
  406, \dodoi{10.1088/0004-637X/693/1/406}

\bibitem[{{Oliver} {et~al.}(1996){Oliver}, {Masheder}, \&
  {Thaddeus}}]{1996A&A...315..578O}
{Oliver}, R.~J., {Masheder}, M.~R.~W., \& {Thaddeus}, P. 1996, \aap, 315, 578

\bibitem[{{Ostriker} {et~al.}(2001){Ostriker}, {Stone}, \&
  {Gammie}}]{2001ApJ...546..980O}
{Ostriker}, E.~C., {Stone}, J.~M., \& {Gammie}, C.~F. 2001, \apj, 546, 980,
  \dodoi{10.1086/318290}

\bibitem[{{Park} {et~al.}(2023){Park}, {Currie}, {Thomas}, {Rosolowsky},
  {Dempsey}, {Kim}, {Rigby}, {Su}, {Eden}, {Colombo}, {Parsons}, \&
  {Moore}}]{2023ApJS..264...16P}
{Park}, G., {Currie}, M.~J., {Thomas}, H.~S., {et~al.} 2023, \apjs, 264, 16,
  \dodoi{10.3847/1538-4365/ac9b59}

\bibitem[{{Penzias} \& {Burrus}(1973)}]{1973ARA&A..11...51P}
{Penzias}, A.~A., \& {Burrus}, C.~A. 1973, \araa, 11, 51,
  \dodoi{10.1146/annurev.aa.11.090173.000411}

\bibitem[{{Pety}(2005)}]{2005sf2a.conf..721P}
{Pety}, J. 2005, in SF2A-2005: Semaine de l'Astrophysique Francaise, ed.
  F.~{Casoli}, T.~{Contini}, J.~M. {Hameury}, \& L.~{Pagani}, 721

\bibitem[{{Pichardo} {et~al.}(2000){Pichardo}, {V{\'a}zquez-Semadeni}, {Gazol},
  {Passot}, \& {Ballesteros-Paredes}}]{2000ApJ...532..353P}
{Pichardo}, B., {V{\'a}zquez-Semadeni}, E., {Gazol}, A., {Passot}, T., \&
  {Ballesteros-Paredes}, J. 2000, \apj, 532, 353, \dodoi{10.1086/308546}

\bibitem[{{Reed} {et~al.}(1995){Reed}, {Hester}, {Fabian}, \&
  {Winkler}}]{1995ApJ...440..706R}
{Reed}, J.~E., {Hester}, J.~J., {Fabian}, A.~C., \& {Winkler}, P.~F. 1995,
  \apj, 440, 706, \dodoi{10.1086/175308}

\bibitem[{{Reid} {et~al.}(2019){Reid}, {Menten}, {Brunthaler}, {Zheng}, {Dame},
  {Xu}, {Li}, {Sakai}, {Wu}, {Immer}, {Zhang}, {Sanna}, {Moscadelli}, {Rygl},
  {Bartkiewicz}, {Hu}, {Quiroga-Nu{\~n}ez}, \& {van
  Langevelde}}]{2019ApJ...885..131R}
{Reid}, M.~J., {Menten}, K.~M., {Brunthaler}, A., {et~al.} 2019, \apj, 885,
  131, \dodoi{10.3847/1538-4357/ab4a11}

\bibitem[{{Rigby} {et~al.}(2016){Rigby}, {Moore}, {Plume}, {Eden}, {Urquhart},
  {Thompson}, {Mottram}, {Brunt}, {Butner}, {Dempsey}, {Gibson}, {Hatchell},
  {Jenness}, {Kuno}, {Longmore}, {Morgan}, {Polychroni}, {Thomas}, {White}, \&
  {Zhu}}]{2016MNRAS.456.2885R}
{Rigby}, A.~J., {Moore}, T.~J.~T., {Plume}, R., {et~al.} 2016, \mnras, 456,
  2885, \dodoi{10.1093/mnras/stv2808}

\bibitem[{{Roman-Duval} {et~al.}(2016){Roman-Duval}, {Heyer}, {Brunt}, {Clark},
  {Klessen}, \& {Shetty}}]{2016ApJ...818..144R}
{Roman-Duval}, J., {Heyer}, M., {Brunt}, C.~M., {et~al.} 2016, \apj, 818, 144,
  \dodoi{10.3847/0004-637X/818/2/144}

\bibitem[{{Rosolowsky} \& {Leroy}(2006)}]{2006PASP..118..590R}
{Rosolowsky}, E., \& {Leroy}, A. 2006, \pasp, 118, 590, \dodoi{10.1086/502982}

\bibitem[{{Rosolowsky} {et~al.}(2008){Rosolowsky}, {Pineda}, {Kauffmann}, \&
  {Goodman}}]{2008ApJ...679.1338R}
{Rosolowsky}, E.~W., {Pineda}, J.~E., {Kauffmann}, J., \& {Goodman}, A.~A.
  2008, \apj, 679, 1338, \dodoi{10.1086/587685}

\bibitem[{{Sanders} {et~al.}(1986){Sanders}, {Clemens}, {Scoville}, \&
  {Solomon}}]{1986ApJS...60....1S}
{Sanders}, D.~B., {Clemens}, D.~P., {Scoville}, N.~Z., \& {Solomon}, P.~M.
  1986, \apjs, 60, 1, \dodoi{10.1086/191086}

\bibitem[{{Sanders} {et~al.}(1984){Sanders}, {Solomon}, \&
  {Scoville}}]{1984ApJ...276..182S}
{Sanders}, D.~B., {Solomon}, P.~M., \& {Scoville}, N.~Z. 1984, \apj, 276, 182,
  \dodoi{10.1086/161602}

\bibitem[{{Schneider} {et~al.}(2006){Schneider}, {Bontemps}, {Simon}, {Jakob},
  {Motte}, {Miller}, {Kramer}, \& {Stutzki}}]{2006A&A...458..855S}
{Schneider}, N., {Bontemps}, S., {Simon}, R., {et~al.} 2006, \aap, 458, 855,
  \dodoi{10.1051/0004-6361:20065088}

\bibitem[{{Schneider} {et~al.}(2011){Schneider}, {Bontemps}, {Simon},
  {Ossenkopf}, {Federrath}, {Klessen}, {Motte}, {Andr{\'e}}, {Stutzki}, \&
  {Brunt}}]{2011A&A...529A...1S}
---. 2011, \aap, 529, A1, \dodoi{10.1051/0004-6361/200913884}

\bibitem[{{Schuller} {et~al.}(2009){Schuller}, {Menten}, {Contreras},
  {Wyrowski}, {Schilke}, {Bronfman}, {Henning}, {Walmsley}, {Beuther},
  {Bontemps}, {Cesaroni}, {Deharveng}, {Garay}, {Herpin}, {Lefloch}, {Linz},
  {Mardones}, {Minier}, {Molinari}, {Motte}, {Nyman}, {Reveret}, {Risacher},
  {Russeil}, {Schneider}, {Testi}, {Troost}, {Vasyunina}, {Wienen}, {Zavagno},
  {Kovacs}, {Kreysa}, {Siringo}, \& {Wei{\ss}}}]{2009A&A...504..415S}
{Schuller}, F., {Menten}, K.~M., {Contreras}, Y., {et~al.} 2009, \aap, 504,
  415, \dodoi{10.1051/0004-6361/200811568}

\bibitem[{{Schuller} {et~al.}(2021){Schuller}, {Urquhart}, {Csengeri},
  {Colombo}, {Duarte-Cabral}, {Mattern}, {Ginsburg}, {Pettitt}, {Wyrowski},
  {Anderson}, {Azagra}, {Barnes}, {Beltran}, {Beuther}, {Billington},
  {Bronfman}, {Cesaroni}, {Dobbs}, {Eden}, {Lee}, {Medina}, {Menten}, {Moore},
  {Montenegro-Montes}, {Ragan}, {Rigby}, {Riener}, {Russeil}, {Schisano},
  {Sanchez-Monge}, {Traficante}, {Zavagno}, {Agurto}, {Bontemps}, {Finger},
  {Giannetti}, {Gonzalez}, {Hernandez}, {Henning}, {Kainulainen}, {Kauffmann},
  {Leurini}, {Lopez}, {Mac-Auliffe}, {Mazumdar}, {Molinari}, {Motte}, {Muller},
  {Nguyen-Luong}, {Parra}, {Perez-Beaupuits}, {Schilke}, {Schneider}, {Suri},
  {Testi}, {Torstensson}, {Veena}, {Venegas}, {Wang}, \&
  {Wienen}}]{2021MNRAS.500.3064S}
{Schuller}, F., {Urquhart}, J.~S., {Csengeri}, T., {et~al.} 2021, \mnras, 500,
  3064, \dodoi{10.1093/mnras/staa2369}

\bibitem[{{Shan} {et~al.}(2012){Shan}, {Yang}, {Shi}, {Yao}, {Zuo}, {Lin},
  {Chen}, {Zhang}, {Duan}, {Cao}, {Li}, {Li}, {Liu}, \&
  {Zhong}}]{2012ITTST...2..593S}
{Shan}, W., {Yang}, J., {Shi}, S., {et~al.} 2012, IEEE Transactions on
  Terahertz Science and Technology, 2, 593, \dodoi{10.1109/TTHZ.2012.2213818}

\bibitem[{{Solomon} {et~al.}(1987){Solomon}, {Rivolo}, {Barrett}, \&
  {Yahil}}]{1987ApJ...319..730S}
{Solomon}, P.~M., {Rivolo}, A.~R., {Barrett}, J., \& {Yahil}, A. 1987, \apj,
  319, 730, \dodoi{10.1086/165493}

\bibitem[{{Solomon} {et~al.}(1979){Solomon}, {Sanders}, \&
  {Scoville}}]{1979IAUS...84...35S}
{Solomon}, P.~M., {Sanders}, D.~B., \& {Scoville}, N.~Z. 1979, in IAU
  Symposium, Vol.~84, The Large-Scale Characteristics of the Galaxy, ed. W.~B.
  {Burton}, 35

\bibitem[{{Su} {et~al.}(2014){Su}, {Fang}, {Yang}, {Zhou}, \&
  {Chen}}]{2014ApJ...788..122S}
{Su}, Y., {Fang}, M., {Yang}, J., {Zhou}, P., \& {Chen}, Y. 2014, \apj, 788,
  122, \dodoi{10.1088/0004-637X/788/2/122}

\bibitem[{{Su} {et~al.}(2016){Su}, {Sun}, {Li}, {Zhang}, {Zhou}, {Fang},
  {Yang}, \& {Chen}}]{2016ApJ...828...59S}
{Su}, Y., {Sun}, Y., {Li}, C., {et~al.} 2016, \apj, 828, 59,
  \dodoi{10.3847/0004-637X/828/1/59}

\bibitem[{{Su} {et~al.}(2018){Su}, {Zhou}, {Yang}, {Chen}, {Chen}, \&
  {Zhang}}]{2018ApJ...863..103S}
{Su}, Y., {Zhou}, X., {Yang}, J., {et~al.} 2018, \apj, 863, 103,
  \dodoi{10.3847/1538-4357/aad04e}

\bibitem[{{Su} {et~al.}(2017){Su}, {Zhou}, {Yang}, {Chen}, {Chen}, {Liu},
  {Wang}, {Li}, \& {Zhang}}]{2017ApJ...836..211S}
---. 2017, \apj, 836, 211, \dodoi{10.3847/1538-4357/aa5cb7}

\bibitem[{{Su} {et~al.}(2019){Su}, {Yang}, {Zhang}, {Gong}, {Wang}, {Zhou},
  {Wang}, {Chen}, {Sun}, {Chen}, {Xu}, \& {Jiang}}]{2019ApJS..240....9S}
{Su}, Y., {Yang}, J., {Zhang}, S., {et~al.} 2019, \apjs, 240, 9,
  \dodoi{10.3847/1538-4365/aaf1c8}

\bibitem[{{Su} {et~al.}(2020){Su}, {Yang}, {Yan}, {Gong}, {Chen}, {Zhang},
  {Sun}, {Zhang}, {Chen}, {Zhou}, {Wang}, {Wang}, {Xu}, \&
  {Jiang}}]{2020ApJ...893...91S}
{Su}, Y., {Yang}, J., {Yan}, Q.-Z., {et~al.} 2020, \apj, 893, 91,
  \dodoi{10.3847/1538-4357/ab7fff}

\bibitem[{{Su} {et~al.}(2021){Su}, {Yang}, {Yan}, {Zhang}, {Wang}, {Sun},
  {Chen}, {Wang}, {Zhou}, {Chen}, {Jiang}, \& {Wang}}]{2021ApJ...910..131S}
---. 2021, \apj, 910, 131, \dodoi{10.3847/1538-4357/abe5ab}

\bibitem[{{Su} {et~al.}(2022){Su}, {Zhang}, {Yang}, {Yan}, {Sun}, {Wang},
  {Zhang}, {Chen}, {Chen}, {Zhou}, \& {Yuan}}]{2022ApJ...930..112S}
{Su}, Y., {Zhang}, S., {Yang}, J., {et~al.} 2022, \apj, 930, 112,
  \dodoi{10.3847/1538-4357/ac63b3}

\bibitem[{{Sun} {et~al.}(2018){Sun}, {Lu}, {Yang}, {Su}, {Zhang}, {Zhou}, \&
  {Lin}}]{2018AcASn..59....3S}
{Sun}, J.~X., {Lu}, D.~R., {Yang}, J., {et~al.} 2018, Acta Astronomica Sinica,
  59, 3

\bibitem[{{Sun} {et~al.}(2017){Sun}, {Su}, {Zhang}, {Xu}, {Chen}, {Yang},
  {Jiang}, \& {Fang}}]{2017ApJS..230...17S}
{Sun}, Y., {Su}, Y., {Zhang}, S.-B., {et~al.} 2017, \apjs, 230, 17,
  \dodoi{10.3847/1538-4365/aa63ea}

\bibitem[{{Sun} {et~al.}(2015){Sun}, {Xu}, {Yang}, {Li}, {Du}, {Zhang}, \&
  {Zhou}}]{2015ApJ...798L..27S}
{Sun}, Y., {Xu}, Y., {Yang}, J., {et~al.} 2015, \apjl, 798, L27,
  \dodoi{10.1088/2041-8205/798/2/L27}

\bibitem[{{Sun} {et~al.}(2020){Sun}, {Yang}, {Xu}, {Zhang}, {Su}, {Wang},
  {Chen}, {Lu}, {Sun}, {Ju}, {Zhang}, {Zhou}, \& {Jiang}}]{2020ApJS..246....7S}
{Sun}, Y., {Yang}, J., {Xu}, Y., {et~al.} 2020, \apjs, 246, 7,
  \dodoi{10.3847/1538-4365/ab5b97}

\bibitem[{{Sun} {et~al.}(2021){Sun}, {Yang}, {Yan}, {Lin}, {Zhang}, {Su}, {Xu},
  {Chen}, {Wang}, \& {Zhou}}]{2021ApJS..256...32S}
{Sun}, Y., {Yang}, J., {Yan}, Q.-Z., {et~al.} 2021, \apjs, 256, 32,
  \dodoi{10.3847/1538-4365/ac11fe}

\bibitem[{{Sun} {et~al.}(2024){Sun}, {Yang}, {Zhang}, {Yan}, {Su}, {Chen},
  {Zhou}, {Xu}, {Wang}, {Wang}, {Jiang}, {Sun}, {Lu}, {Ju}, \&
  {Zhang}}]{2024ApJ...977L..35S}
{Sun}, Y., {Yang}, J., {Zhang}, S., {et~al.} 2024, \apjl, 977, L35,
  \dodoi{10.3847/2041-8213/ad9605}

\bibitem[{{Thaddeus}(1977)}]{1977IAUS...75...37T}
{Thaddeus}, P. 1977, in IAU Symposium, Vol.~75, Star Formation, ed. T.~{de
  Jong}, A.~{Maeder}, \& S.~B. {Pikel'Ner}, 37

\bibitem[{{Ulich} \& {Haas}(1976)}]{1976ApJS...30..247U}
{Ulich}, B.~L., \& {Haas}, R.~W. 1976, \apjs, 30, 247, \dodoi{10.1086/190361}

\bibitem[{{Umemoto} {et~al.}(2017){Umemoto}, {Minamidani}, {Kuno}, {Fujita},
  {Matsuo}, {Nishimura}, {Torii}, {Tosaki}, {Kohno}, {Kuriki}, {Tsuda},
  {Hirota}, {Ohashi}, {Yamagishi}, {Handa}, {Nakanishi}, {Omodaka}, {Koide},
  {Matsumoto}, {Onishi}, {Tokuda}, {Seta}, {Kobayashi}, {Tachihara}, {Sano},
  {Hattori}, {Onodera}, {Oasa}, {Kamegai}, {Tsuboi}, {Sofue}, {Higuchi},
  {Chibueze}, {Mizuno}, {Honma}, {Muller}, {Inoue}, {Morokuma-Matsui},
  {Shinnaga}, {Ozawa}, {Takahashi}, {Yoshiike}, {Costes}, \&
  {Kuwahara}}]{2017PASJ...69...78U}
{Umemoto}, T., {Minamidani}, T., {Kuno}, N., {et~al.} 2017, \pasj, 69, 78,
  \dodoi{10.1093/pasj/psx061}

\bibitem[{Virtanen {et~al.}(2020)Virtanen, Gommers, Oliphant, Haberland, Reddy,
  Cournapeau, Burovski, Peterson, Weckesser, Bright, {van der Walt}, Brett,
  Wilson, Millman, Mayorov, Nelson, Jones, Kern, Larson, Carey, Polat, Feng,
  Moore, {VanderPlas}, Laxalde, Perktold, Cimrman, Henriksen, Quintero, Harris,
  Archibald, Ribeiro, Pedregosa, {van Mulbregt}, \& {SciPy 1.0
  Contributors}}]{2020SciPy-NMeth}
Virtanen, P., Gommers, R., Oliphant, T.~E., {et~al.} 2020, Nature Methods, 17,
  261, \dodoi{10.1038/s41592-019-0686-2}

\bibitem[{{Wang} {et~al.}(2017){Wang}, {Yang}, {Xu}, {Li}, {Su}, \&
  {Zhang}}]{2017ApJS..230....5W}
{Wang}, C., {Yang}, J., {Xu}, Y., {et~al.} 2017, \apjs, 230, 5,
  \dodoi{10.3847/1538-4365/aa6c6b}

\bibitem[{{Wang} {et~al.}(2023{\natexlab{a}}){Wang}, {Feng}, {Yang}, {Chen},
  {Su}, {Yan}, {Du}, {Ma}, \& {Cai}}]{2023AJ....165..106W}
{Wang}, C., {Feng}, H., {Yang}, J., {et~al.} 2023{\natexlab{a}}, \aj, 165, 106,
  \dodoi{10.3847/1538-3881/acafee}

\bibitem[{{Wang} {et~al.}(2023{\natexlab{b}}){Wang}, {Feng}, {Yang}, {Chen},
  {Su}, {Yan}, {Du}, {Ma}, \& {Cai}}]{2023AJ....166..121W}
---. 2023{\natexlab{b}}, \aj, 166, 121, \dodoi{10.3847/1538-3881/acebdd}

\bibitem[{{Wilson} {et~al.}(1970){Wilson}, {Jefferts}, \&
  {Penzias}}]{1970ApJ...161L..43W}
{Wilson}, R.~W., {Jefferts}, K.~B., \& {Penzias}, A.~A. 1970, \apjl, 161, L43,
  \dodoi{10.1086/180567}

\bibitem[{{Xiong} {et~al.}(2017){Xiong}, {Chen}, {Yang}, {Fang}, {Zhang},
  {Zhang}, {Du}, \& {Long}}]{2017ApJ...838...49X}
{Xiong}, F., {Chen}, X., {Yang}, J., {et~al.} 2017, \apj, 838, 49,
  \dodoi{10.3847/1538-4357/aa6443}

\bibitem[{{Xiong} {et~al.}(2019){Xiong}, {Chen}, {Zhang}, {Yang}, {Fang},
  {Zhang}, {Guo}, \& {Sun}}]{2019ApJ...880...88X}
{Xiong}, F., {Chen}, X., {Zhang}, Q., {et~al.} 2019, \apj, 880, 88,
  \dodoi{10.3847/1538-4357/ab2a70}

\bibitem[{{Yan} {et~al.}(2020){Yan}, {Yang}, {Su}, {Sun}, \&
  {Wang}}]{2020ApJ...898...80Y}
{Yan}, Q.-Z., {Yang}, J., {Su}, Y., {Sun}, Y., \& {Wang}, C. 2020, \apj, 898,
  80, \dodoi{10.3847/1538-4357/ab9f9c}

\bibitem[{{Yan} {et~al.}(2019){Yan}, {Yang}, {Sun}, {Su}, \&
  {Xu}}]{2019ApJ...885...19Y}
{Yan}, Q.-Z., {Yang}, J., {Sun}, Y., {Su}, Y., \& {Xu}, Y. 2019, \apj, 885, 19,
  \dodoi{10.3847/1538-4357/ab458e}

\bibitem[{{Yan} {et~al.}(2021{\natexlab{a}}){Yan}, {Yang}, {Sun}, {Su}, {Xu},
  {Wang}, {Zhou}, \& {Wang}}]{2021A&A...645A.129Y}
{Yan}, Q.-Z., {Yang}, J., {Sun}, Y., {et~al.} 2021{\natexlab{a}}, \aap, 645,
  A129, \dodoi{10.1051/0004-6361/202039768}

\bibitem[{{Yan} {et~al.}(2021{\natexlab{b}}){Yan}, {Yang}, {Yang}, {Sun}, \&
  {Wang}}]{2021ApJ...910..109Y}
{Yan}, Q.-Z., {Yang}, J., {Yang}, S., {Sun}, Y., \& {Wang}, C.
  2021{\natexlab{b}}, \apj, 910, 109, \dodoi{10.3847/1538-4357/abe628}

\bibitem[{{Yan} {et~al.}(2022){Yan}, {Yang}, {Su}, {Sun}, {Zhou}, {Xu}, {Wang},
  {Zhang}, \& {Chen}}]{2022AJ....164...55Y}
{Yan}, Q.-Z., {Yang}, J., {Su}, Y., {et~al.} 2022, \aj, 164, 55,
  \dodoi{10.3847/1538-3881/ac77ea}

\bibitem[{{Yan} {et~al.}(2024){Yan}, {Yang}, {Su}, {Sun}, {Zhang}, {Zhou},
  {Wang}, {Ao}, {Chen}, \& {Wang}}]{2024ApJ...968L..14Y}
---. 2024, \apjl, 968, L14, \dodoi{10.3847/2041-8213/ad509d}

\bibitem[{{Yang}(1999)}]{1999AcApS..19...55Y}
{Yang}, J. 1999, Acta Astrophysica Sinica, 19, 55

\bibitem[{Yang {et~al.}(2025)Yang, Yan, Su, Zhang, Zhou, Sun, Ao, Chen, \&
  Chen}]{scienceDBMWISP}
Yang, J., Yan, Q.-Z., Su, Y., {et~al.} 2025, {MWISP Data Release 1}, V1,
  \dodoi{10.57760/sciencedb.27351}

\bibitem[{{Yoda} {et~al.}(2010){Yoda}, {Handa}, {Kohno}, {Nakajima}, {Kaiden},
  {Yonekura}, {Ogawa}, {Morino}, \& {Dobashi}}]{2010PASJ...62.1277Y}
{Yoda}, T., {Handa}, T., {Kohno}, K., {et~al.} 2010, \pasj, 62, 1277,
  \dodoi{10.1093/pasj/62.5.1277}

\bibitem[{{Yuan} {et~al.}(2021){Yuan}, {Yang}, {Du}, {Liu}, {Zhang}, {Lin},
  {Sun}, {Yan}, {Ma}, {Su}, {Sun}, \& {Zhou}}]{2021ApJS..257...51Y}
{Yuan}, L., {Yang}, J., {Du}, F., {et~al.} 2021, \apjs, 257, 51,
  \dodoi{10.3847/1538-4365/ac242a}

\bibitem[{{Yuan} {et~al.}(2024){Yuan}, {Yang}, {Chen}, {Su}, {Zhang}, {Zhou},
  {Chen}, {Yan}, {Fang}, {Du}, {Sun}, {Wang}, \& {Xu}}]{2024AJ....167..207Y}
{Yuan}, L., {Yang}, J., {Chen}, X., {et~al.} 2024, \aj, 167, 207,
  \dodoi{10.3847/1538-3881/ad323a}

\bibitem[{{Zhang} {et~al.}(2020){Zhang}, {Yang}, {Xu}, {Chen}, {Su}, {Sun},
  {Zhou}, {Li}, \& {Lu}}]{2020ApJS..248...15Z}
{Zhang}, S., {Yang}, J., {Xu}, Y., {et~al.} 2020, \apjs, 248, 15,
  \dodoi{10.3847/1538-4365/ab879a}

\bibitem[{{Zhang} {et~al.}(2024){Zhang}, {Su}, {Chen}, {Fang}, {Yan}, {Zhang},
  {Sun}, {Wang}, {Feng}, {Ma}, {Zhang}, {Zhuang}, {Zhou}, {Chen}, \&
  {Yang}}]{2024AJ....167..220Z}
{Zhang}, S., {Su}, Y., {Chen}, X., {et~al.} 2024, \aj, 167, 220,
  \dodoi{10.3847/1538-3881/ad2fcb}

\bibitem[{{Zhou} {et~al.}(2025){Zhou}, {Yang}, {Sun}, {Yan}, {Yuan}, {Su},
  {Chen}, \& {Zhang}}]{2025arXiv250814547Z}
{Zhou}, X., {Yang}, J., {Sun}, Y., {et~al.} 2025, arXiv e-prints,
  arXiv:2508.14547, \dodoi{10.48550/arXiv.2508.14547}

\bibitem[{{Zhou} {et~al.}(2023){Zhou}, {Su}, {Yang}, {Chen}, {Sun}, {Jiang},
  {Wang}, {Wang}, {Zhang}, {Xu}, {Yan}, {Yuan}, {Chen}, {Ao}, \&
  {Ma}}]{2023ApJS..268...61Z}
{Zhou}, X., {Su}, Y., {Yang}, J., {et~al.} 2023, \apjs, 268, 61,
  \dodoi{10.3847/1538-4365/acee7f}

\bibitem[{{Zhuang} {et~al.}(2024){Zhuang}, {Su}, {Zhang}, {Chen}, {Yan},
  {Feng}, {Sun}, {Xu}, {Sun}, {Zhou}, {Wang}, \& {Yang}}]{2024ApJ...966..202Z}
{Zhuang}, Z., {Su}, Y., {Zhang}, S., {et~al.} 2024, \apj, 966, 202,
  \dodoi{10.3847/1538-4357/ad3552}

\end{thebibliography}





%

\end{document}